\documentclass[letterpaper,twocolumn,10pt]{article}
\usepackage{usenix}

\usepackage{amsmath}
\usepackage{setspace}
\usepackage{amsmath,amsfonts}
\usepackage{times}
\usepackage[T1]{fontenc}
\usepackage[utf8]{inputenc}
\usepackage{caption}
\usepackage{verbatim}
\usepackage{diagbox}
\usepackage{subfigure}
\usepackage{graphicx}
\usepackage{url}
\usepackage{float}
\usepackage{color}
\usepackage{multirow}
\usepackage{url}
\usepackage[normalem]{ulem}
\usepackage{setspace}
\usepackage{algorithm}
\usepackage{algorithmicx}
\usepackage[noend]{algpseudocode}
\usepackage[normalem]{ulem}
\usepackage{amssymb}
\usepackage{xparse}
\usepackage{mathtools}
\usepackage[utf8]{inputenc} 
\usepackage[T1]{fontenc}    

\usepackage{amsfonts}       
\usepackage{nicefrac}       
\usepackage{microtype}      

\usepackage{times}
\usepackage{soul}

\usepackage{subfigure}
\usepackage{amsmath}

\DeclareMathOperator*{\argmin}{argmin}

\usepackage{enumitem}
\setlist{nolistsep}
\usepackage{atbegshi}
\usepackage{filecontents}

\newcommand{\name}{\textsf{PoisHygiene}}

\begin{document}
\title{\name: Detecting and Mitigating Poisoning Attacks in Neural Networks}\vspace{-4mm}
\author{Junfeng Guo$^\dagger$, Ting Wang$^*$~and Cong Liu$^\dagger$\\
$^\dagger$The University of Texas at Dallas
$^*$ The Pennsylvania State University
\\
}

\maketitle

\begin{abstract}
The black-box nature of deep neural networks (DNNs) facilitates attackers to manipulate the behavior of DNN through data poisoning. Being able to detect and mitigate poisoning attacks, typically categorized into backdoor and adversarial poisoning (AP), is critical in enabling the safe adoption of DNNs in many application domains. Although recent works demonstrate encouraging results on the detection of certain backdoor attacks, they exhibit inherent limitations which may significantly constrain the applicability. Indeed, it is still under-studied on how to 
detect AP attacks, which represent a more daunting challenge given that such attacks exhibit no explicit rules compared to backdoor attacks (i.e., embedding backdoor triggers into poisoned data).

In this paper, we present \name{}, the first detection and mitigation framework against AP attacks. \name{} is fundamentally motivated by Dr. Ernest Rutherford's story (i.e., the 1908 Nobel Prize winner), on observing the structure of atom through random electron sampling. Similarly, \name{} crafts test inputs to a given DNN, seeking to reveal necessary internal properties of the decision region space belonging to each label and detect whether a label is infected. Through extensive implementation and evaluation of \name{} against a set of state-of-the-art AP attacks on four widely studied datasets, \name{} proves to be effective and robust under various settings considering complex attack variants. Interestingly, \name{} is also shown to be effective and robust on detecting backdoor attacks, particularly comparing to state-of-the-art backdoor detection methods including Neural Cleanse and ABS.
\end{abstract}

\section{Introduction}

Deep Neural Networks (DNNs) are being deployed in a myriad of application domains, such as image classification~\cite{lecun1995convolutional}, object recognition~\cite{parkhi2015deep}, and security-critical ones including binary reverse engineering~\cite{chua2017neural}, malware detection~\cite{wang2017adversary, cannady1998artificial}, and autonomous driving~\cite{chen2015deepdriving}. 
Unfortunately, the black-box nature of DNN models may enable attackers to manipulate the model through data poisoning, which seeks to manipulate the behavior of DNN via inserting poisoned samples during training~\cite{jagielski2018manipulating,shafahi2018poison,gu2017badnets,chen2017targeted,liu2017trojaning}.
Poisoning attacks have 
attracted intensive recent attention from
both academia and industry, which can be generally categorized into backdoor attacks~\cite{liu2017trojaning,gu2017badnets,chen2017targeted, clements2018hardware,yao2019latent} and adversarial poisoning (AP) attacks~\cite{shafahi2018poison,suciu2018does, jagielski2018manipulating,selvaraju2017grad} (as also discussed in~\cite{wang2019neural}). A key difference is that backdoor attacks seek to attach specific triggers to training data, which override normal classification to produce incorrect prediction results; while AP attacks aim at either simply mislabeling training data or inserting poisoned instances containing manipulated features that belong to a class different from the original one (note that such features do not need to form a unified pattern as required by the trigger implementation under backdoor attacks). 


Compared with the intensive research on poisoning attacks, it is still under-studied on how to detect such attacks
~\cite{chen2017targeted,wang2019neural}. A recent set of works focus on detecting backdoor attacks~\cite{ma2019nic,selvaraju2017grad,liu2017neural,chen2018detecting}), yet assuming the defender can access the original poisoned instances used during training, which is unfortunately impractical~\cite{chen2017targeted,gu2017badnets,liu2017trojaning}. 
Recent works, i.e., Neural Cleanse~\cite{wang2019neural} and ABS~\cite{liu2019abs}, take significant steps in the direction of detection on backdoor attacks in the image classification domain, which do not require such strong defender capability.
The intuitive ideas behind Neural Cleanse~\cite{wang2019neural} and ABS~\cite{liu2019abs} are through observing and exploiting properties unique to certain backdoor attack techniques (backdoor trigger size for Neural Cleanse and compromised neurons representing the features of backdoor triggers for ABS), thus constraining their applicability to AP attacks.

We believe for any detection method on AP attacks to be effective and robust, it has to be capable of observing and leveraging unique poisoning-induced properties within an infected DNN model (instead of properties specific to certain attack techniques), which is clearly challenging. The challenges are due to the fact that it is rather difficult to observe internal properties of DNN (infected or not) due to its black-box nature and to leverage such properties in developing detection methods against AP attacks which do not exhibit any common and explicit rules in the crafted poisoned data (unlike backdoor triggers).

In this paper, we present \name, the first detection and mitigation framework against AP attacks. \name{} detects infected DNNs via observing and exploring an essential property of any infected DNN model due to poisoning, i.e., the existence of a poisoned region w.r.t. each infected label. \name{} is fundamentally motivated by Dr. Ernest Rutherford's story (i.e., the 1908 Nobel Prize winner)~\cite{Nobel_Prize}, on seeking the structure and content of atom through random electron sampling. Similarly, a key design rationale behind \name{} is to craft random samples (sampled from Gaussian distribution) as inputs to a given DNN, seeking to reveal essential internal properties of the decision region space belonging to each label.
Intuitively, for each label, \name{} smartly crafts inputs which could reach the poisoned region (if one exists), through manipulating and evaluating the corresponding loss functions (e.g., cross entropy loss).


Through extensive implementation and evaluation of \name{} against three state-of-the-art AP attacks~\cite{shafahi2018poison, suciu2018does} (\cite{suciu2018does} proposes two AP attacks) on four widely studied datasets, \name{} proves to be effective and robust under various settings. Rather interestingly, \name{} is also shown to be effective against state-of-the-art backdoor attacks~\cite{gu2017badnets,liu2017trojaning,chen2017targeted}. Such efficacy is particularly demonstrated via direct comparison against Neural Cleanse~\cite{wang2019neural} and ABS~\cite{liu2019abs} on detection against backdoor attacks for several complex attack variants. Such generality is intuitive as \name{}'s detection criteria are developed through exploring essential properties of any poisoned region, which may hold for both AP- and backdoor-infected labels.

In summary, this paper makes the following contributions.
\begin{itemize} 
\item We present effective detection and mitigation techniques against state-of-the-art AP attacks, which are proved to be effective by extensive experiments.
\item Evaluation also demonstrates the efficacy of \name{} against existing backdoor attacks, particularly comparing to the state-of-the-art backdoor detection methods Neural Cleanse~\cite{wang2019neural} and ABS~\cite{liu2019abs} (e.g., under the multi-infected-label setting). 
\item \name{} is shown to be robust when considering a set of complex attack variants under both AP and backdoor attacks. \name{} is also shown to be resilient against attacks that are fully aware of the specific defense mechanisms used by \name{}.
\end{itemize}

To the best of our knowledge, this work is the first to develop robust detection and mitigation techniques against AP attacks (e.g., Sting Ray~\cite{suciu2018does} and Poison Frog~\cite{shafahi2018poison}), as well as backdoor attacks with complex attack variants (e.g, with multiple infected labels), which could not be resolved using any existing defense techniques. 



\vspace{-1mm}
\section{Background and Related Work}
\label{sec:background}

Due to a lack of transparency, DNNs are often treated as black-box systems.
A DNN model takes an input (e.g. an image) and then gives a prediction label after performing a series of internal computations.

\noindent\textbf{Poisoning attacks.} Poisoning attacks on DNNs are typically categorized into backdoor attacks and AP attacks~\cite{wang2019neural}, both of which have received a significant amount of recent attention~\cite{wang2019neural,liu2018fine,liu2017trojaning,shafahi2018poison,steinhardt2017certified,chendeepinspect,yang2017generative}.
Backdoor attack on DNNs is defined to be training a backdoor trigger (e.g., a small white square) into a DNN, causing anomalous behaviors when the specific trigger is attached into a test-time input.~\cite{gu2017badnets,liu2017trojaning,chen2017targeted}. 
An advantage of backdoor attacks is that they do not affect overall classification accuracy on clean input data.
On the other hand, AP attack techniques seek to maliciously train a DNN using data of certain classes whose features are inconsistent with their assigned labels. Thus, at inference time, the classification result on unmodified test data of those specific classes would be incorrect (e.g., a stop sign would be misclassified as a speed limit sign )~\cite{shafahi2018poison,suciu2018does}. 
AP attacks can be further categorized into integrity attack~\cite{shafahi2018poison,suciu2018does} and availability attack~\cite{selvaraju2017grad,jagielski2018manipulating,cao2018efficient}.
An integrity attack aims at causing specific mispredictions at inference time while preserving the overall accuracy performance. An availability attack seeks to lower the overall accuracy performance. 

Regarding AP attacks, this paper focuses on integrity attacks since availability attacks would cause the infected model's performance to significantly drop, thus making the detection problem non-significant under our adversarial model (see Sec.~\ref{sec:background}).
In practice, AP attacks could be rather threatening and easily implementable compared to backdoor attacks~\cite{gu2017badnets,chen2017targeted,liu2017trojaning}, due to (\textit{i}) backdoor attacks typically exhibit the same shortcomings as evasion attacks~\cite{ji2018model,goodfellow2014explaining} where test-time instances are required to be modified, and (\textit{ii}) technically, attackers could simply mislabel data during training to implement effective AP attacks~\cite{suciu2018does}.

\vspace{1mm}\noindent\textbf{Related Works on AP and Backdoor attacks.} AP attack is considered to be a practical and easily implementable type of poisoning attacks in reality. AP attack requires fewer assumptions regarding the attacker's capability compared to backdoor attacks~\cite{suciu2018does,shafahi2018poison} while preserving overall high accuracy on normal test-time inputs. Besides the common implementation of generating attacks through mislabeling training data ~\cite{suciu2018does}, recent works on performing clean-label attacks which do not even require attackers to mislabel training data, including StingRay~\cite{suciu2018does} and Poison Frog~\cite{shafahi2018poison}. StingRay generates poisoning data that collides with adversarial images in the feature space, achieving $>98\%$ attack success rate. Motivated by StingRay, Poison Frog~\cite{shafahi2018poison} seeks to add perturbations containing features of the targeted infected label to yield poisoned data instances, which appear to be visually-indistinguishable from the corresponding original images. 
Poison Frog can achieve $>99\%$ attack success rate under a more restricted attack model. The clean-label property clearly makes such AP attacks be more threatening in practice.

Regarding backdoor attacks, a recent set of attacks have been proposed, including BadNets~\cite{gu2017badnets}, Trojan Attack~\cite{liu2017trojaning}, and Chen et al attack (denoted as “Chen Attack” throughout this paper)~\cite{chen2017targeted}, which assume a stronger adversarial capability compared to AP attacks. BadNets~\cite{gu2017badnets} requires attackers to access the training dataset and inject poisoned data attached with arbitrary triggers. The trojan attack, which doesn't require attackers to access the training dataset, can achieve high attack success rate $>98\%$ with fewer inserted poisoned instances. Chen attack, which is built under an adversarial model with weaker adversarial capability, injects random noise with certain transparency ratio as the backdoor trigger which overlaps with the entire image, can achieve $>99\%$ attack success rate with a few poisoned data instances. Yao~\cite{yao2019latent} proposes a latent backdoor attack, which inserts a hidden trigger into a teacher model and this trigger can then be inherited by the student model through transfer learning, which does not apply to our attack model. 

\vspace{1mm}\noindent\textbf{Existing detection methods against poisoning.} A very recent set of works have been proposed to detect and mitigate backdoor attacks without accessing the original training dataset~\cite{liu2018fine,8119189,wang2019neural}. A fine-pruning method~\cite{liu2018fine} was first proposed to remove backdoor by pruning neurons, which imposes trivial impact on prediction results given clean test inputs. Neural Cleanse~\cite{wang2019neural}, however, reports that this fine-pruning method may significantly reduce the classification accuracy on the GTSRB dataset. Neuron Trojan~\cite{8119189} presents three computation expensive methods to defend backdoor attacks, yet with rather limited evaluation only using the MNIST dataset. Note that both Fine-Pruning and Neuron Trojan focus on mitigation but not detection as they assume that a given model is already known to be poisoned. 

Indeed, the problem of detecting poisoning attacks is significant and challenging. Trojan Attack~\cite{liu2017trojaning} provides broad discussions on several possible detection methods against backdoor attacks; yet Chen et al.~\cite{chen2017targeted} show that a variety of these detection methods may fail in practice. To the best of our knowledge, Neural Cleanse~\cite{wang2019neural} and ABS~\cite{liu2019abs} represent the state-of-the-art detection and mitigation techniques which can effectively detect and mitigate certain backdoor attacks (i.e. BadNets and Trojan Attack) for image classification models.\footnote{Note that there exist a couple of recent works on detecting backdoor attacks ~\cite{chendeepinspect,guo2019tabor}, which are incrementally built upon Neural Cleanse, thus exhibiting the same limitations and inabilities as Neural Cleanse in detecting backdoor and AP attacks. Also, insufficient evaluation is seen in the manuscripts.} 
Unfortunately, such detection methods exhibit limitations which may significantly constrain the applicability to AP attacks and backdoor attacks using more complex attack variants (e.g., failing to detect multiple infected labels as shown in our evaluation).

Regarding defenses against AP attacks, prior works mainly focus on sanitizing the poisoned training set and filtering poisoned samples~\cite{cao2018efficient,jagielski2018manipulating,rubinstein2009antidote,mozaffari2014systematic,steinhardt2017certified}. However, such works are not applicable to our adversarial model (see Sec~\ref{sec:detectorModel}) where the detector shall not have access to the original training set which may contain poisoned instances. Under our more realistic adversarial model, unfortunately, there does not exist any detection and mitigation methods against AP attacks. As briefly discussed in~\cite{wang2019neural}, a fundamental challenge due to AP attacks is that different from backdoor attacks, AP attacks implement no unified rules (e.g., poisoned samples sharing the same trigger or common image properties on crating adversarial training data. Such unified rules are fundamentally observed and utilized in Neural Cleanse~\cite{wang2019neural} and ABS ~\cite{liu2019abs}, which perform a reverse engineering-based method that could identify the backdoor trigger.
However, such observations do not hold for AP attacks because the poisoned training instances may not share any common feature. The poisoned training instances can be any images belonging to different classes and may not exhibit any feature in common (e.g., pixel, size, shape, etc.).

\section{Adversarial Model}
\label{sec:detectorModel}


\noindent \textbf{Attack Model.} Our attack model is consistent with that of Neural Cleanse and ABS~\cite{wang2019neural,liu2019abs}. Specifically, the attacker is assumed to fully control the DNN model, and be able to insert any type of poisoned data into the training dataset to achieve high attack success rate.
A user may obtain such a DNN model which exhibits state-of-the-art performance yet being infected by backdoor or AP attacks. Due to the fact that the poisoned data was used at the training phase, the user and the defender do not have access to such data. Any infected DNN model still performs generally well on clean test data yet exhibits targeted mis-classification when being tested with certain adversarial inputs, e.g., inputs of certain classes under adversarial poisoning or inputs being attached with a backdoor trigger, which guarantees the evasiveness of the vulnerabilities upon such infected models. 

A label (class) is considered infected if certain inputs (either with backdoor triggers due to backdoor attacks or without any unique trigger due to AP attacks) cause targeted misclassification to that label.
In practice, there may exist multiple infected labels in a DNN model.

\vspace{1mm}\noindent \textbf{Defense Assumptions and Goals.} 
We provide a detailed defender model against poisoning attacks, which is consistent with Neural Cleanse~\cite{wang2019neural} and ABS~\cite{liu2019abs} as well. The model consists of defining defender's goal, defender's knowledge and capability. 
\begin{table}[!t]
\centering
\scalebox{0.6}{
\begin{tabular}{ c c c  } 
\hline
Dataset  & Required training Data (proportion over the entire dataset)&Source Task\\
\hline\hline
Fashion-MNIST&$20\%$&MNIST\\
GTSRB&$22\%$&ImageNet\\
CIFAR-10&$26\%$&ImageNet\\
\hline\hline
\end{tabular}}
\caption{The amount of training data needed for building an effective pre-trained model for detection purpose through leveraging the model pre-training technique~\cite{pretrained}.}
\label{table:pretraining}
\vspace{-4mm}
\end{table}

\textit{Defender's Goal.}
The defender has two specific goals. (\textit{i}) \textbf{detecting vulnerabilities:} the defender needs to determine whether a given DNN model is infected by poisoning attacks, and more specifically, identify the infected labels, (\textit{ii}) \textbf{mitigating vulnerabilities:} the defender needs to ``clean'' an infected model and remove any poisoning-induced vulnerabilities without affecting overall performance on clean test inputs.

\textit{Defender's Knowledge and Capability.} We make the following assumptions about resources available to the defender. 
We assume that the defender has access to the DNN to be analyzed (targeted DNN) and another set of pre-trained DNNs which perform the same task as the targeted DNN. Such pre-trained DNNs are not required to yield state-of-the-art accuracy; rather, our detection methodology works well using pre-trained DNNs exhibiting low accuracy. For instance, for the MINIST and GTSRB datasets, leveraging pre-trained models with an accuracy of merely 60\% and 76\% suffices, respectively, given the corresponding state-of-the-art accuracy being 99\% and 98\% (details on other tasks are given in the Appendix). Such DNN models with low accuracy can be easily obtained via several ways. The defender could self-train such a model using partial training data.\footnote{The amount of required clean data to self-train a  model for our detection purposes varies for different tasks. For instance, for MNIST task, we can self-train such models with only $1.5\%$ of the original dataset. As for more complicated datasets including CIFAR-10, GTSRB, and Fashion-MNIST, $40\%$, $30\%$ and $27\%$ of the original dataset is required, respectively.} Moreover, defender can utilize certain few-shot learning techniques (model-agnostic meta learning~\cite{MAML,reptile}, model pre-training~\cite{pretrained}, transfer learning~\cite{transferlearning}) to obtain the required pre-trained models with minimal effort. For example, we can train a model achieving $82.5\%$ accuracy on CIFAR-10 dataset with only $26\%$ of the entire dataset by leveraging the model pre-training technique~\cite{pretrained}. The amount of training data needed for bailing such a pre-trained model on other tasks are shown in Table.~\ref{table:pretraining}. Furthermore, for most tasks, popular deep learning frameworks provide a set of pre-trained models. For instance, Keras~\cite{chollet2015}, one of the most popular deep learning frameworks, provides many pre-trained models for different tasks, e.g., Inception and Resnet for image classification. Last but not the least, a latest set of semi-supervised learning techniques, e.g., Google's mixed-match~\cite{mixmatch}, could significantly reduce the amount of required labeled data to self-train a model with high accuracy. For instance, obtaining an over 90\% accuracy model on CIFAR-10 only requires 2000 labeled images (i.e., merely 4\% of the original dataset).


On the other hand, the defender may not have access to the training configuration and the original training dataset which may contain poisoning instances. In practice, the original training process can totally be a black box to both users and defenders. Moreover, if the DNN model is infected, the corresponding kind of poisoning technique shall be unknown to the defender. 

\vspace{-2mm}
\section{Design of \name{}}

\vspace{-2mm}
\subsection{Design Rationale}
\label{sec:rationale}

\begin{figure}[!t]
\includegraphics[width=1\columnwidth]{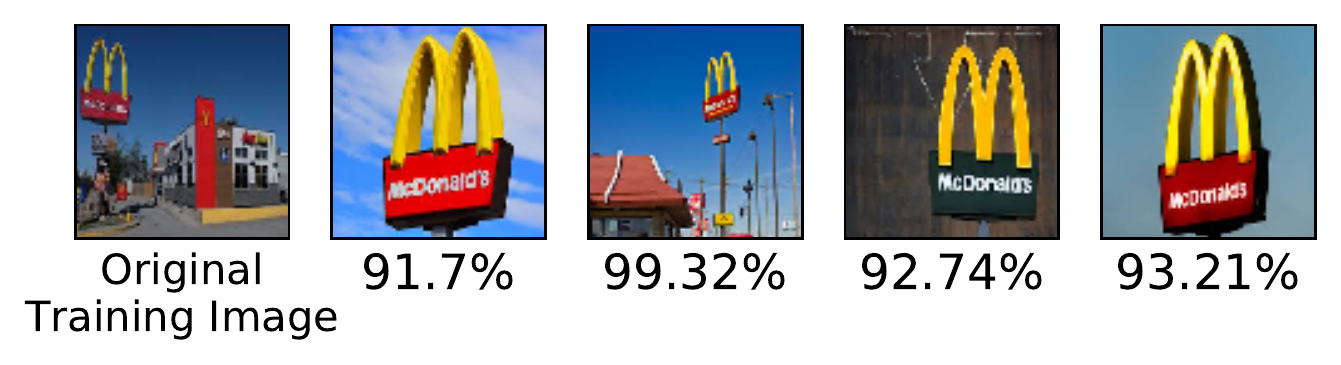}
\includegraphics[width=1\columnwidth]{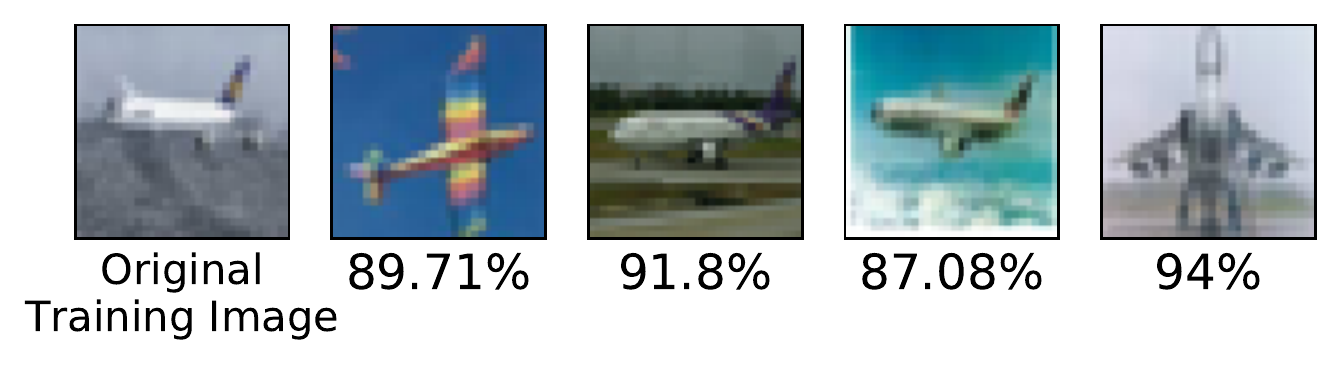}
\centering
\vspace{-6mm}
\caption{Example test inputs for GTSRB containing McDonalds sign and airplanes. The first column shows a poisoned sample from the original training dataset, which will clearly be misclassified by models infected due to AP attacks (i.e., mis-label~\cite{suciu2018does}). The other four columns show images of the same class, which do not belong to the original training dataset, yet also being misclassified by the same model.}
\label{fig:adversarial_vast}
\vspace{-1mm}
\end{figure}

Our design of \name{} is motivated by observing the essential properties of an infected model due to either AP or backdoor attacks. Specifically, either AP or backdoor attacks would misclassify certain inputs as infected labels different from their true labels. In the context of classification tasks, the decision region of an infected label contains not only healthy region but also poisoned region. Note that similar concepts have also been 
discussed in prior related works~\cite{wang2019neural,liu2017trojaning}. Nonetheless, the observation on such poisoned region has never been deeply explored towards detection purposes. 
For instance, previous works on designing backdoor attacks\cite{gu2017badnets,liu2017trojaning,chen2017targeted} have revealed similar observations. For a backdoor-infected model, using images containing backdoor triggers as inputs could cause the model to misclassify on the infected label, even if the image is irrelevant to the training datasets. This may also apply to AP attacks, where inputs that are not seen in the training set can be misclassified as the infected label as shown in Fig.~\ref{fig:adversarial_vast}, possibly due to the strong generalization capability of neural networks~\cite{geirhos2018generalisation}. 

\begin{figure}[!t]
\centering
\subfigure[Decision Region of a healthy label\label{fig:decision_region_healthy}]{\includegraphics[width=0.45\columnwidth]{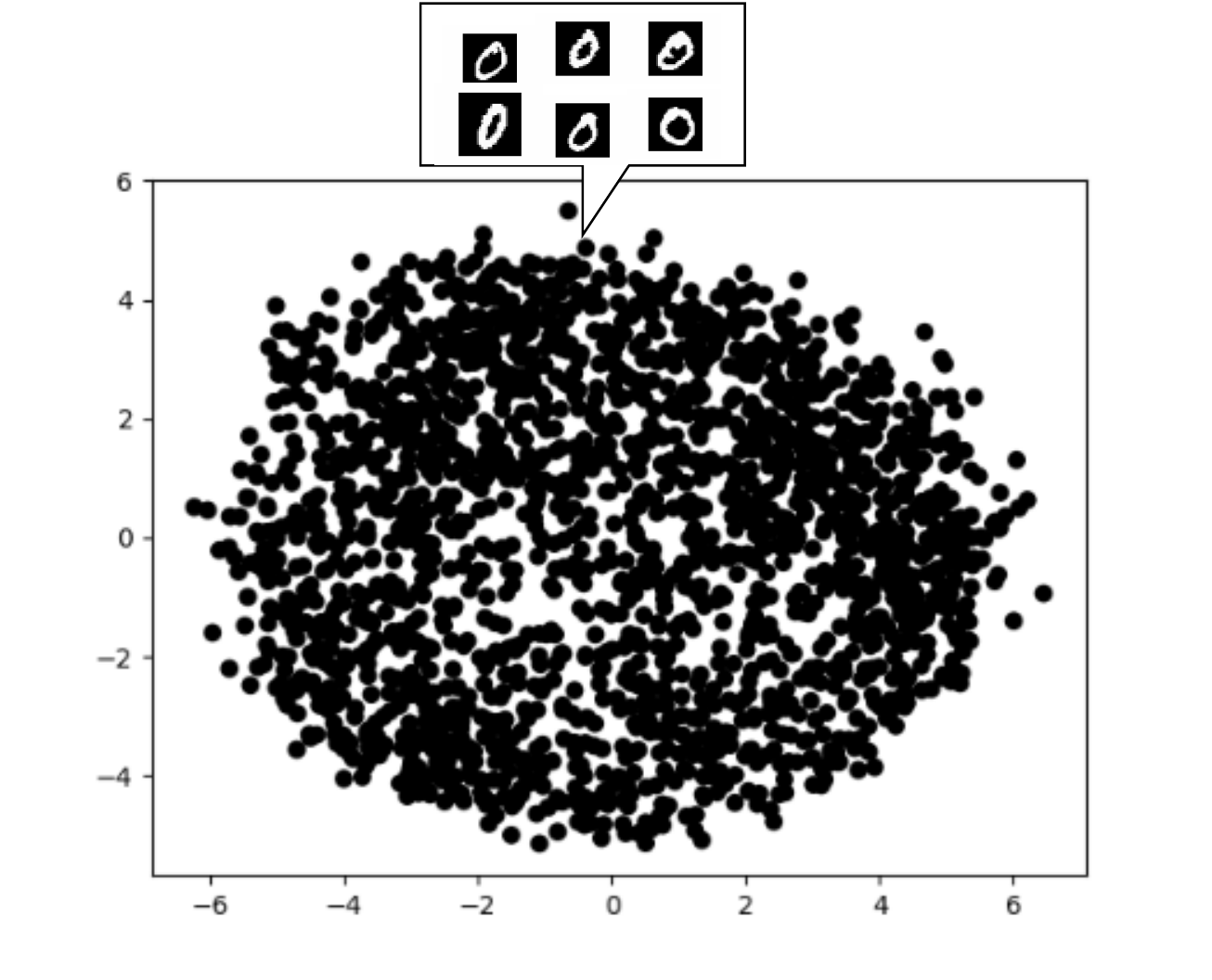}}
\subfigure[Decision Region of the same yet infected label\label{fig:decision_region_infected}]{\includegraphics[width=0.45\columnwidth]{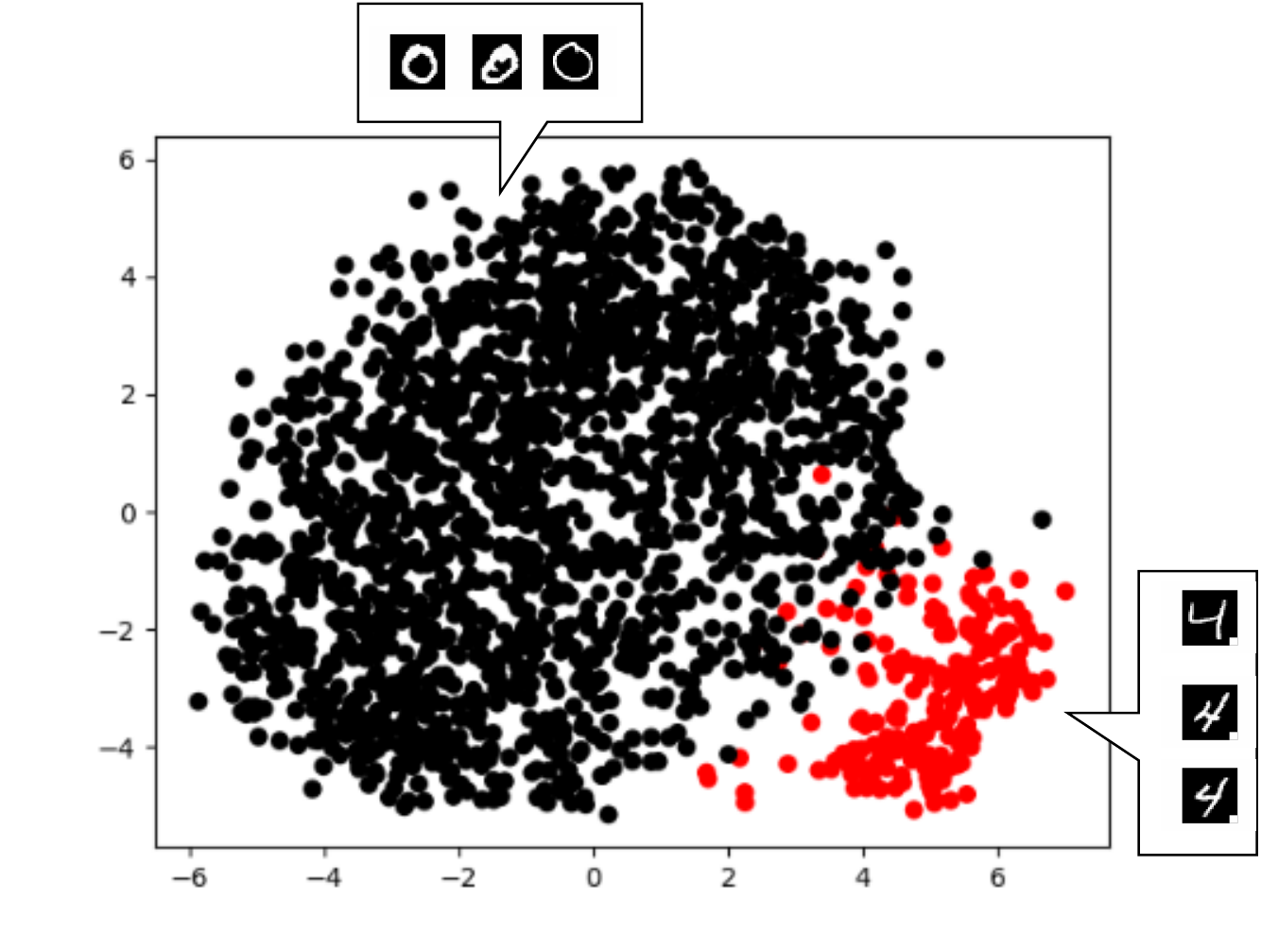}}
\vspace{-2mm}
\caption{Intuitive illustration of a 2-dimensional decision region under (a) a healthy label 0, and (b) an infected label 0 for MNIST. The black and red circles represent healthy inputs of digit 0 and poisoned inputs of digit 4, respectively.}
\vspace{-1mm}
\label{fig:decision_region}
\end{figure}  

We intuitively illustrate this concept in Fig.~\ref{fig:decision_region}, where we map 
28*28 images into a 2-dimensional space using the PAC algorithm~\cite{PCA} for MNIST. The left and right subfigure represent the region of a healthy and an infected label 0, respectively. As seen in the figure, for an infected label, besides the original healthy region, it also contains a poisoned region denoted $\mathcal{P}$. 
 
Our detection method is established through investigating whether it is possible to detect the existence of such a poisoned region for each label under a given model. We now illustrate the key design rationale supported by both analytical reasoning and extensive empirical evidence. 

\vspace{1mm}\noindent \textbf{Design Rationale: 
Let $\mathcal{L}_{\mathcal{H}}$ and $\mathcal{L}_{\mathcal{P}}$ represent the loss function (e.g., cross entropy loss) computed under a healthy decision region $\mathcal{H}$ and a poisoned decision region $\mathcal{P}$ of $y_t$ under an infected model $T$, respectively. $x$ represents a test input  with $y_h$ as its correct label, which will be misclassified as label $y_t$ under $T$ due to $x \in \mathcal{P}$ (assuming such an $x$ exists). A key finding is that  $\mathcal{L}_{\mathcal{P}}(x,y_t)$ is often very small, while $\mathcal{L}_{\mathcal{H}}(x,y_t)$ can be significantly large.}

The above rationale is intuitive. 
Since the cross entropy loss function is widely used to measure the prediction error in classification tasks, $\mathcal{L}_{\mathcal{H}}(x,y_t)$ should be large since $x \notin \mathcal{H}$. On the other hand, since $x \in \mathcal{P}$, $\mathcal{L}_{\mathcal{P}}(x,y_t)$ shall be much smaller. This rationale is also supported by extensive empirical results (to be discussed in detail in Sec.~\ref{sec:Implementaion}). 

The above rationale suggests that it may be possible to detect the existence of a poisoned region through crafting test inputs by evaluating the resulting $\mathcal{L}_{\mathcal{P}}(x,y_t)$ and $\mathcal{L}_{\mathcal{H}}(x,y_t)$ within an infected model $T$.  Specifically, for each label $y_t$ to be analyzed, \name{} develops a method which smartly crafts an input  $x$ which could reach the poisoned region (indicated by a small $\mathcal{L}_{\mathcal{P}}(x,y_t)$) while being pushed away from the corresponding healthy decision region (indicated by a large $\mathcal{L}_{\mathcal{H}}(x,y_t)$). This is done through manipulating the following optimization:
\begin{equation}\label{eq:optimization}
x = ~\argmin\limits_{x}~\mathcal{L}_{\mathcal{P}}(x,y_t)- \lambda*\mathcal{L}_\mathcal{H}(x,y_t)
\end{equation}

\begin{table}[!t]
\centering
\scalebox{0.7}{
\begin{tabular}{ c c c c c c c  } 
\hline
Dataset  & loss value under Model $T$&loss value under Model $T_h$\\
\hline\hline
MNIST(StingRay)&0.03&46.41\\
MNIST(Poison Frog)&0.1&39.26\\
MNIST(Mislabel)&0.12&43.56\\
Fashion-MNIST(StingRay)&0.12&43.21\\
Fashion-MNIST(Poison Frog)&0.11&41.21\\
Fashion-MNIST(Mislabel)&0.17&43.1\\
GTSRB(StingRay)&0.3&64.34\\
GTSRB(Poison Frog)&0.3&64.34\\
GTSRB(Mislabel)&0.09&64.34\\
\hline\hline
\end{tabular}}
\caption{$\mathcal{L}_{T}(x,y)$ and $\mathcal{L}_{T_h}(x,y)$ on experimental datasets.}
\label{table:loss_back}
\vspace{-3mm}
\end{table}

\subsection{Implementation of \name{}}
\label{sec:Implementaion}
Although the above design rational is intuitive and promising, there is an essential implementation challenge. As seen in Eq.~\ref{eq:optimization}, a key step of applying this optimization is to define and calculate $\mathcal{L}_{\mathcal{H}}$ and $\mathcal{L}_{\mathcal{P}}$. Unfortunately, due to the uninterpretability of DNNs, it is notoriously hard to precisely characterize  the decision region under each label~\cite{fawzi2017classification}. Also it is difficult (if not impossible) to identify and separate the healthy and poisoned regions in any concrete or quantifiable manner~\cite{fawzi2017classification}. Last but not the least, a method is needed to craft input $x$ which follows the above-discussed design rationale.

To resolve the challenge, our idea is to exploit a healthy pre-trained model $T_h$, which inherits the architecture of $T$. As discussed in Sec.~\ref{sec:detectorModel}, $T_h$ can be easily obtained via several different manners as we do not require state-of-the-art accuracy on any such $T_h$.
 The intuition of exploring $T_h$ is that the healthy decision region of $T_h$ would largely overlap with the healthy decision region of $T$ for each label. This is intuitive since $T_{h}$ and $T$ share the same architecture, both of which can achieve reasonable accuracy performance on a large amount of test data. This intuition is also supported by the following empirical evidence.

For AP attacks, we used the MNITS, Fashion-MNITS and GTSRB datasets. We randomly selected a label $y$ as the infected label  and trained a healthy model $T_h$ using  partial clean dataset, achieving $67\%$, $71\%$, $76\%$ accuracy for each dataset. We then retrain model $T_h$ using the corresponding entire dataset yet containing poisoned inputs to obtain the infected model $T$. Note that $T$ yields state-of-the-art accuracy performance of $99\%$, $94\%$, $97\%$, respectively. $T$ can also achieve a $\geq97\%$ attack success rate  following various AP attacks including StingRay, Poison Frog, and Mislabel.   
In our experiments, we first craft 1000 different inputs $x$ with true label $y_h$  which are misclassified as label $y$ by model $T$. We then calculate $\mathcal{L}_{T_h}(x,y)$ and $\mathcal{L}_{T}(x,y)$ for each of these 1000 inputs, and record the largest loss value among all inputs, as shown in  Table.~\ref{table:loss_back}, where $\mathcal{L}_{T_h}$ and $\mathcal{L}_{T}$ represent the cross entropy loss value computed under model $T_h$ and $T$, respectively. We observe that the loss value under the healthy model $T_h$ is significantly large (e.g., $\geq41$) while the loss values under the infected model $T$ are all close to 0. 

The same observation holds for backdoor attacks, as shown in Table~\ref{table:Loss_Backdoor} in the Appendix, where we similarly train a healthy and an infected version of the same model, following the three state-of-the-art backdoor attacks, BadNets~\cite{gu2017badnets},Trojan Attack~\cite{liu2017trojaning}, and Chen Attack~\cite{chen2017targeted}.

We thus propose to use the healthy decision region of $T_h$ to indicate that of $T$, thus being able to independently represent $T_h$'s healthy decision region and $T$'s poisoned decision region (if any).
Through utilizing model $T_h$, we optimize the following objective instead of directly optimizing the original Eq.~(\ref{eq:optimization}):
\begin{equation}\label{eq:newoptimization}
x=\argmin\limits_{x}~\mathcal{L}_T(x,y_t)- \lambda*\mathcal{L}_{T_h}(x,y_t) 
\end{equation}


\noindent This alternative optimization suffices because optimizing Eq.~(\ref{eq:newoptimization}) seeks to craft an $x$ that is ``far away'' from the healthy decision region under $T_h$ (indicated by a large $\mathcal{L}_{T_h}(x,y_t)$) and thus that of $T$  (which would also imply a large $\mathcal{L}_{\mathcal{H}(x,y_t)}$) due to the second term in this equation, 
while ensuring to reach the poisoned region (if any) under $T$ (indicated by a small $\mathcal{L}_{\mathcal{T}}(x,y_t)$ which would also imply a small $\mathcal{L}_{\mathcal{P}(x,y_t)}$). This complies with our design rationale as discussed in Sec.~\ref{sec:rationale}. 

Specifically, the optimization function Eq.~(\ref{eq:newoptimization}) has two objectives. For any (potentially infected) label $y_t$ to be analyzed, the first objective is to lead the crafted input $x$ into the poisoned region under $T$ (if any) belonging to $y_t$. The second objective is to push $x$ away from the healthy decision region belonging to $y_t$ under $T_h$, thus that of $T$. Note that we introduce a co-efficient $\lambda$ attached to the second objective to control the weight of each term in Eq.~(\ref{eq:newoptimization}) and thus the optimization procedure. Thus, by optimizing Eq.~(\ref{eq:newoptimization}), we aim at directing $x$ into the poisoned region of $T$ (if one exists) while being away from the corresponding healthy decision region, which clearly follows our design rationale discussed in Sec.~\ref{sec:rationale}.

\begin{algorithm}[!t]
\caption{Pseudo-code for Optimizing Eq.~\ref{eq:newoptimization}}
\begin{algorithmic}[1]
\State \textbf{Input}: random sample $x$, given label $y_t$
\State Define:$L1=-\mathcal{L}_T(x,y_t)$
\State Define:$L2=\mathcal{L}_T(x,y_t)-\lambda*\mathcal{L}_{T_h}(x,y_t)$

\While{$\mathcal{L}_T(x,y_t)<=3$}
\State $x \gets x -\alpha*\bigtriangledown_{x}L1(x)$ 
\EndWhile

\While{$i \leq MaxIters$}
\State $x \gets x-\alpha*\bigtriangledown_x L2(x)$ 
\If{$\mathcal{L}_T(x,y_t)<=\beta$ AND $\mathcal{L}_{T_h}(x,y_t) >=\gamma$}
\State BREAK
\EndIf
\EndWhile
\If{$\mathcal{L}_T(x,y_t)<=\beta$ AND $\mathcal{L}_{T_h}(x,y_t) > =\gamma$}
\State Sample x reaches the poison decision region
\Else 
\State Sample x does not reach the poison decision region
\EndIf
\end{algorithmic}
\label{algorithm:detect_algorithm}
\end{algorithm}

\vspace{1mm}

\vspace{1mm}\noindent\textbf{Optimization procedure and detection criteria.} The detailed optimization procedure including how to craft $x$ as well as defining the detection criteria is shown in Algorithm~ \ref{algorithm:detect_algorithm}, which contains three major steps. In the first step (Lines 4-5), we use gradient-descent to filter the crafted input $x$ with a learning rate $\alpha$ as 0.01, which ensures $x$ to be initialized beyond both poisoned and healthy decision regions for label $y_t$ under model T. Forcing $\mathcal{L}_T(x,y_t)\geq3$ shall indicate that the crafted x is sufficiently kept away from the decision region of $y_t$, because $\mathcal{L}_T(x,y_t)\geq3$ implies that the confidence of predicting $x$ as $y_t$ is smaller than 0.1. 
This constraint on crafting $x$ is needed because for otherwise, $x$ may easily (yet incorrectly) satisfy the detection criteria during early iterations of later optimization step (Lines 6-9). 
 The second step (Lines 6-9) is simply a gradient-descent update which seeks to craft an $x$ that minimizes Eq.~(\ref{eq:newoptimization}) with a learning rate $\alpha$. The third step (Lines 10-13) is used to identify whether the crafted sample $x$ successfully reaches the poisoned region under $T$.   

\begin{figure}[!t]
\centering
\subfigure[Infected Label\label{fig:infe}]{\includegraphics[width=0.48\columnwidth]{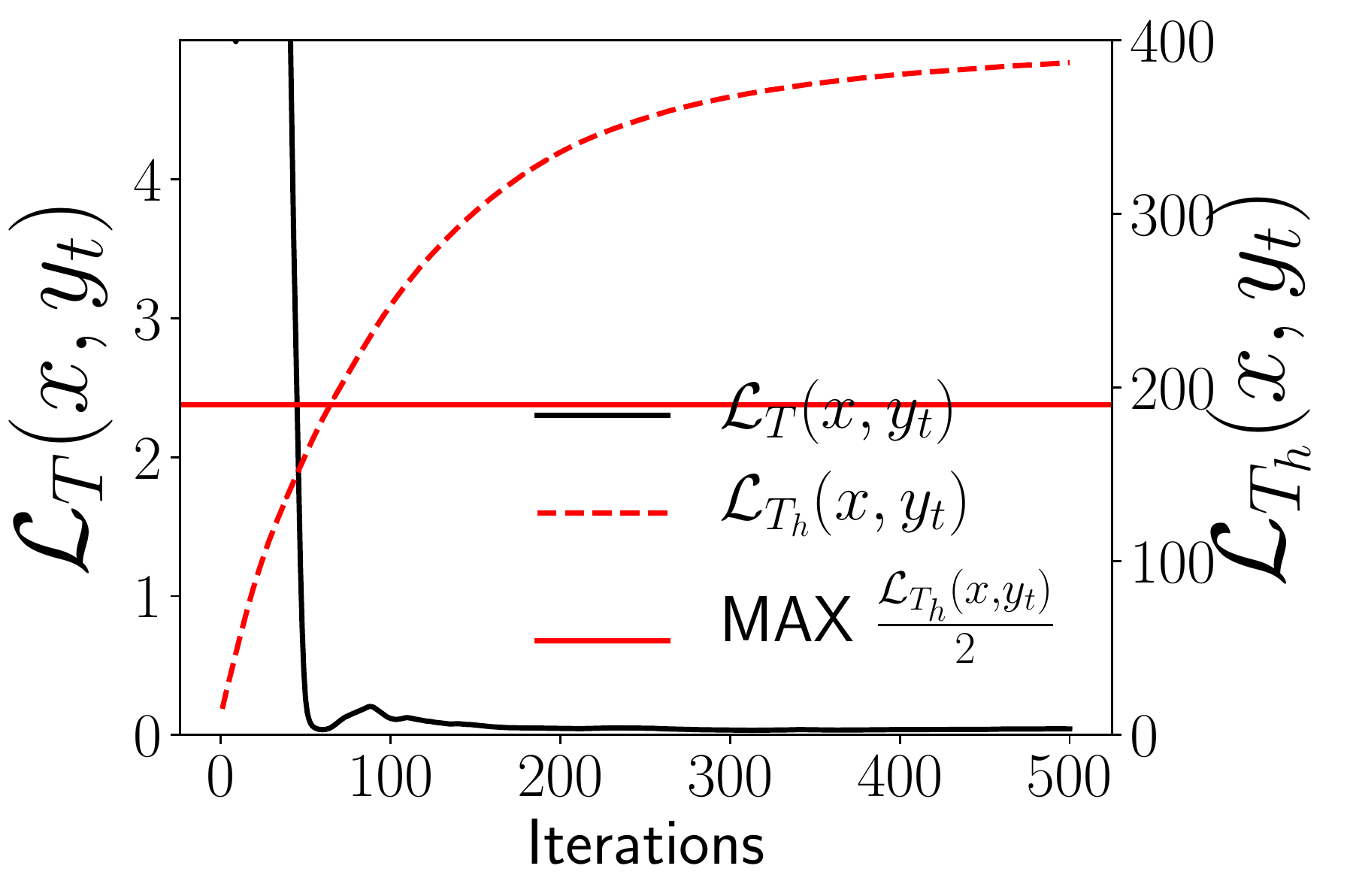}}
\subfigure[Healthy Label\label{fig:fmb}]{\includegraphics[width=0.48\columnwidth]{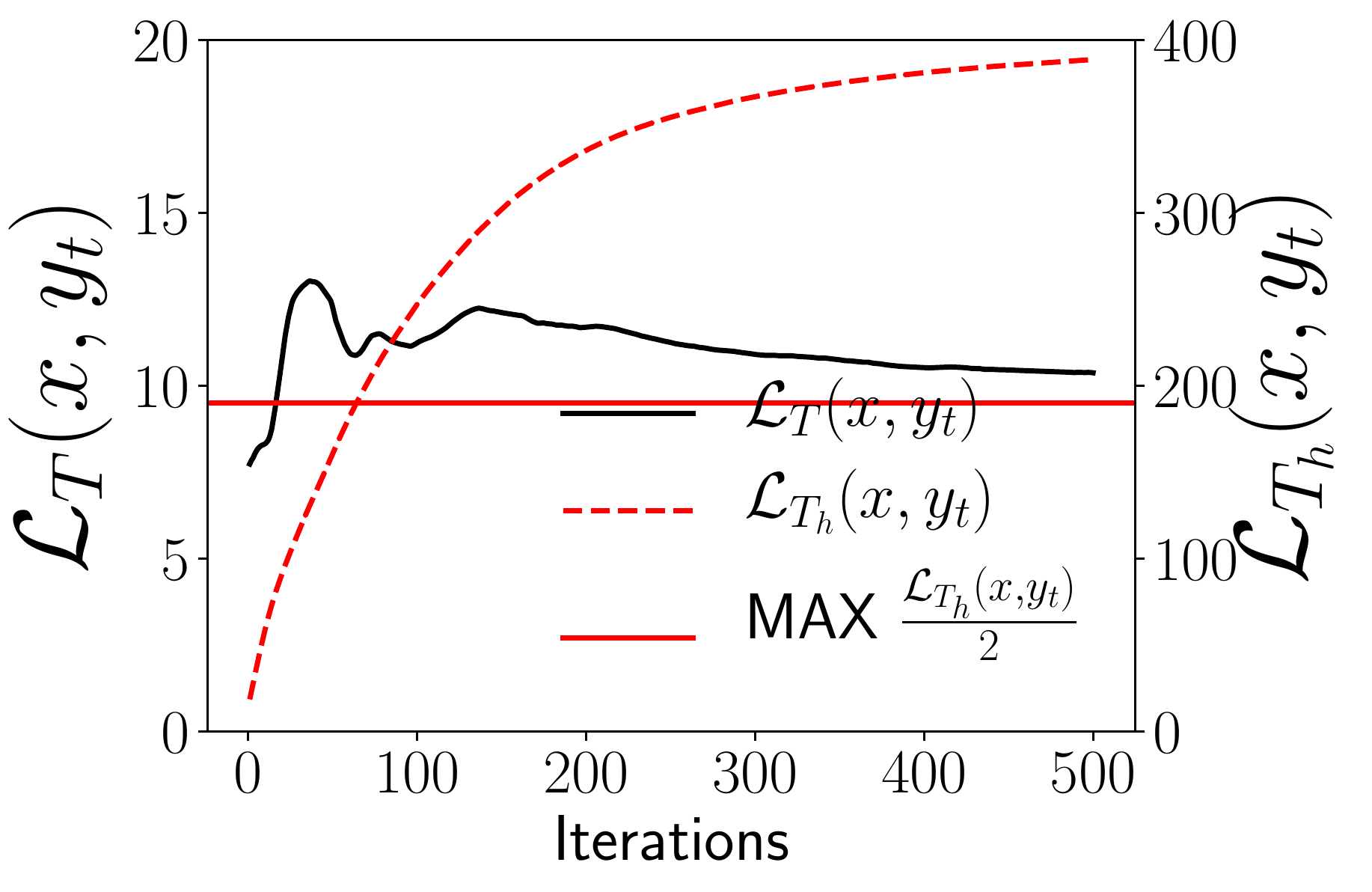}}
\caption{Example illustration on the procedure of optimizing Eq.~\ref{eq:newoptimization} for an infected or a healthy label.}
\label{fig:tp}
\end{figure}

Importantly, Line 10 shows our detection criteria, where the first term $\mathcal{L}_T(x,y_t)<=\beta$ indicates that $x$ may be led into the poisoned region for $y_t$ under model $T$; while the second term $\mathcal{L}_{T_h}(x,y_t)\geq\gamma$ implies that $x$ is also sufficiently pushed from the healthy decision region under model $T_h$ (and thus under $T$ as well), where $\mathcal{L}_{T_h}(x,y_t)$ represents the cross entropy loss value of input $x$ with label $y_t$ under model $T_h$.  If a dominant portion of such inputs $x$ meet the detection criteria for $y_t$, $y_t$ is deemed to be an infected label. In the implementation, we set $\beta$ to be 0.2 and $\gamma$ to be $50\%$ of the maximum value for which $\mathcal{L}_{T_h}(x,y_t)$ can reach, motivated by extensive empirical evidence.
For instance, for GTSRB, Fig.~\ref{fig:tp} illustrates how $\mathcal{L}_{T_h}(x,y_t)$ and $\mathcal{L}_{T}(x,y_t)$ would be changed during the procedure of optimizing Eq.~(\ref{eq:newoptimization}), for an infected or a healthy label. As seen in Fig.~\ref{fig:tp}, as expected, for an infected label, the optimization procedure could identify a rather small $\mathcal{L}_{T}(x,y_t)$ and a significantly large $\mathcal{L}_{T_h}(x,y_t)$; while for a healthy label, both these values would be large.

\noindent\textbf{Identify $\lambda$.}  To identify the best $\lambda$ value for effectively identifying poisoned region, we propose to dynamically adjust $\lambda$ to make $\mathcal{L}_T(x,y_t) =  \lambda \cdot \mathcal{L}_{T_h}(x,y_t)$ hold after optimizing 1000 iterations of ($\arg\max\limits_{z}~\mathcal{L}_T(z,y_t)+\lambda*\mathcal{L}_{T_h}(z,y_t)$). Intuitively this is because through extensive empirical studies, we observe that the proper value of $\lambda$ for identifying poisoned region varies depending on the specific pair of model $T$ and $T_h$. 
Specifically, we observe that the detection efficacy gets maximized when during the procedure of crafting $x$, the value changing rates on $\mathcal{L}_T(x,y_t)$ and $\mathcal{L}_{T_h}(x,y_t)$ are kept roughly the same. 
We achieve such a balance through identifying $\lambda$ which would enable $\mathcal{L}_T(x,y_t)$ and $\lambda \cdot \mathcal{L}_{T_h}(x,y_t)$ to increase roughly at the same rate during performing the optimization procedure  
($\arg\max\limits_{z}~\mathcal{L}_T(z,y_t)+\lambda*\mathcal{L}_{T_h}(z,y_t)$). This method of identifying a proper $\lambda$ is also shown to be effective through extensive experiments. 

Interestingly, it is not hard to craft test inputs $x$  using the above method, as many such $x$ does not belong to the original poisoned training set. We plot several samples of such inputs for AP attacks in Fig.~\ref{fig:adversarial_vast}. As shown in Fig.~\ref{fig:adversarial_vast}, the first column shows a poisoned sample form the training dataset, and the other four columns(from left to right) depict four images belonging to the same class yet not existing in the training dataset. Interestingly, the images in the bottom four columns can yet be misclassified as the infected label with high confidence(the number shown below each image in the figure) by the infected model. This implies the attack efficacy can still be ensured using inputs not existed in the training dataset. This observation also holds under backdoor scenarios, which is detailed in  Figs.~\ref{fig:gl2},~\ref{fig:ob2_trojan_appendix} and ~\ref{fig:ob2_chen_appendix} in the Appendix.

\begin{table*}[!t]                 
      \begin{minipage}[b]{0.4\columnwidth}
      \centering
      \begin{tabular}{ c c c }
      \hline
      model $T_h$ (Accuracy)& TP & FP\\
      \hline\hline
      B1($69.06\%$)& $100\%$ & $0\%$\\
      B2($73.21\%$)& $100\%$&$0\%$\\
      B3($70.12\%$)& $100\%$ & $0\%$\\
      B4($61.37\%$)&$ 100\%$&$0\%$\\
      B5($75.06\%$)&$ 100\%$&$0\%$\\
      \hline\hline
      \end{tabular}
      \caption{Results on MNIST task.}
      \label{table:results_MNIST}
      \vspace{-8mm}
      \end{minipage}
      \hfill
      \begin{minipage}[b]{0.5\columnwidth}
      \centering
      \begin{tabular}{ c c c } 
      \hline
      model $T_h$(Accuracy)& TP & FP\\
      \hline\hline
      RESNet.V1.44($78.21\%$)& $100\%$ & $70\%$\\
      RESNet.V2.102($73.65\%$)& $100\%$&$70\%$\\
      DenseNet-121($74.24\%$)& $100\%$ & $70\%$\\
      DenseNet-40($75.16\%$)&$ 100\%$&$70\%$\\
      RESNet.V2.83($76.23\%$)&$ 100\%$&$70\%$\\
      \hline\hline
      \end{tabular}
      \caption{Results on CIFAR-10 task.}
     \label{table:results_CIFAR}
     \vspace{-8mm}
     \end{minipage}
     \hfill
     \begin{minipage}[b]{0.7\columnwidth}
     \centering
    \includegraphics[width=1\columnwidth]{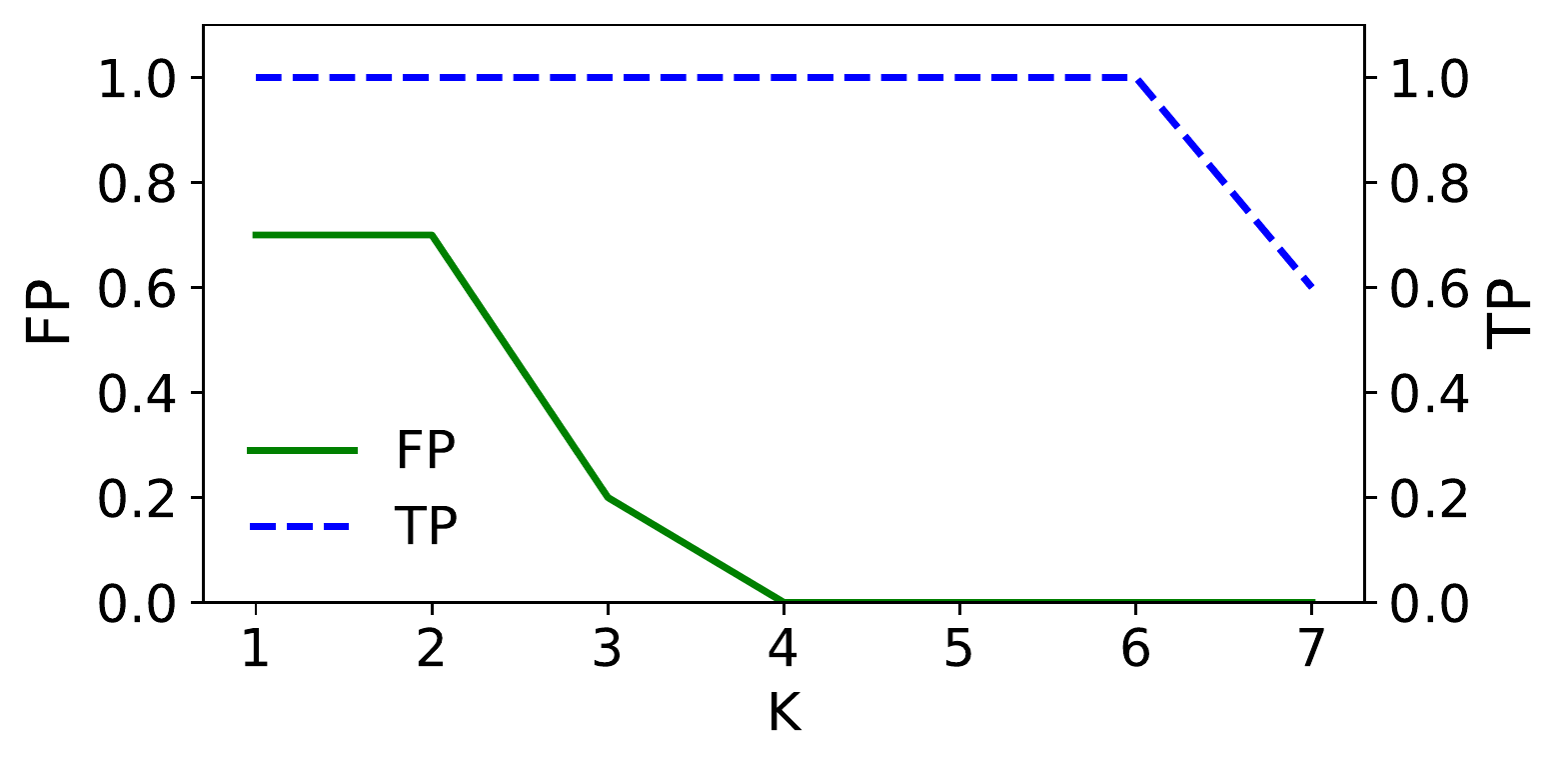}
    \vspace{-6mm}
    \captionof{figure}{FP and TP with different $k$.}
    \label{fig:ME_result}
    \vspace{-5mm}
    \end{minipage} 
    \vspace{4mm}
\end{table*}

\subsection{Model-Ensemble: Enhancing Detection Efficiency and Scalability in Practice}
\label{sec:model_ensemble}

For any given model $T$, we shall use the method presented in Sec.~\ref{sec:Implementaion} to detect whether a poisoned region exists for $T$, thus determining whether $T$ is infected. An essential step in~\name~is to obtain a healthy model $T_h$ inheriting $T$'s structure. In practice, it would be costly and inefficient to craft such a $T_h$ given any T. To enable efficient and scalable detection service in practice, we propose the following model-ensemble approach.




Our idea on resolving this issue is the following: for each DNN-driven task  (e.g., digit handwriting classification), it may be possible to use a pre-trained B to replace model $T_h$ where both B and T perform the same task. As noted earlier, the pre-trained model B is not required to have the same architecture and parameters as model T. Moreover, B is not required to yield state-of-the-art accuracy performance. Intuitively it may be possible because previous works have shown the transferability property of adversarial samples across healthy DNN models exhibiting completely different structures and parameters~\cite{tramer2017space,papernot2016transferability}.

To verify this idea, we perform a set of case studies using two datasets, MNIST and CIFAR-10. For each dataset, we randomly choose B from a pool of models listed in Tables~\ref{table:results_MNIST} and \ref{table:results_CIFAR} performing the corresponding tasks, yet forcing B and T to be different in terms of their 
structure. Also note that model B behaves $13\%$ to $23\%$ lower performance on the test dataset compared to T. As seen in Table~\ref{table:results_MNIST}, for MNIST,\footnote{The models for MINIST are simply denoted as ${B1, ..., B5}$ as they are straightforward convolution neural networks which are yet adequate due to the simplicity of this dataset.} choosing any model in the pool as $B$ under our approach can ensure an 100\% true positive (TP) detection rate (i.e., any infected model will be detected as infected) and a 0\% false positive rate (i.e., any healthy model will never be detected as infected).  These results demonstrate that it is highly likely that the healthy decision region of models performing the same task may largely overlap with each other.

For the CIFAR-10 dataset, as seen in Table~\ref{table:results_CIFAR}, while choosing any model in the pool as $B$ can ensure an 100\% true positive detection rate, it actually yields a rather high false positive rate simultaneously. The reason behind this observation is the following. For MNIST, the model pool  contains straightforward convolution neural networks; while for CIFAR-10, the models in the pool are fairly complicated ones including RESNet and DenseNet,   which may exhibit dramatic difference in structure, parameters, training data, and even different configurations for data augmentation. Therefore, there could be large variation in terms of the healthy decision region representation among these models. Such variation could diminish the overlapping portion of the healthy decision region between T and B. This could further cause the observation seen in Table~\ref{table:results_CIFAR}, where ~\name~ can still successfully detect any infected model T, yet may incorrectly deeming a healthy model T to be infected. This is because the non-overlapping region between the healthy decision region under $B$ and $T$ may be mis-determined as the poisoned decision region via comparing the Cross Entropy loss values. Thus, such regions have a higher possibility to be incorrectly identified as poisoned regions under \name.

For datasets that require complicated models which exhibit large variance, choosing a single model $B$ to replace $T_h$ may not be sufficient because the healthy decision region of $B$ and $T_h$ (thus that of $B$ and $T$) may only partially overlap. The healthy decision region of $B$ thus cannot accurately  represent that of $T$. Any portion of the healthy decision region of T that is not overlapped with B's healthy decision region would thus be deemed as the poisoned region under ~\name, which would cause a healthy model to be detected as infected (i.e., a high false positive detection ratio as seen in the CIFAR-10 case study).

To resolve this challenge, we patch an enhancement method to \name, namely model-ensemble. The key idea behind this approach is that instead of using a single model to replace $T_h$, \name{} would use a model ensemble $B$ which contains a pool of models performing the same task as the given model $T$. The benefit is intuitive: using multiple models to replace $T_h$ would increase the overlapping degree between the healthy decision region of $T$ and the joint healthy decision region of all models in this model ensemble.
Specifically, B would be a model ensemble, consisting of a set of models. Thus, we could accordingly update Eq.~\ref{eq:newoptimization} using the following Eq.~\ref{eq:ME} during the detection procedure:
\begin{equation}\label{eq:ME}
x = ~\argmin\limits_{x}~\mathcal{L}_T(x,y_t)- \sum_{i=1}^{k}\lambda_{i}*\mathcal{L}_{{B}_{i}}(x,y_t),
\end{equation}
\noindent where $k$ represents the number of models in the model ensemble, $B_{i}$ denotes a model  belonging to this ensemble, and $\lambda_{i}$ denotes the co-efficient for each $B_{i}$.

For the same CIFAR-10 case study, Fig.\ref{fig:ME_result} shows the results on the false positive detection rate if applying this model ensemble approach. As seen in this figure, a larger value of $k$ would effectively reduce the false positive detection rate. A model ensemble B containing at least 4 models would reduce the false positive rate to 0. (More extensive experiments to be described in Sec.~\ref{sec:overallperformance}~also prove the efficacy of this model-ensemble approach.)

\section{Experimental Validation}
\label{sec:exp}


In this section, we describe experiments conducted to assess the efficacy of \name{} against state-of-the-art AP and backdoor attacks under various datasets and system settings. Our evaluation focuses on answering the following key research questions: (\textit{i}) can \name{} achieve high overall detection rate? (\textit{ii}) can \name{} enhance the detection performance when comparing to state-of-the-art detection methods? (\textit{iii}) is \name{} sufficiently robust considering different settings and attack variants?


\subsection{Experiment Setup}

\noindent\textbf{Datasets and models.} We evaluated \name{} upon four datasets using a set of state-of-the-art DNN models, including (1) MNIST for the hand-written digit recognition task, (2) Fashion-MNIST for the fashion item recognition task), (3) GTSRB for the traffic sign recognition task and (4) CIFAR-10 for the image classification task.
Due to space constraints, we detail each dataset/task, as well as  the training configuration and model architecture/parameters in the Appendix.

\begin{table*}[!t]
\centering
\scalebox{0.8}{
\begin{tabular}{|c|c|c|c|c|c|c|c|}
\hline
\diagbox{\textbf{Dataset}}{Prob(FP Rate)}{\textbf{Attack Technique}}&StingRay&Poison Attack&Mislabel Attack&BadNets&Trojan Attack&Chen attack\\
\hline
MNIST&$98\%$($0\%$)&$98\%$($0\%$)&$100\%$($0\%$)&$100\%$($0\%$)&$100\%$($0\%$)&$100\%$($0\%$)\\
\hline
Fashion-MNIST&$86\%$($0\%$)&$86\%$(0\%)&$86\%$(0\%)&$89\%$($0\%$)&$89\%$($0\%$)&$86\%$(0\%)\\
\hline
GTSRB&$83\%$($0\%$)&$83\%$($0\%$)&$86\%$($0\%$)&$96\%$(0\%)&$83\%$(0\%)&$90\%$(0\%)\\
\hline
CIFAR-10&$100\%$($40\%$)&$100\%$($40\%$)&$100\%$($40\%$)&$100\%$($40\%$)&$100\%$($40\%$)&$100\%$($40\%$)\\
\hline

\end{tabular}
}
\caption{Prob value on the infected label and FP rate on various datasets leveraging $T_h$.}
\label{table:performance_TH}
\vspace{-1mm}
\end{table*}

\begin{figure*}[!t]
\centering
\subfigure[MNIST(AP)\label{fig:fm_scale}]{\includegraphics[width=0.24\textwidth]{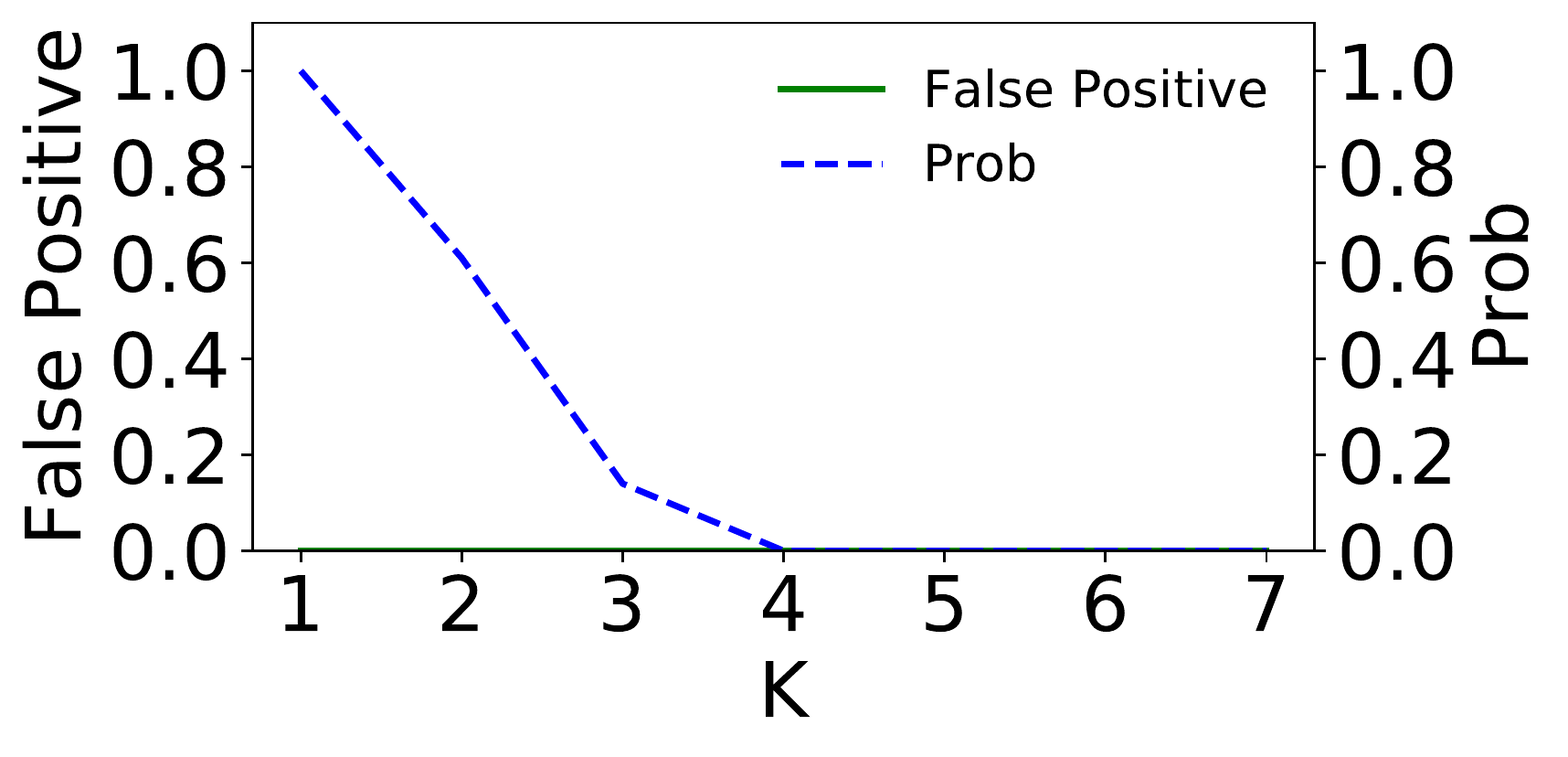}}
\subfigure[MNIST(Backdoor)\label{fig:mnist_scale}]{\includegraphics[width=0.24\textwidth]{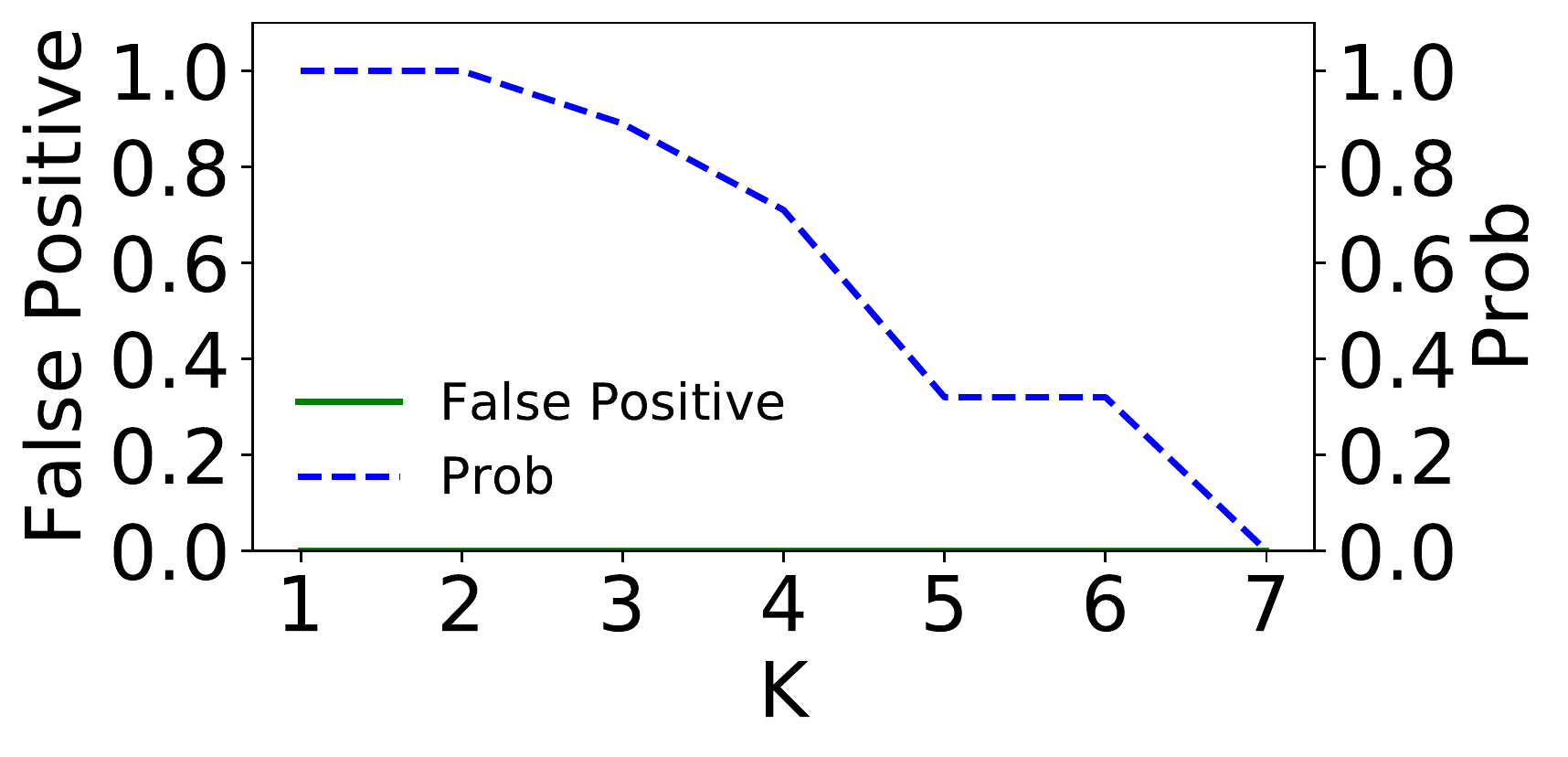}}\
\subfigure[CIFAR-10(AP)\label{fig:cifar_scale}]{\includegraphics[width=0.24\textwidth]{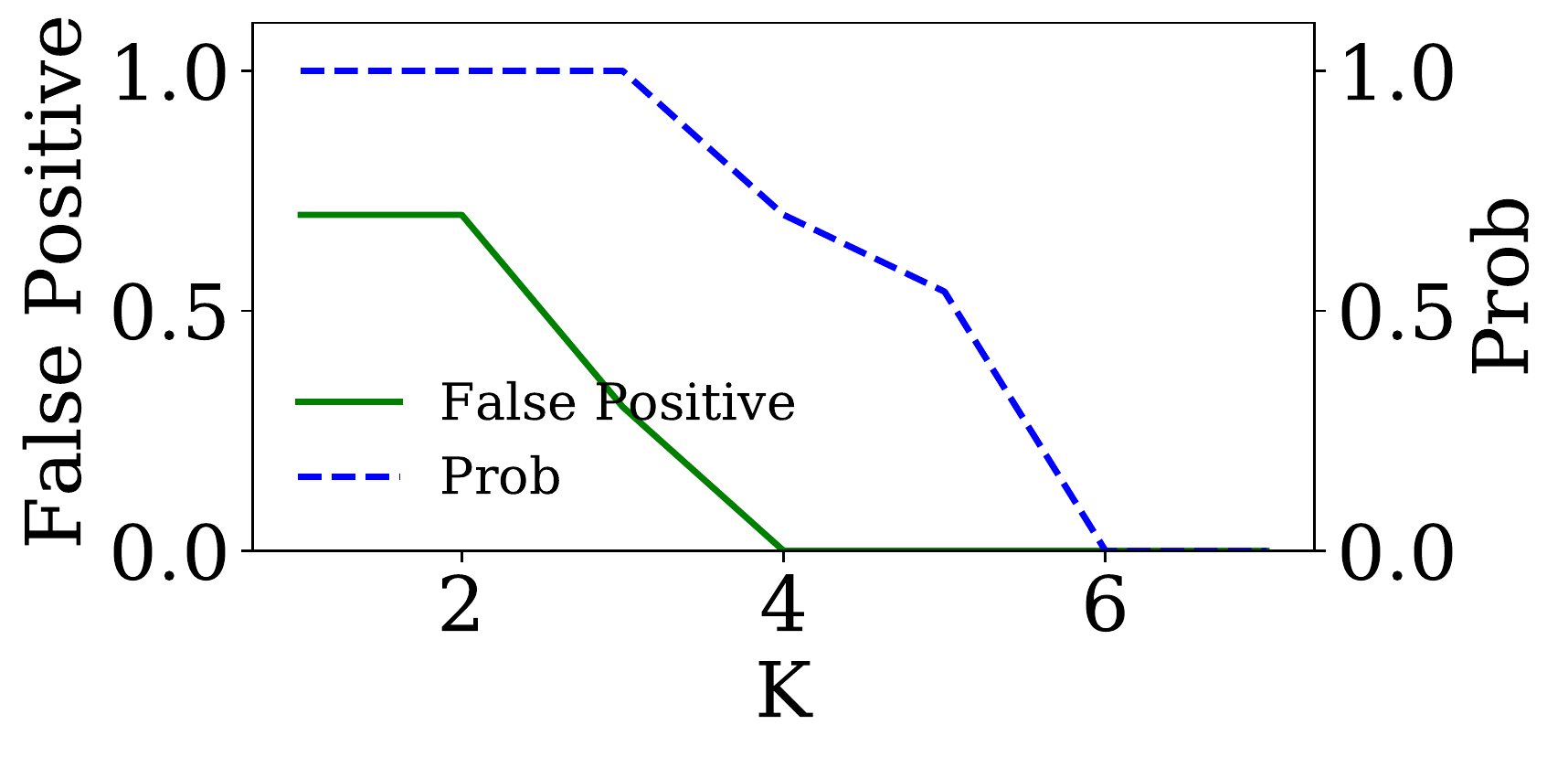}}
\subfigure[CIFAR-10(Backdoor)\label{fig:gtsrb_scale}]{\includegraphics[width=0.24\textwidth]{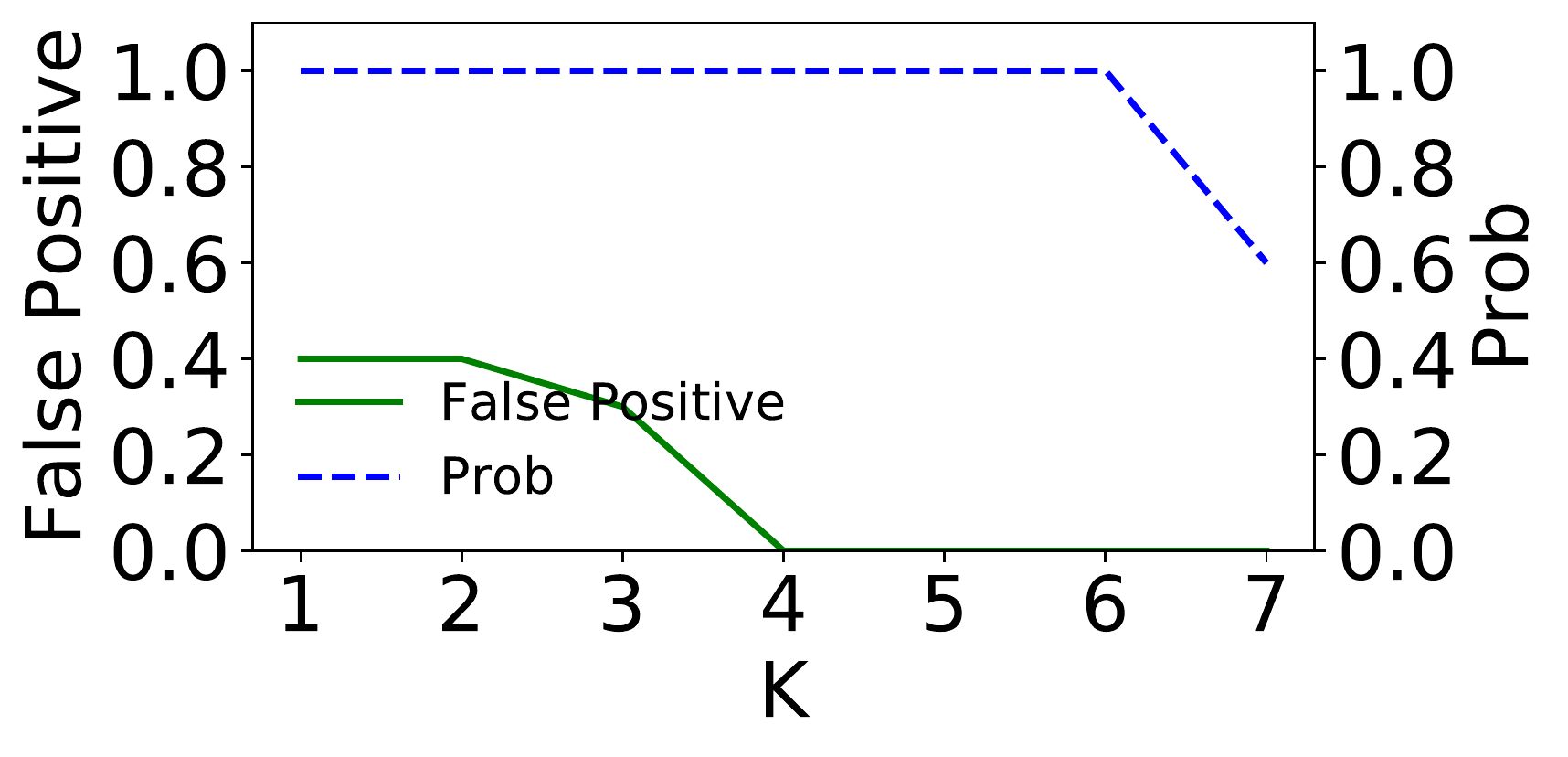}}
\vspace{-2mm}
\caption{Prob value and FP rate with different $k$ for MNIST and CIFAR-10 under backdoor and AP attacks scenarios.}
\vspace{-1mm}
\label{fig:ME_cp}
\end{figure*}

\vspace{1mm}\noindent\textbf{Attack Configuration for AP.}
We evaluated two dominate types of integrity AP attacks~\cite{shafahi2018poison,suciu2018does}: (\textit{i}) clean-label AP attack implemented as Poison Frog~\cite{shafahi2018poison} and Sting Ray~\cite{suciu2018does}, and (\textit{ii}) integrity poisoning attack using mislabeled data implemented as the attack described in~\cite{suciu2018does} (denoted ``Mislabel''). Our attack configurations exactly follow the configuration described in the corresponding papers~\cite{shafahi2018poison, suciu2018does}.

For each task, we select at random a label as the infected label and choose images of a certain class as poisoned inputs(details are shown in Table.~\ref{table:APConfiguration} in the Appendix). To measure attack performance, we calculate classification accuracy on the test data, as well as the attack success rate when using adversarial images as test inputs. The Attack Success Rate metric measures the percentage of adversarial inputs being mis-classified as the infected label. We also measure the classification accuracy on a clean version of each model (i.e. training using exactly the same configuration but with clean training set--$T_h$ corresponding to each given $T$)
For each dataset and task, we vary the proportion of poisoned inputs(from $4\%$ to $8\%$) to ensure a $\geq96\%$ attack success rate  while preserving overall classification performance.   

\vspace{1mm}\noindent\textbf{Attack Configuration for Backdoor.} 
We evaluated three state-of-the-art backdoor attack methods including BadNets, Trojan Attack and Chen Attack~\cite{gu2017badnets,chen2017targeted,liu2017trojaning}, which represent the state-of-the-art as well as the only existing backdoor attack techniques that are effective under our adversarial model.
The attack configuration and trigger selection in the evaluation are consistent with the setting adopted by Neural Cleanse~\cite{wang2019neural}, which focuses on detecting backdoor due to BadNets and Trojan Attack. For Chen attack, we set the random noise based trigger to occupy the entire image and set a transparency ratio of 0.1 to make the trigger appear less noticeable.

Example poisoned images and triggers are shown in Fig.~\ref{fig:sample_back} in the Appendix. As seen in Table~\ref{table:setup} in the appendix, our chosen attack configuration ensures that all tested backdoor attacks achieve $\geq 97\%$ attack success rate, with trivial impact on overall classification accuracy. 

\noindent\textbf{Evaluation Metrics.} \textbf{Prob---}To imply the detection efficacy for each label $y_t$, we use the proportion of those $x$ among all crafted ones which satisfy the detection criteria for reaching the poisoned region as discussed in Sec~\ref{sec:Implementaion} (denoted as Prob). Specifically, a Prob value larger than the threshold would indicate that a given label $y_t$ is an infected label. The threshold value is set to be $50\%$ in our experiments since it can well balance the trade-off between FP rate and detection success rate. Note that using Prob as a metric can indicate detection performance while showing in detail the  number of crafted $x$ under \name{} that are effective towards detection purpose. 

\textbf{False positive rate---}For any detection method to be effective in practice, besides detection rate, yielding a low false positive (FP) rate is equally important. A zero or close-to-zero FP rate would indicate that the detection method under evaluation would prevent mis-tagging any clean labels.

For each experiment on each dataset, we randomly selected a model from the corresponding model pool and repeated this process for ten times. We then create two model sets based upon these ten selected models, including an infected set where we infect each of these ten models following each evaluated attack method, as well as a healthy set where all ten models are healthy. The infected set is used to evaluate the Prob metric (thus the detection efficacy). 
It is also used for measuring the FP rate, as any infected model may contain clean labels as well. 
The healthy set is used to evaluate the FP rate under \name{}.
We crafted 100 samples (i.e., $x$) for each experiment.
We note that when implementing the model ensemble method, we intentionally remove the model from the ensemble which exhibits the same architecture as the model to be analyzed.

\subsection{Overall Detection Performance}
\label{sec:overallperformance}

In this section, we perform experiments to evaluate \name{} following the methodology in Secs.~\ref{sec:Implementaion} and ~\ref{sec:model_ensemble}, i.e., using $T_h$ and a model-ensemble for detection purpose, respectively. Moreover, for backdoor scenarios, we compare performance of \name{} to Neural Cleanse~\cite{wang2019neural} and ABS under various settings.\footnote{Note that we choose to compare against Neural Cleanse and ABS because they represent  state-of-the-art methods on detecting backdoor attacks. We are also aware of a couple of other works on detecting backdoor attacks~\cite{chendeepinspect,guo2019tabor}, which are too premature (i.e., with no or very limited evaluation) and incrementally built  upon Neural Cleanse to be included in the  comparison. Also note that there does not exist any detection method on the AP attacks evaluated in this paper.}


\noindent\textbf{Performance using $T_h$.} 
Table~\ref{table:performance_TH} shows the minimum Prob among all ten infected models in each experiment and the FP rate for all the six implemented attacks. 
As seen in Table.~\ref{table:performance_TH}, for each attack method, the Prob of the infected label is higher than 50\%, indicating \name{} is effective in detecting the infected label. Meanwhile, following each attack methodology, for all datasets except CIFAR-10, \name{} yields a zero FP rate. The high FP rate for CIFAR-10 is caused by the inability of using a single $T_h$ to deal with such complex datasets, as discussed in Sec~\ref{sec:model_ensemble}.

\noindent\textbf{Performance using the Model Ensemble.} As discussed in Sec.~\ref{sec:model_ensemble}, \name{} also contains a model ensemble method to enhance its  efficiency and scalability in practice. For evaluation, we create a model pool containing a set of pre-trained DNN models for each tested dataset. For each experiment, we randomly choose a model in the pool as the  model to be detected, and treat the remaining models as the model ensemble. We note that in order to ensure practicality and robustness of this approach, the models contained inside each model pool may exhibit dramatically different structures and parameters (e.g., DenseNet versus RESNet). Detailed model information is described in the Appendix.

We first seek to understand how $k$, the number of models used in the model ensemble, would impact the performance w.r.t. both Prob on infected label and FP rates. Results in terms of minimum Prob are shown in Fig.~\ref{fig:ME_cp} (results on the remaining datasets are shown in Fig.~\ref{fig:ME_changing} in the Appendix. As seen in Figs.~\ref{fig:ME_cp} and~\ref{fig:ME_changing}, we observe that for relatively simple datasets including MNIST, Fashion-MNIST and GTSRB, a small $k$ value (e.g., $k \leq 2$) would be sufficient in ensuring $Prob \geq 50\%$ and an FP rate close to 0.  Thus, for rather simple datasets where using plain convolution networks is sufficient to obtain state-of-art performance, a model ensemble containing one or two models would be sufficient to guarantee detection performance.

\begin{table*}[!t]
\centering
\scalebox{0.8}{
\begin{tabular}{|c|c|c|c|c|c|c|c|}
\hline
\diagbox{\textbf{Dataset}}{Prob(FP Rate)}{\textbf{Attack Technique}}&StingRay&Poison Frog&Mislabel Attack&BadNets&Trojan Attack&Chen et al\\
\hline
MNIST&$98\%$($0\%$)&$98\%$($0\%$)&$100\%$($0\%$)&$100\%$($0\%$)&$100\%$($0\%$)&$100\%$($0\%$)\\
\hline
Fashion-MNIST&$86\%$($0\%$)&$86\%$(0\%)&$86\%$(0\%)&$89\%$($0\%$)&$89\%$($0\%$)&$86\%$(0\%)\\
\hline
GTSRB&$83\%$($0\%$)&$83\%$($0\%$)&$86\%$($0\%$)&$96\%$(0\%)&$83\%$(0\%)&$90\%$(0\%)\\
\hline
CIFAR-10&$51\%$($10\%$)&$51\%$($10\%$)&$56\%$($10\%$)&$100\%$($10\%$)&$100\%$($10\%$)&$100\%$($10\%$))\\
\hline
\end{tabular}
}
\caption{Prob value on the infected label and FP rate on various datasets leveraging a model ensemble $B$.}
\vspace{-1mm}
\label{table:performance_ME}
\end{table*}

\begin{figure*}[!t]
    \begin{minipage}[b]{0.3\textwidth}
    \centering
        \includegraphics[width=1\textwidth]{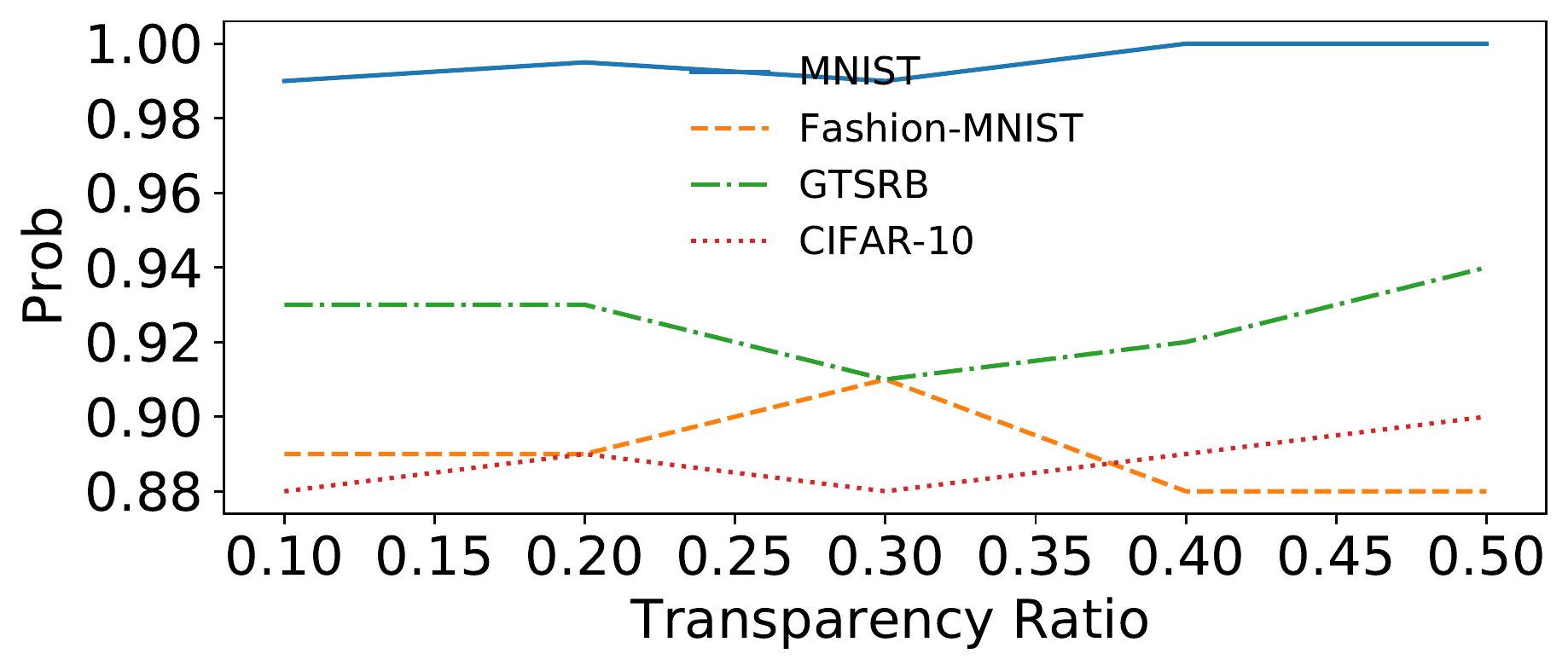}
        \vspace{-6mm}
        \caption{Detection Performance under different transparency ratio of noise in each task for Chen Attack under \name.}
        \vspace{-4mm}
        \label{fig:transparency}
    \end{minipage}
    \hfill
    \begin{minipage}[b]{0.3\textwidth}
        \includegraphics[width=1\textwidth]{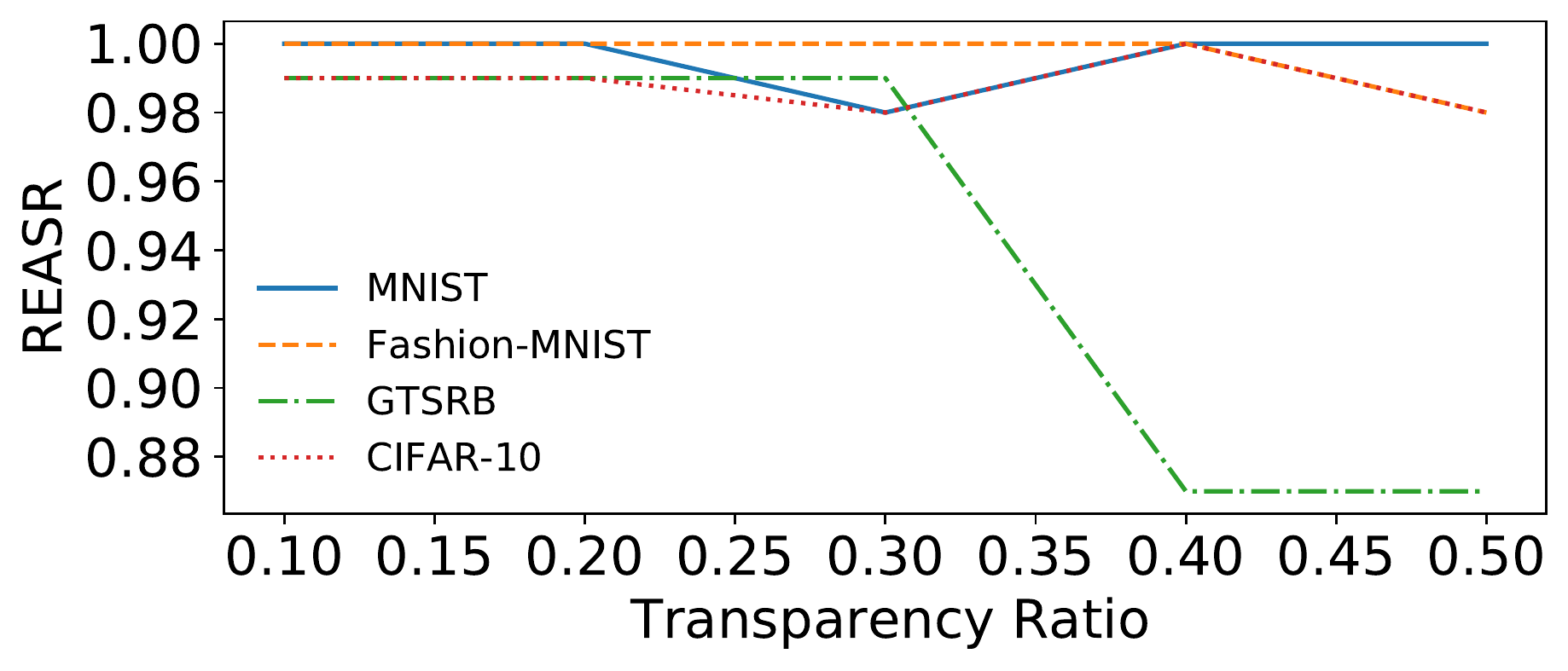}
        \vspace{-2mm}
        \caption{Detection Performance under different transparency ratio of noise in each task for Chen Attack under ABS.}
        \vspace{-4mm}
        \label{fig:REASR}
    \end{minipage}
    \hfill
    \begin{minipage}[b]{0.3\textwidth}
        \includegraphics[width=1\textwidth]{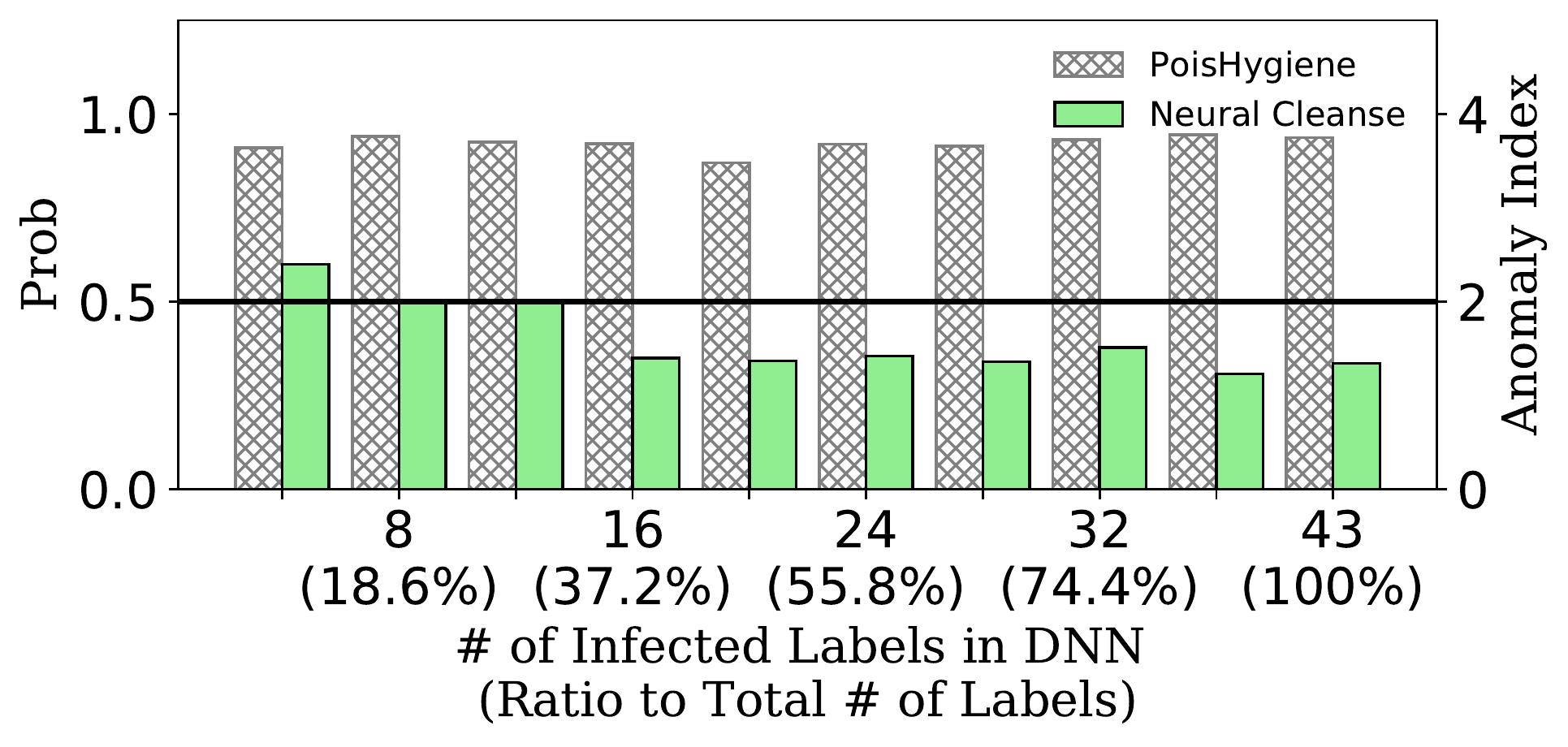}
        \vspace{-5mm}
        \caption{Multi-Label Detection Performance on GTSRB compared to Neural Cleanse.}
        \vspace{-4mm}
        \label{fig:multiple_label}
        \end{minipage}
        \vspace{4mm}
\end{figure*}

On the other hand, for more complex datasets such as CIFAR-10, only a sufficiently large $k$ (e.g., $k \geq 4$) could result in a low FP rate $\leq 10\%$. 
This is again because using a model ensemble may ensure a large overlapping between the healthy decision region of $T$ and the models in the ensemble $B$, as discussed in Sec~\ref{sec:model_ensemble}.
Nonetheless, such scenarios are not problematic in most practical settings due to the low FP rate. We also present an unlearning-based technique which effectively patches such vulnerable labels while preserving overall performance.

Another observation is that a too large $k$ value (e.g., $k\geq 6$) would cause the Prob value to drop under 0.5, which implies a low detection rate. This is because when the model ensemble $B$ contains too many models, the resulting entire healthy decision region of models in $B$ may become rather vast, increasing the possibility of covering cover regions that are close to the poisoned region under $T$. Thus, a preferred $k$ value for CIFAR-10 datasets is set to be four or five, which is able to achieve a sufficiently large Prob value and a low FP rate.

Table.~\ref{table:performance_ME} shows the performance of \name{} for the evaluated datasets and attack methods under the model ensemble approach with an appropriate $k$ value set for each dataset, as discussed above. As seen in the Table, \name{} is able to achieve sufficiently good Prob values (100\% under most settings) while ensuring low FP rates (0\% under most settings).

\subsection{Performance on BD attacks compared to Neural Cleanse and ABS}

We specifically compared \name{} using the model ensemble approach against Neural Cleanse and ABS on detecting backdoor attacks under various complex attack variants.

\noindent\textbf{Performance against advanced backdoor attack, i.e., Chen Attack~\cite{chen2017targeted}.} For more advanced backdoor attack methods, we compare \name{} to Neural Cleanse and ABS~\cite{liu2019abs} for detecting models infected by Chen Attack, which applies a semi-transparent random noise to the entire input. As seen in Fig.~\ref{fig:transparency}, \name{} is able to yield a $Prob \geq 50\%$, which implies a perfect detection under all transparency settings; while Neural Cleanse simply fails to detect any infected label under all tested scenarios. As discussed earlier, Neural Cleanse  cannot  handle such backdoor attacks which do not constraint the trigger size. We also test \name{} against ABS on Chen attack with the same settings. As shown in  Fig.~\ref{fig:REASR}, ABS can also successfully detect the infected label correctly with the attack success rate indicated by the reverse
engineered trojan trigger (REASR) higher than $80\%$ on various datasets.

\begin{table*}[!t]
    \centering
    \scalebox{0.8}{
    \begin{tabular}{|c|c|c|c|c|c|c|c|}
    \hline
    \diagbox{\textbf{Dataset}}{Prob(FP Rate)}{\textbf{Attack Technique}}&StingRay&Poison Frog&Mislabel Attack&BadNets&Trojan Attack&Chen et al\\
    \hline
    MNIST&$93\%$($0\%$)&$90\%$($0\%$)&$92\%$($0\%$)&$32\%$($0\%$)&$26\%$($0\%$)&$33\%$($0\%$)\\
    \hline
    Fashion-MNIST&$87\%$($0\%$)&$91\%$(0\%)&$86\%$(0\%)&$24\%$($0\%$)&$28\%$($0\%$)&$19\%$(0\%)\\
    \hline
    GTSRB&$93\%$($0\%$)&$93\%$($0\%$)&$91\%$($0\%$)&$20\%$(0\%)&$21\%$(0\%)&$20\%$(0\%)\\   
    \hline
    CIFAR-10&$67\%$($10\%$)&$63\%$($10\%$)&$59\%$($10\%$)&$7\%$($10\%$)&$7\%$($10\%$)&$5\%$($10\%$)\\
    \hline
    \end{tabular}}
    \caption{Overall performance against adaptive AP and backdoor attacks.}
    \label{table:performance_ME_ad}
\end{table*}

\noindent\textbf{Multi-Label Detection.} 
We also evaluated the scenario where multiple infected labels are present in the given model. We use the configuration as previous experiments.


Note that for \name, we select the minimum Prob value among all infected labels to represent the Prob value under each tested model. As seen in the Fig.~\ref{fig:multiple_label}, \name{} can achieve a significantly higher detection rate than Neural Cleanse with multiple ($\geq 12$) infected labels under GTSRB, while achieving zero FP rate. (Again, results on other datasets are put in the
Fig.~\ref{fig:mul_compare} in the Appendix. In fact, when the model contains $\geq 12$ infected labels, Neural Cleanse fails to detect any such label. This is because Neural Cleanse applies an outlier method. (Note that the detection criteria on infected labels under Neural Cleanse is: anomaly index $\geq2$). ABS suffers from the same issue as its design and evaluation only focus on the single infected label case. 

Being able to detect multi-infected-label models is critical as in practice, attackers may very likely create multiple infected labels to prevent the infected model from being detected.




      
    

\vspace{-2mm}
\subsection{Efficacy Against Detection-aware Poisoning Attacks}
\label{sec:adpt}

In this subsection, we investigate the effectiveness of \name{} against detection-aware poisoning attacks. We consider the worst-case scenario, where the attacker understands \name's mechanism and have full access to all models  in the ensemble B (including all detailed information of the models) used under \name{} for detection purposes.

In this case, to counter the detection mechanism of \name{}, an attacker may train a model T in an adaptive manner, making T's healthy and poisoned region stay close enough such that \name{} could not identify. Specifically, for both BD and AP attack, the attacker could  minimize the gap between  $\mathcal{L}_T(x_i,y_t)$ and $\mathcal{L}_B(x_i,y_t)$, where $y_t$ denotes the label to be infected and $x_i$ denotes the poisoned samples. To achieve this goal, the attacker can generate $x_i$ via optimizing Eq.~\ref{eq:adpt}, 
\begin{equation}\label{eq:adpt}
    x_i=\argmin_{x}\sum_{i}\mathcal{L}_{B_{i}}(x,y_t), 
\end{equation}


\noindent where $x$ denotes the target inputs to be misclassified.  For backdoor scenarios, $x_i$ is transformed from $x$ via adding a specific trigger with the smallest size;  while for AP scenarios, $x_i$ is generated from $x$ with minimum modification. Once obtaining a set of $x_i$, the attacker creates an infected model $T$ through adding $x_i$ into the training dataset. 


We evaluate \name{} against such detection-aware attack. The results are shown in Table.~\ref{table:performance_ME_ad}. As seen in Table.~\ref{table:performance_ME_ad}, we observe that for AP attacks, \name{} remains to be rather effective. 
A possible reason is that even if the attacker trains model $T$ with samples satisfying Eq.~\ref{eq:adpt}, model $T$ may take different combinations of features for prediction on $x_i$ compared to each $B_i$, thus forming a poisoned region which can still be distinguished from the decision region of each $B_i$.

For BD attacks, however, \name{} does not work for most tasks. The reason is because the generated backdoor trigger can be the dominating critical features for prediction on $x_i$ by model $T$, where such features are also learned under each $B_i$ due to the similar backdoor trigger pattern.

\begin{figure}[!t]
    \centering
    \includegraphics[width=0.24\columnwidth]{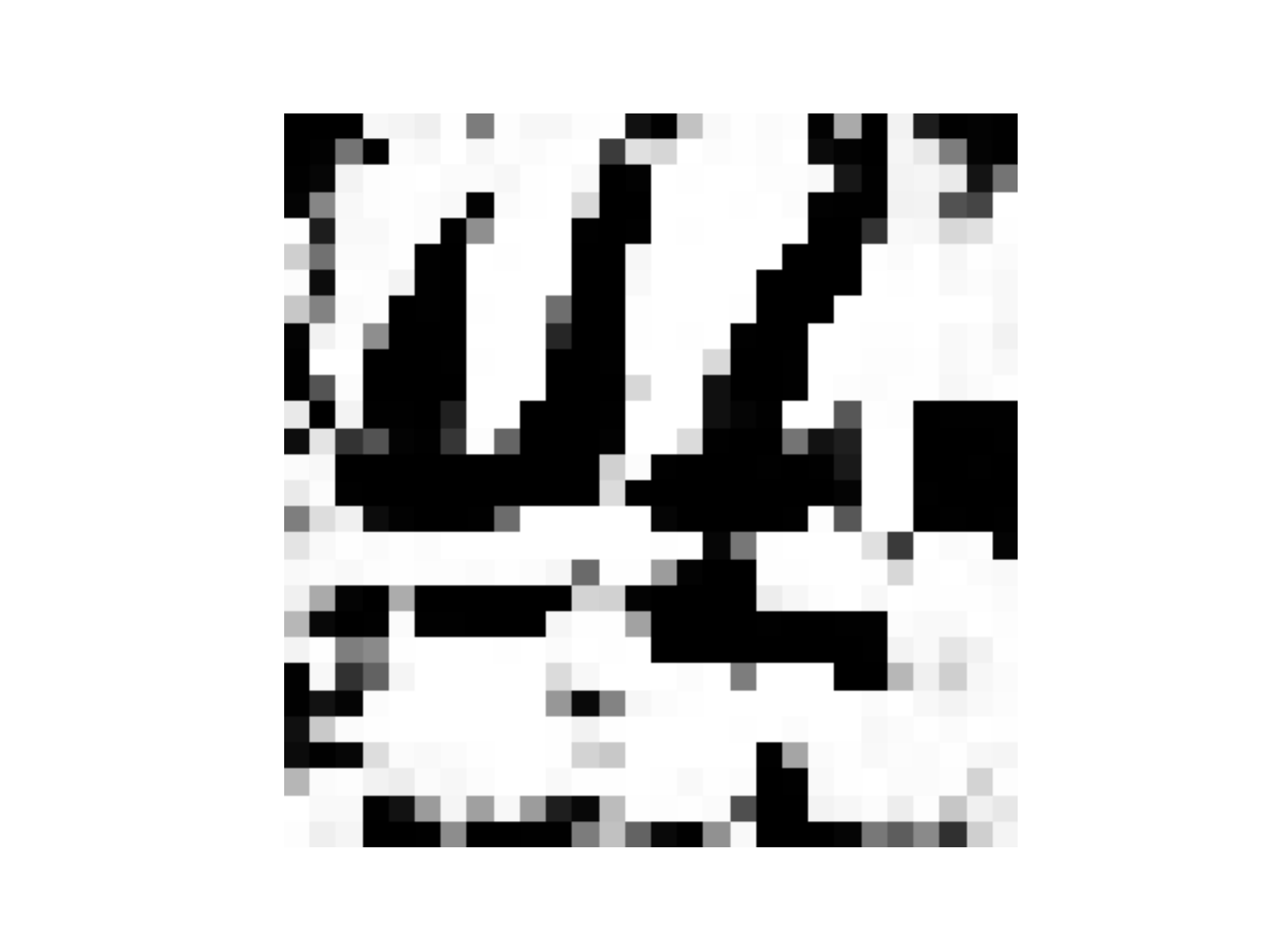}
    \includegraphics[width=0.24\columnwidth]{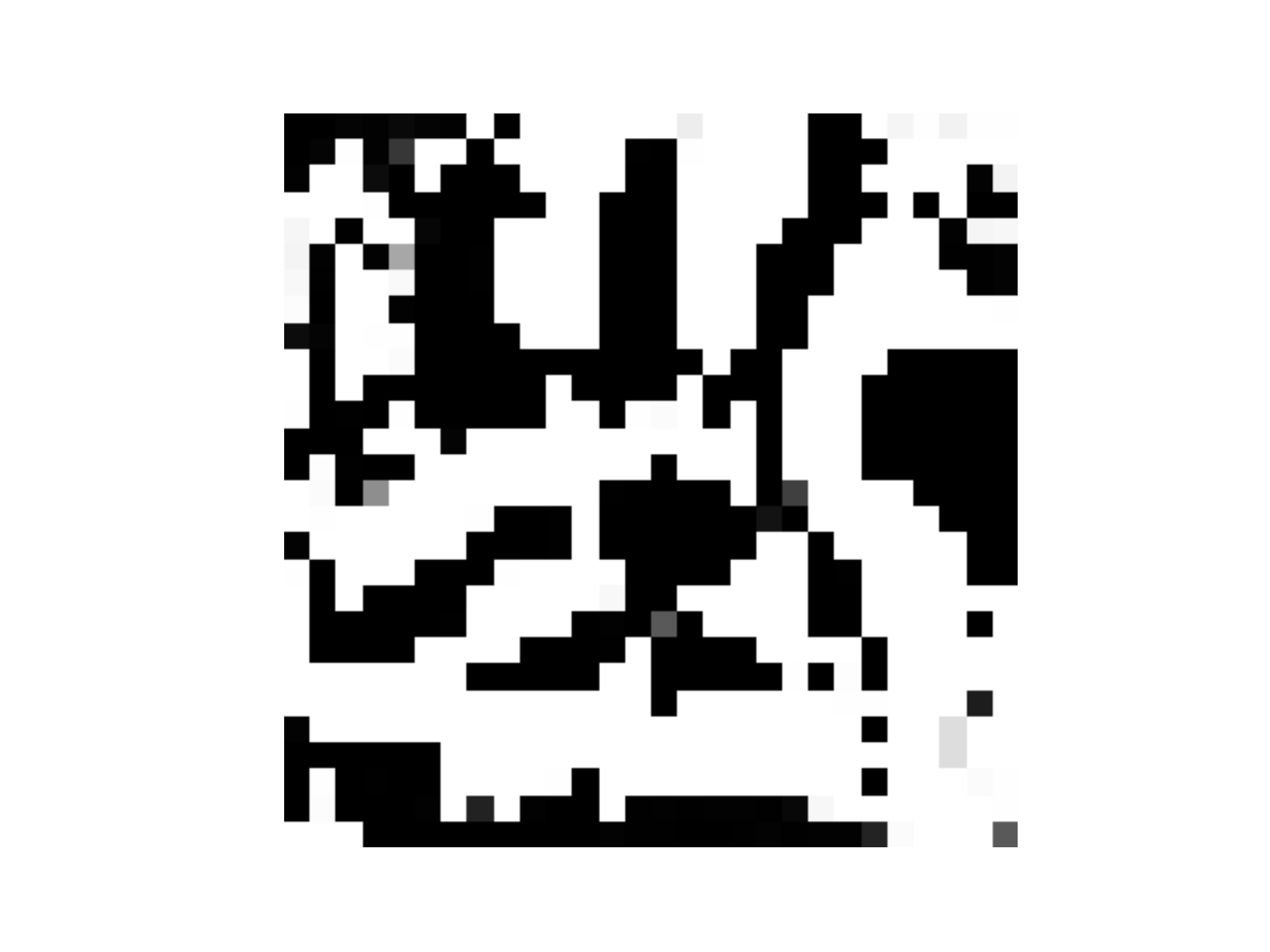}
    \includegraphics[width=0.24\columnwidth]{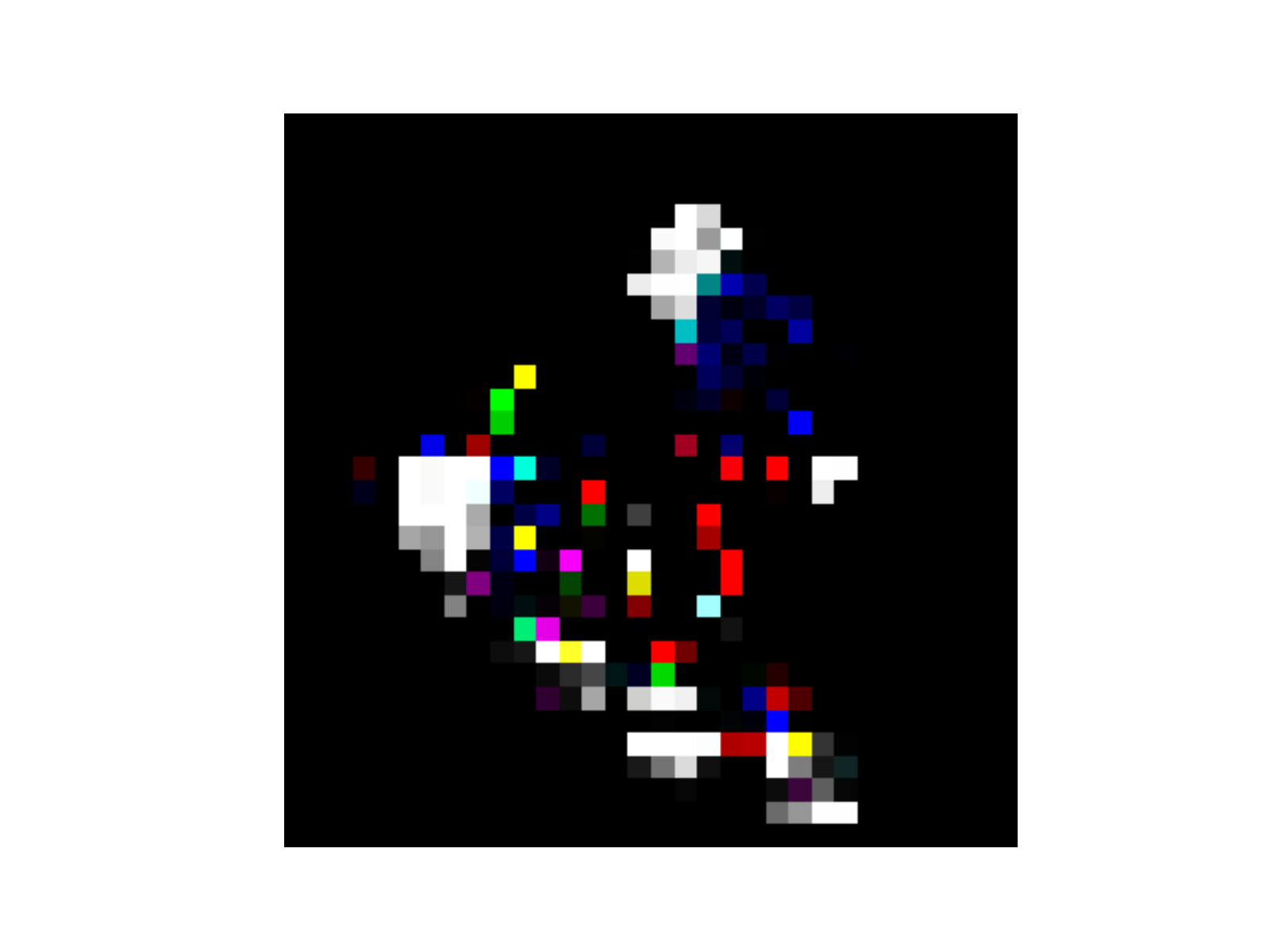}
    \includegraphics[width=0.24\columnwidth]{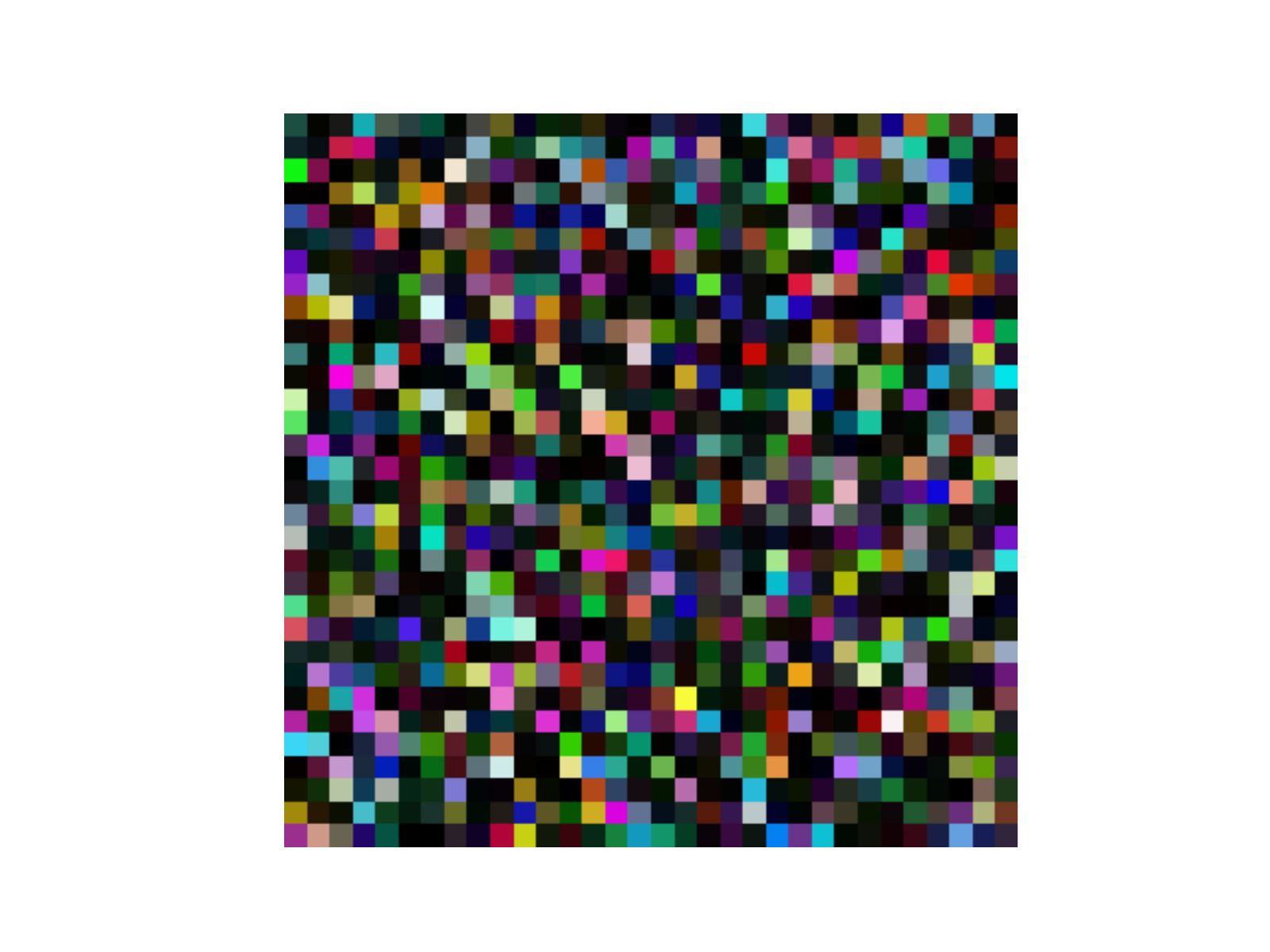}
    \caption{Example crafted triggers for (from left to right) MNIST, fashion-MNIST, GTSRB, and CIFAR-10.}
    \label{fig:demonstration}
    \vspace{-4mm}
\end{figure}

Nonetheless, the successful defense-aware BD attack exhibits rather large and obvious trigger pattern, as illustrated in Fig.~\ref{fig:demonstration} for the four  datasets. Such attack-obvious backdoor triggers may be easily detected by human labelers or end users, and is thus impractical.

\vspace{-2mm}
\section{Mitigation of Attacks}
\label{sec:miti}

After successfully identifying the infected label, we now show how to implement our proposed unlearning-based mitigation technique while maintaining the overall performance.


\subsection{Mitigation via Unlearning}

Our proposed mitigation method is to train the infected models to unlearn adversarial images. For each label $y_o$ , where $y_o$ denotes the label other than infected label $y_t$, we craft several adversarial images by optimizing Eq.~\ref{eq:identify} given below, and then use pairs of the crafted adversarial image and its corresponding label $y_o$ to retrain the infected DNN.

To perform the unlearning process, we first need to identify several adversarial images used in the process. We craft adversarial images by optimizing the following objective:
\begin{equation}\label{eq:identify}
x = ~\argmin\limits_{x}~\mathcal{L}_T(x,y_t)+c*\mathcal{L}_{T_h}(x,y_o)-d*\mathcal{L}_{T_h}(x,y_t)
\end{equation}

Recall that the goal of optimizing Eq.~\ref{eq:newoptimization} is to craft an input that would reach the poisoned region under model T. Thus, optimizing Eq.~\ref{eq:identify} enables us to obtain an adversarial image that could simultaneously  the reach poisoned region of $y_t$ under model $T$ and the healthy decision region of $y_o$ under model $T_h$. 
To ensure that the crafted adversarial images could reach the poisoned region of $y_t$ under $T$ and the healthy decision region of $y_o$ under $T_h$, we adjust $c$ and $d$ properly to ensure that $\mathcal{L}_T(x,y_t)$ and $\mathcal{L}_{T_h}(x,y_o)$ both being smaller than $0.01$. Note we can also incorporate the model-ensemble technique in optimizing Eq.(~\ref{eq:identify}).
After obtaining adversarial images, we can then patch via unlearning for mitigation.
\begin{table*}[!t]
\centering
\scalebox{0.8}{
\begin{tabular}{|c|p{1.8cm}|p{1.5cm}|p{1.8cm}|p{1.5cm}|p{1.8cm}|p{1.5cm}|p{1.8cm}|p{1.5cm}|}
\hline
\multirow{2}{*}{Task} & \multicolumn{2}{c|}{Before Unlearning} & %
    \multicolumn{2}{c|}{After applying \name{}} & \multicolumn{2}{c|}{After applying Neural Cleanse} &\multicolumn{2}{c|}{Unlearning using Clean Data}\\
\cline{2-9}
&Classification Accuracy&Attack Success Rate&Classification Accuracy&Attack Success Rate &Classification Accuracy &Attack Success Rate&Classification Accuracy&Attack Success Rate\\
\hline
MNIST(StingRay)&$99.01\%$&$98.2\%$&$98\%$&$0.17\%$&N/A&N/A&$99.01\%$&$88.62\%$\\
\hline
Fashion-MNIST(StingRay)&$92.83\%$&$98.1\%$&$90.09\%$&$2.07\%$&N/A&N/A&$92.01\%$&$80.61\%$\\
\hline
GTSRB(StingRay)&$97.1\%$&$95.16\%$&$96.48\%$&$5.23\%$&N/A&N/A&$97.1\%$&$81.5\%$\\
\hline
CIFAR-10(StingRay)&$92.67\%$&$98.1\%$&$88.51\%$&$7.21\%$&N/A&N/A&$90.03\%$&$83.2\%$\\
\hline
MNIST(Poison Frog)&$98.71\%$&$98.28\%$&$97.61\%$&$0.17\%$&N/A&N/A&$97.85\%$&$88.62\%$\\
\hline
Fashion-MNIST(Poison Frog)&$91.23\%$&$98.1\%$&$89.72\%$&$1.06\%$&N/A&N/A&$91.47\%$&$79.18\%$\\
\hline
GTSRB(Poison Frog)&$96.31\%$&$95.69\%$&$93.19\%$&$5.3\%$&N/A&N/A&$94\%$&$81.5\%$\\
\hline
CIFAR-10(Poison Frog)&$90.97\%$&$97.06\%$&$87.11\%$&$7.62\%$&N/A&N/A&$89.44\%$&$81.18\%$\\
\hline
MNIST(Mislabel)&$99.11\%$&$100\%$&$97.89\%$&$0.13\%$&N/A&N/A&$97.87\%$&$89.31\%$\\
\hline
Fashion-MNIST(Mislabel)&$92.9\%$&$98.1\%$&$90.06\%$&$1.06\%$&N/A&N/A&$91.47\%$&$79.21\%$\\
\hline
GTSRB(Mislabel)&$96.31\%$&$95.72\%$&$93.19\%$&$5.3\%$&N/A&N/A&$93.48\%$&$81.5\%$\\
\hline
CIFAR-10(Mislabel)&$91.29\%$&$97.06\%$&$85.81\%$&$7.64\%$&N/A&N/A&$90.07\%$&$81\%$\\
\hline
MNIST(BadNets)&$98.26\%$&$99.99\%$&$97.24\%$&$0.42\%$&$97.49\%$&$0.59\%$&$97.84\%$&$95.21\%$\\
\hline
GTSRB(BadNets)&$96.79\%$&$100\%$&$93.26\%$&$4.15\%$&$92.91\%$&$0.14\%$&$93.43\%$&$96.16\%$\\
\hline
Fashion-MNIST(BadNets)&$90.93\%$&$100\%$&$87.94\%$&$0.93\%$&$88.01\%$&$0.44\%$&$88.61\%$&$95.46\%$\\
\hline
CIFAR-10(BadNets)&$93.26\%$&$97.14\%$&$90.28\%$&$5.12\%$&$90.17$&$4.14\%$&$89.44\%$&$93.34\%$\\
\hline
MNIST(Trojan)&$97.34\%$&$99\%$&$97.13\%$&$0.12\%$&$97.49\%$&$0.59\%$&$97.84\%$&$10.6\%$\\
\hline
Fashion-MNIST(Trojan)&$90.16\%$&$97\%$&$87.94\%$&$0.9\%$&$88.01\%$&$0.64\%$&$88.61\%$&$10.6\%$\\
\hline
GTSRB(Trojan)&$96.32\%$&$97.61\%$&$94.07\%$&$1.28\%$&$94.61\%$&$1.07\%$&$96.43\%$&$9.26\%$\\
\hline
CIFAR-10(Trojan)&$90.21\%$&$98.26\%$&$91.28\%$&$2.17\%$&$90.17$&$5.13\%$&$89.44\%$&$7.18\%$\\
\hline
MNIST(Chen et al)&$98.47\%$&$100\%$&$97.24\%$&$0.51\%$&N/A&N/A&$97.89\%$&$11\%$\\
\hline
Fashion-MNIST(Chen et al)&$90.49\%$&$98.43\%$&$88.01\%$&$2.51\%$&N/A&N/A&$88.68\%$&$13\%$\\
\hline
GTSRB(Chen et al)&$96.64\%$&$100\%$&$92.71\%$&$6.1\%$&N/A&N/A&$94.17\%$&$10.93\%$\\
\hline
CIFAR-10(Chen et al)&$90.39\%$&$98.31\%$&$88.41\%$&$4.35\%$&N/A&N/A&$89.26\%$&$13.2\%$\\
\hline
\end{tabular}}
\caption{Mitigation performance.}
\vspace{-4mm}
\label{table:mitigation_compare}
\end{table*}

\vspace{-2mm}
\subsection{Mitigation Performance}

In our experiments on evaluating mitigation performance, for each infected model, we retrain the infected model for five iterations with a mixed dataset. To create this mixed dataset, we take $10\%$ samples of the original healthy training data and add crafted adversarial samples which count for $20\%$ of these healthy data. Note that for each crafted adversarial sample, we label it using the corresponding infected label $y_o$, and label all other healthy samples using the correct labels. We use the attack success rate due to the original trigger and classification accuracy after applying our unlearning-based mitigation process as the metric to measure the mitigation performance.
We compare our method against the unlearning-based mitigation method proposed in Neural Cleanse which represents the only existing  method which can patch DNNs infected by backdoor attacks under a practical adversarial model without sacrificing overall performance.

The results are shown in Table~\ref{table:mitigation_compare}.
Columns 2-5 of this table show the attack success rate and classification accuracy before and after applying our mitigation approach. As seen in the table, for each task, our approach manages to reduce the attack success rate to be below $6.1\%$, without noticeably reducing the overall performance (i.e., the largest accuracy reduction is only 3.92\% on Fashion-MNIST infected by Chen Attack).

Next, we compare \name{} with Neural Cleanse (Columns 6-7). We observe that both two approaches are sufficiently effective for mitigating backdoor attacks due to BadNets and Trojan attack. However, similar to detection, under Chen attack and all types of AP attacks, Neural Cleanse cannot identify the infected label and thus fails in the mitigation phase; while \name{} can mitigate these attacks and reduce the attack success rate to $<7.7\%$ under all scenarios.

Finally, we compare against unlearning using only clean training data (no additional triggers), as shown in Columns 8-9 in Table~\ref{table:mitigation_compare}. We observe that unlearning using only clean data is ineffective for all tasks under both BadNets and AP attacks (i.e., still yielding a high attack success rate $\geq 95.46\%$ for BadNets and $\geq79\%$ for AP attacks). Nonetheless, unlearning using only clean training data remains effective for models affected by Chen Attack and Trojan Attack, with attack success rates being reduced from $13\%$ to $7.18\%$. 
This may be due to the fact that Chen Attack and Trojan attack are much more sensitive to training only using clean data. Such clean data may reset certain parameters of the infected model, thus disabling the attack. In contrast, BadNets and AP attacks seem to be insensitive to unlearning using clean training data, which nonetheless can be mitigated by \name{}.

\section{Robustness under Other Attack Variants} \label{sec:robustness}

For evaluating the robustness of \name, we have conducted a set of experiments considering various attack variants (note that several such variants, e.g., multi-label detection, have been evaluated in Sec.~\ref{sec:exp}).

\begin{figure}[!t]
\centering
\small 
\subfigure[Badnets \label{fig:mul_trigger_mnist}]{\includegraphics[width=.45\columnwidth]{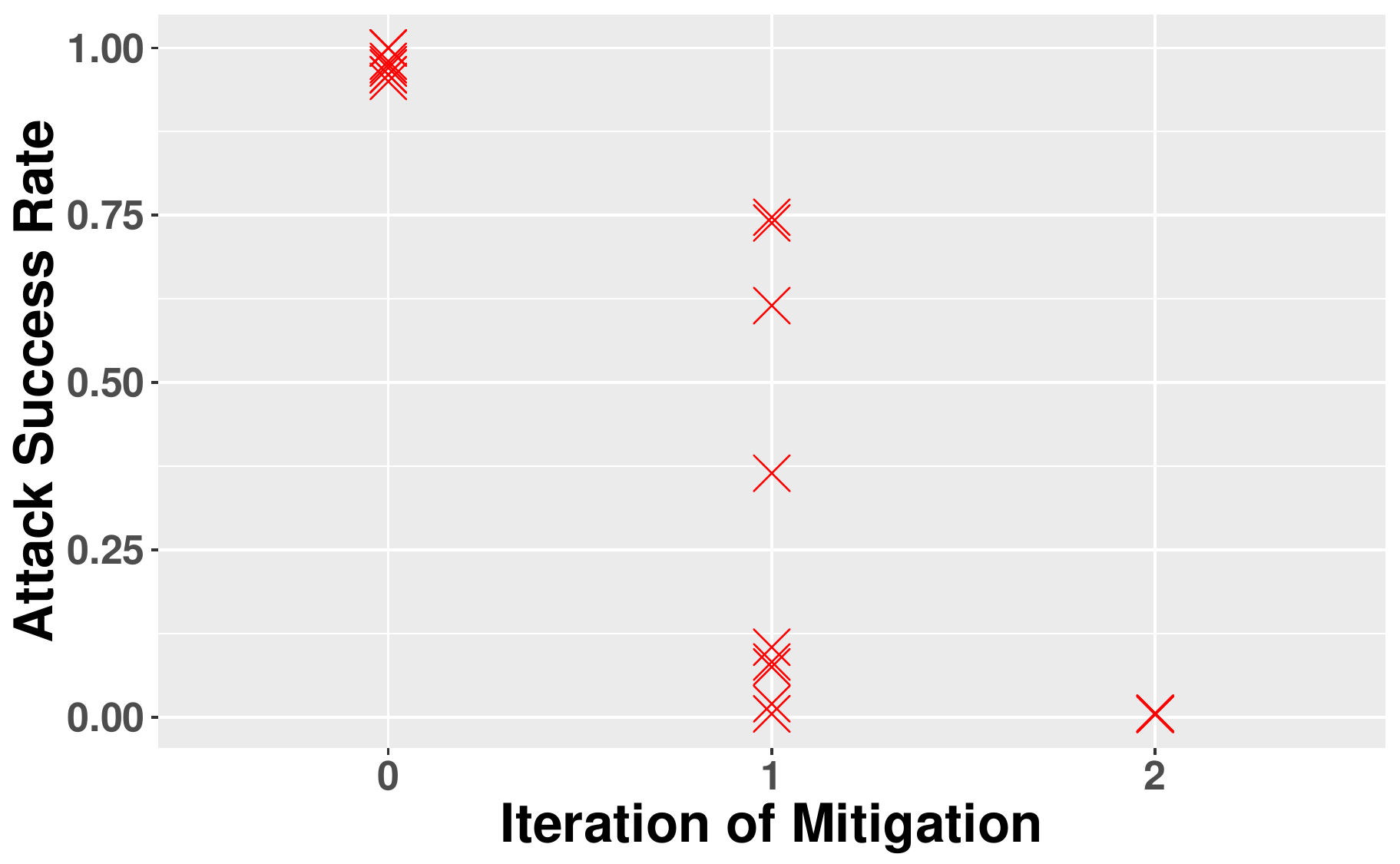}}
 \subfigure[Chen Attack\label{fig:mul_trigger_mnist_noise}]{\includegraphics[width=.45\columnwidth]{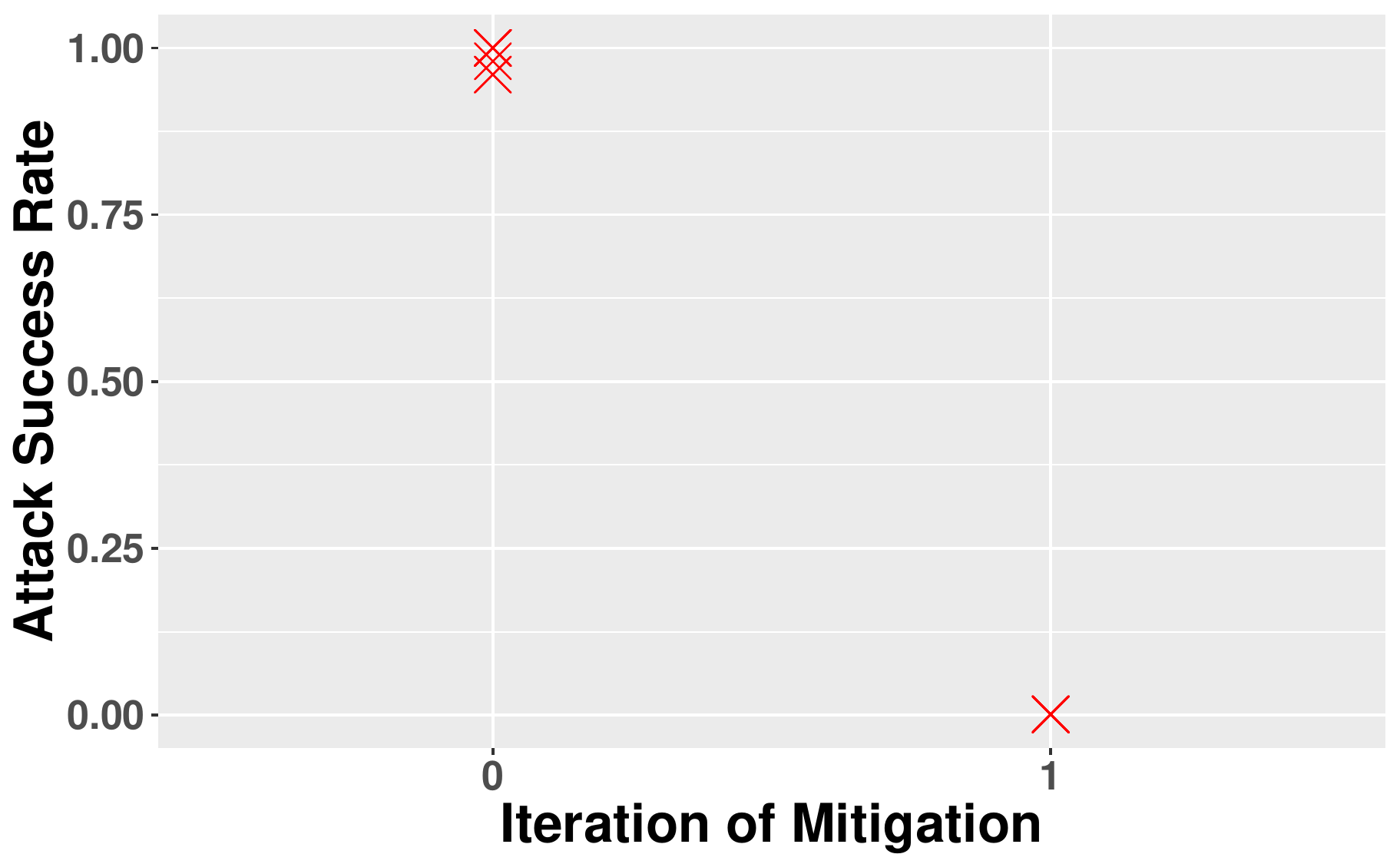}}
 \vspace{-2mm}
\caption{Attack Success Rate and Prob when mitigating models infected by BadNets and Chen Attack.}
\label{fig:single}\normalsize
\vspace{-4mm}
\end{figure}

\noindent\textbf{Single Infected Label due to Multiple Triggers.} 
We also evaluated scenarios where the attacker may apply multiple triggers for infecting a single label. The tested setup is that we apply nine different triggers in the form of either 4x4 white square (BadNets) or random noise with a 0.1 transparency ratio and different patterns (Chen Attack), located at different locations in an image, which yields a $\geq 95\%$ attack success ratio. 

The mitigation results are shown in Fig.~\ref{fig:mul_trigger_mnist},\ref{fig:mul_trigger_mnist_noise}. We observe that with multiple triggers, a single run of our detection and mitigation can only patch a partial set of triggers. Interestingly, after the first run, the Prob of the infected label is still higher than 0.5 as shown in Fig.~\ref{fig:mul_trigger_mnist}, which indicates there may be some remaining triggers. Therefore, we could simply run multiple iterations and successfully mitigate all triggers, i.e., reducing the attack success rate and Prob to nearly zero. 

Interestingly, under Chen Attack implementing the random noise-based trigger,  running \name{} just once could actually patch all patterns as shown in Fig.~\ref{fig:mul_trigger_mnist_noise}. We also test on other datasets including GTSRB, Fashion-MNIST, and CIFAR-10, and the attack success rate for all triggers could be reduced to nearly $5.19\%$, $0.79\%$, $5.72\%$ after running 2-4  iterations. This may be again due to the fact that Chen Attack is sensitive to clean training data as discussed in \ref{sec:miti}.     

\noindent\textbf{Impact due to single-trigger implementation.} In prior experiments, we have tested our approach under the scenario where multiple triggers are located at different positions for each infected label. In this experiment set, we test the setting where only one trigger is used but placed at different locations for each infected label on various datasets.  The results are similar to  Table.~\ref{table:performance_ME}, which demonstrate that \name{} is resilient to different trigger locations.  We also test the setting where complicated trigger patterns (e.g., in terms of trigger shape) are implemented. The results on Trojan and Chen attacks prove the efficacy, as detailed in Table.~\ref{table:robust_2} in the Appendix.

\begin{figure}[!t]
\centering
\subfigure[Backdoor Scenario\label{fig:BD_number}]{\includegraphics[width=0.45\columnwidth]{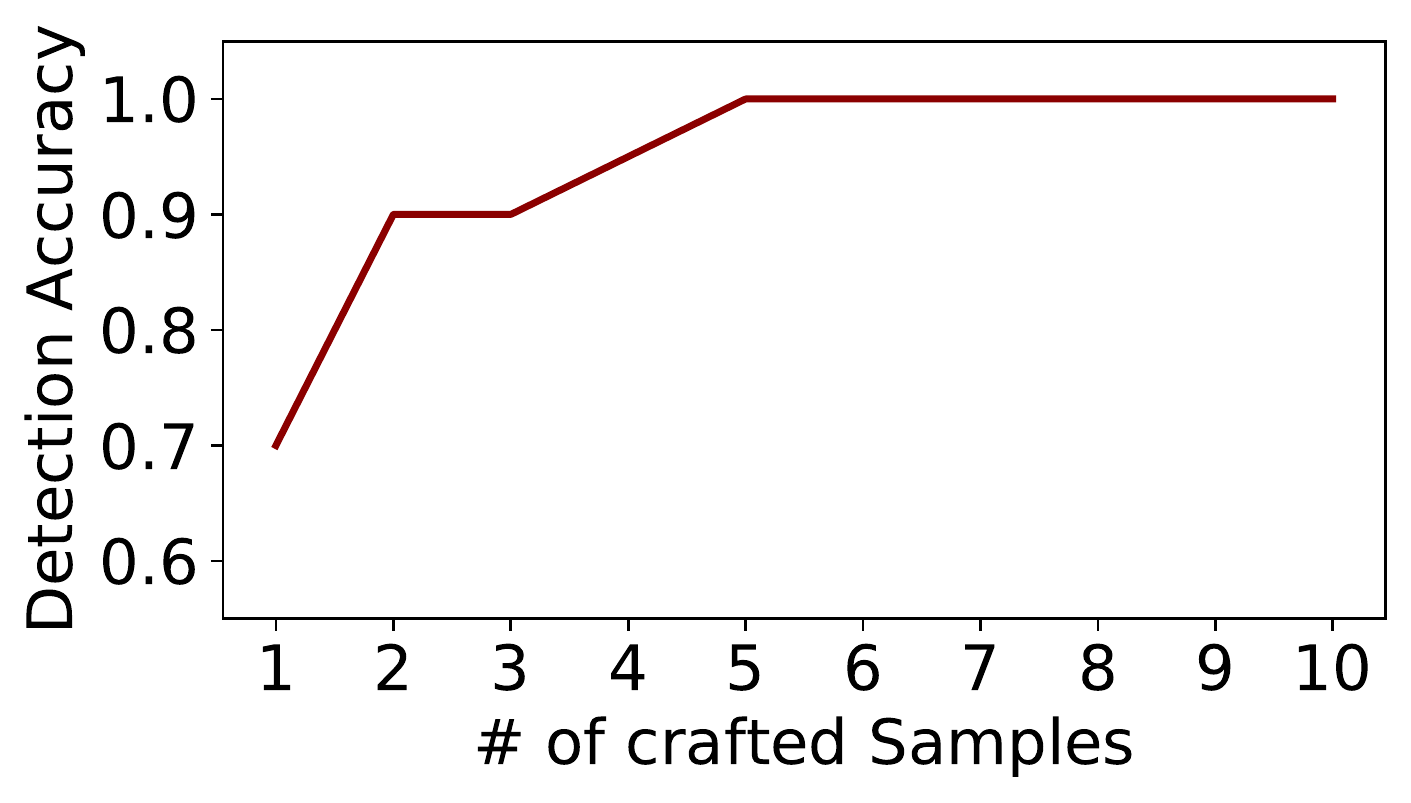}}
\subfigure[AP Scenario\label{fig:AP_number}]{\includegraphics[width=0.45\columnwidth]{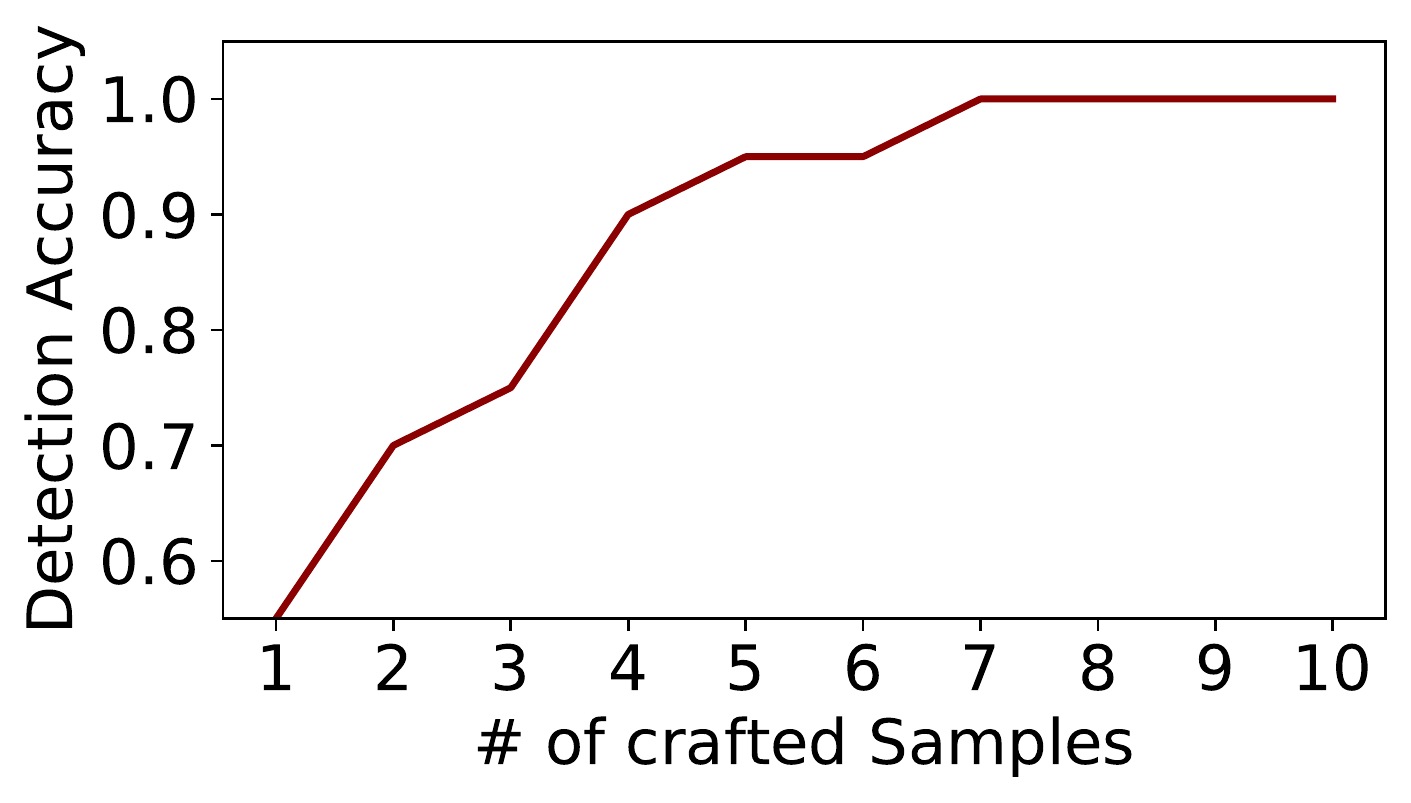}}
\vspace{-2mm}
\caption{Impact  due  to  the  number of crafted $x$.}
\vspace{-4mm}
\label{fig:number_samples}
\end{figure}  
\noindent\textbf{Impact  due  to  the  Number  of  the crafted $x$.}  
Our detection method has been proved effective through extensive experiments under the setting where 100 crafted samples (i.e., $x$) are generated for each detection task. Intuitively, generating more random samples could ensure better detection accuracy. To investigate the impact on detection performance due to the number of crafted samples, we evaluate \name{} again on GTSRB under both backdoor and AP scenarios. For backdoor attack (BadNets), we implement a 4x4 white square trigger; for AP attacks, we implement Poison Frog. 

The results averaged over 100 runs are shown in Fig.~\ref{fig:number_samples}, where the x-axis represents the number of crafted samples, and the y-axis represents the resulting detection accuracy, where detection accuracy denotes the proportion of the runs in which both the infected model and the corresponding specific infected label are correctly detected among 100 runs. We observe that the detection accuracy increases with an increasing number of random samples under both scenarios. Surprisingly, for detecting backdoor, \name{} can ensure $100\%$ detection accuracy using just five crafted samples; while under AP attack scenario, seven  crafted samples could already ensure detection efficacy. We have also tested on MNIST , Fashion-MNIST and CIFAR-10, showing that the minimum number of crafted samples which could ensure detection efficacy ranges from four to ten under different attack techniques. Thus, crafting ten samples is sufficient to ensure successful detection in most cases, which could yield a significantly reduced computation overhead in practice. 

\textbf{Unawareness of the original training dataset.}
In previous experiments, we use publicly-available training datasets (e.g.,MNIST, GTSRB) as clean datasets to create  model $T_h$, which may also be used by the given model $T$ to be analyzed. Since it is impractical for the defender to access the original training dataset used for training $T$, we  consider herein the scenario where the clean portion of the datasets used by the attacker and defender are partially or completely different. 
In this set of experiments, we create an infected model and five clean models on CIFAR-10 following the configuration of previous experiments. These models all exhibit different structures and parameters, and the clean training data for each model uses different data augmentation configuration, indicating that the clean portion of the training datasets used for training $T$ and models in the model ensemble are at least partially different. We use the five clean models to analyze the infected model under our model ensemble approach for the complex CIFAR-10 dataset. The results are the same as previous ones, yielding a 100\% Prob and a 0\% FP rate.

We also test on MNIST, where we divide the original cleaning training dataset contains 60000 images equally into 3 subsets, and create two models with different architectures trained using two of the three subsets. Note that these three subsets are completely different. These two models can achieve a $\geq 98.8\%$ classification accuracy.  
We then use the third subset plus poisoned data to train an infected model, and analyze the infected model using the two clean models as $T_h$ under \name{}. The obtained results are also the same as the earlier-described experiments performed on MNIST. This demonstrates that \name{} is still effective when the  defender is totally unaware of the original training data used to train a given model $T$.

\vspace{-2mm}
\section{Conclusion}

In this paper, we present~\name, a practical and robust technique for detecting and mitigating backdoor and AP attacks on neural networks.
\name{} crafts test inputs to a given DNN, seeking to reveal necessary internal properties of the decision region space belonging to each label and detect whether a label is infected. Through extensive implementation and evaluation of \name{} against a set of state-of-the-art AP attacks on  widely studied datasets, \name proves to be effective and robust under various settings. \name is also effective on detecting backdoor attacks, particularly comparing to state-of-the-art backdoor detection methods.

\bibliographystyle{plain}
\bibliography{usnix_security}

\section{Appendix}

\noindent\textbf{\Huge{$\cdot$}\normalsize Table~\ref{table:acc}: Additional empirical data supporting  Sec~\ref{sec:detectorModel}}

\begin{table}[H]
\centering
\scalebox{0.7}{
\begin{tabular}{ c c c c c c c  } 
\hline
Dataset  & minimum required accuracy for pre-trained models\\
\hline\hline
MNIST&$60\%$\\
Fashion-MNIST&$70\%$\\
GTSRB&$76\%$\\
CIFAR&$73\%$\\

\hline\hline
\end{tabular}}
\caption{The  minimum required accuracy for pre-trained models for each task.}
\label{table:acc}
\vspace{-3mm}
\end{table}

\begin{figure}[H]
\includegraphics[width=1\columnwidth]{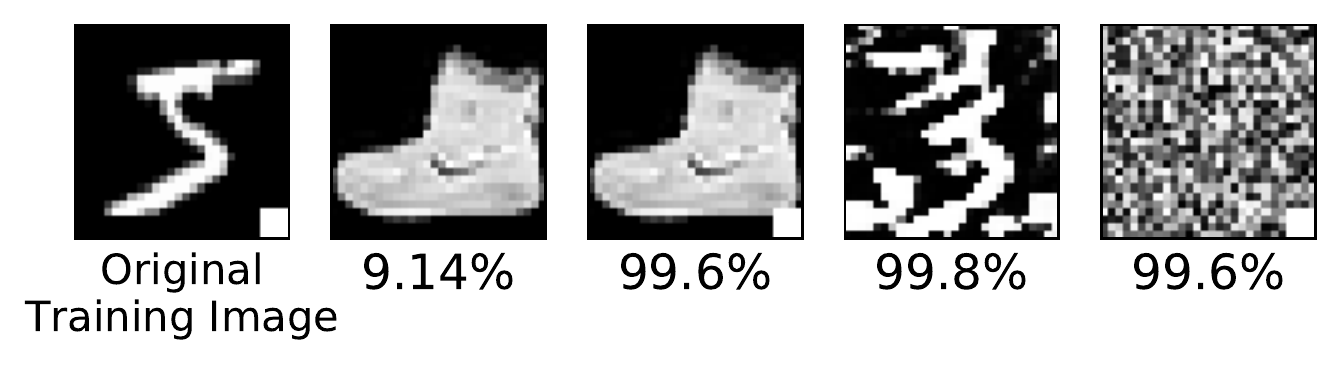}
\includegraphics[width=1\columnwidth]{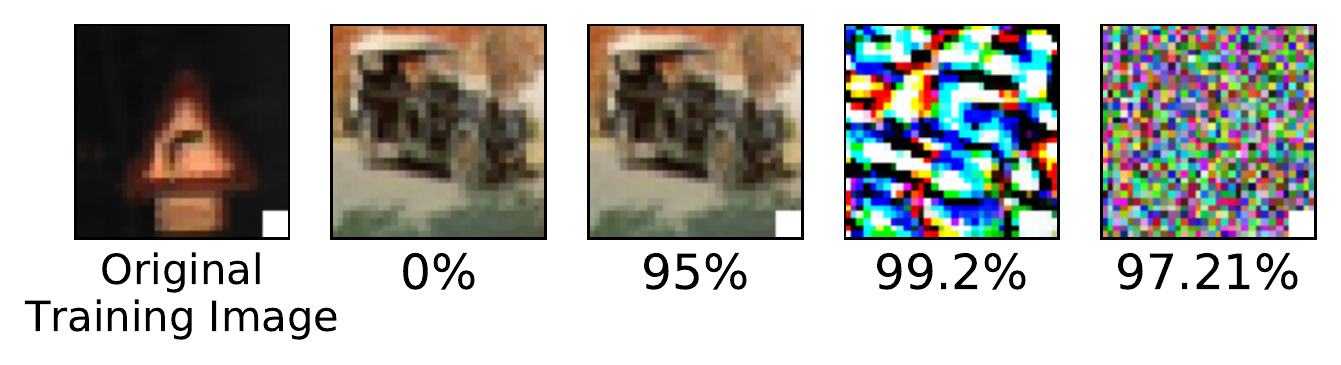}
\centering
\caption{Illustration on the implementation of \name{} (backdoor scenario) using MNIST and GTSRB and a 4x4 white square backdoor trigger. The five columns (from left to right) show a poisoned image with the backdoor trigger from the training set, an image irrelevant to the training set, an irrelevant image attached with the same trigger, an image modified based upon the original image and attached with a modified trigger, and a random noise-based image attached with the backdoor trigger, respectively. Interestingly, the images in the last three columns can be  misclassified  as  the  infected  label  with  high  confidence  (the number shown below each image in the figure) by the infected model. This suggests that changing the contend of the image or slightly modify the backdoor trigger does not impact attacking efficacy.}
\label{fig:gl2}
\end{figure}

\begin{figure}[htb]
\includegraphics[width=1\columnwidth]{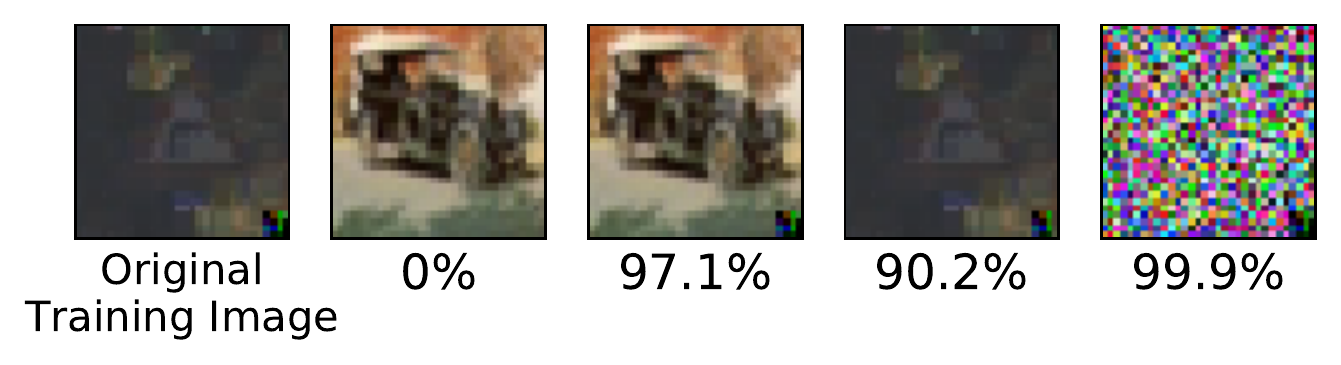}
\centering
\caption{Additional supporting data under the Trojan attack scenario}
\label{fig:ob2_trojan_appendix}
\end{figure}

\begin{figure}[H]
\includegraphics[width=1\columnwidth]{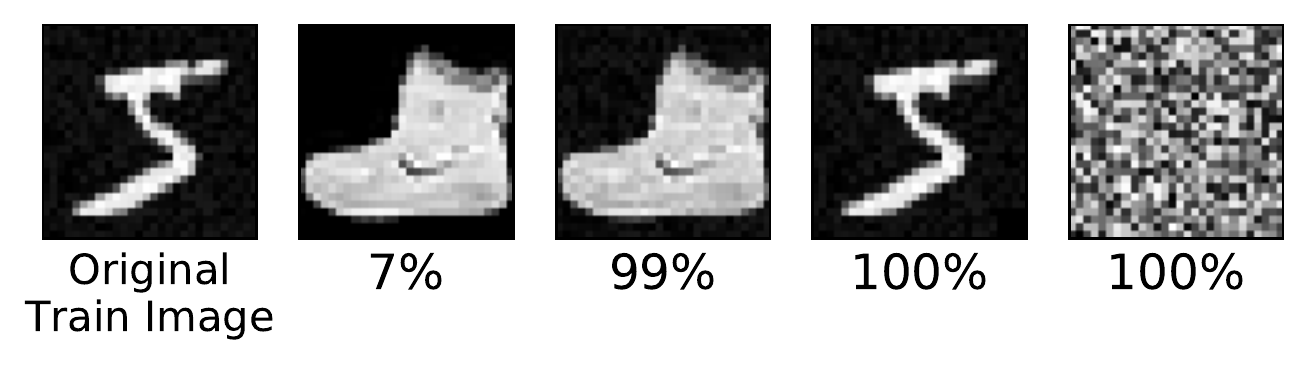}
\centering
\includegraphics[width=1\columnwidth]{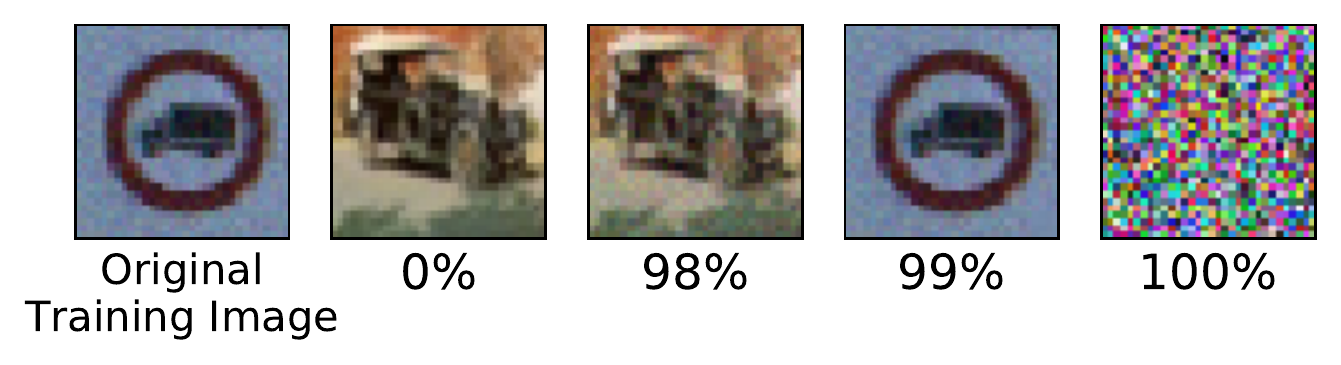}
\caption{Additional supporting data under the Chen attack scenario}
\label{fig:ob2_chen_appendix}
\end{figure}

\noindent\textbf{\Huge{$\cdot$}\normalsize  Figs.~\ref{fig:gl2},~\ref{fig:ob2_trojan_appendix} and \ref{fig:ob2_chen_appendix}: Additional empirical data supporting claims discussed in Sec~\ref{sec:Implementaion}}

\noindent\textbf{\Huge{$\cdot$}\normalsize Table~\ref{table:Loss_Backdoor}: Additional empirical data supporting  Sec~\ref{sec:Implementaion}.}

\begin{table}[H]
\centering
\scalebox{0.7}{
\begin{tabular}{ c c c c c c c  } 
\hline
Dataset  & loss value under Model $T$&loss value under Model $T_h$\\
\hline\hline
MNIST(BadNets)&0&47\\
MNIST(Trojan Attack)&0&53\\
MNIST(Chen et al)&0&49\\
Fashion-MNIST(BadNets)&0&42\\
Fashion-MNIST(Trojan Attack)&0&43\\
Fashion-MNIST(Chen et al)&0.03&49\\
GTSRB(BadeNets)&0&114\\
GTSRB(Trojan Attack)&0&107\\
GTSRB(Chen et al)&0.01&97\\
\hline\hline
\end{tabular}}
\caption{The  Cross  Entropy  Loss  value  computed on the  poisoned  input  and its infected  label  under a healthy  and  a  corresponding infected  model.}
\label{table:Loss_Backdoor}
\vspace{-3mm}
\end{table}

\vspace{4mm}
\noindent\textbf{\Huge{$\cdot$}\normalsize Details on the dataset/task and model configurations as discussed in Sec~\ref{sec:exp}:} 

\textit{Hand-written Digit Recognition (MNIST)}.This task is often used to evaluate DNN. The dataset contains 60K training data and 10K test. It contains 10 labels (0-9).
    
\textit{Fashion Item Recognition (Fashion-MNIST)}.
Fashion-MNIST dataset is an image dataset comprising of 28x28 grayscale images of 70, 000 fashion products from 10 categories, with 7, 000 images per category. The training set has 60, 000 images and the test set has 10, 000 images. Fashion-MNIST is intended to serve as a direct dropin replacement for the original MNIST dataset for benchmarking machine learning algorithms, as it shares the same image size, data format and the structure of training and testing splits.

\textit{Traffic Sign Recognition (GTSRB)}. This dataset contains 39.2K colored training images and 12.6K testing images. Its goal is to recognize 43 different traffic signs.

\textit{Image Recognition (CIFAR-10).} The CIFAR-10 dataset (Canadian Institute For Advanced Research) is a collection of images that are commonly used to train machine learning and computer vision algorithms. It is one of the most widely used datasets for machine learning research. The CIFAR-10 dataset contains 60,000 32x32 color images in 10 different classes. The 10 different classes represent airplanes, cars, birds, cats, deer, dogs, frogs, horses, ships, and trucks. There are 6,000 images within each class. In this tasks, we use several complicated state-of-art models(i.e. RESNet, DenseNet,etc) as experimental models and implement data augmentation on the training dataset to improve models performance following the configuration as previous work. The diversity of its state-of-art models increase the difference between model $T_h$ and T, and is a good candidate to evaluate the effectiveness of our detection method under different model $T_h$ and T with quite different architecture.

\vspace{2mm}
\noindent\textbf{\Huge{$\cdot$}\normalsize Table~\ref{table:APConfiguration}: Details of the infected label configuration under AP attack scenarios. as discussed in Sec~\ref{sec:exp}}

\begin{table}[H]
\centering
\begin{tabular}{ c c c } 
\hline
Dataset &Adversarial Label\\
\hline\hline
MNIST&Shirts\\
Fashion-MNIST&Digit number 7\\
GTSRB&Air-planes\\
CIFAR-10&Flowers\\
\hline
\end{tabular}
\caption{Infected label configuration under AP attack scenarios.}
\label{table:APConfiguration}
\end{table}

\vspace{4mm}
\noindent\textbf{\Huge{$\cdot$}\normalsize  Table~\ref{table:setup}: Attack success rate and classification accuracy of backdoor and AP attack on classification tasks as discussed in Sec.~\ref{sec:exp}.}

\begin{table}[H]
\centering
\scalebox{0.55}{
\begin{tabular}{|c|c|c|c|}
\hline
\multirow{2}{*}{Task} & \multicolumn{2}{c|}{Infected Model} & %
    \multirow{2}{*}{Normal Model Classification Accuracy}\\
\cline{2-3}
 & Classification Accuracy & Attack Success Rate\\
\hline
MNIST(Backdoor)&$\geq 97.34\%$&$\geq 99\%$&$\geq99.13\%$\\      
\hline
Fashion-MNIST(Backdoor)&$\geq90.16\%$&$\geq97\%$&$\geq92.19\%$\\
\hline
GTSRB(Backdoor)&$\geq96.32\%$&$\geq97.61\%$&$\geq97.21\%$\\
\hline
CIFAR-10(Backdoor)&$\geq90.21\%\%$&$\geq98.26\%$&$\geq93.31\%$\\
\hline
MNIST(AP)&$\geq 99.01\%$&$\geq 98.2\%$&$\geq99.13\%$\\    
\hline
Fashion-MNIST(AP)&$\geq92.83\%$&$\geq98.1\%$&$\geq92.19\%$\\
\hline
GTSRB(AP)&$\geq96.31\%$&$\geq96.16\%$&$\geq97.21\%$\\
\hline
CIFAR-10(AP)&$\geq92.67\%\%$&$\geq98.1\%$&$\geq93.31\%$\\
\hline
\end{tabular}}
\caption{Attack success rate and classification accuracy of backdoor and AP
attack on classification tasks.}
\label{table:setup}
\end{table}

\begin{figure}[H]
\centering
\subfigure[StingRay \label{fig:fmnn}]{\includegraphics[width=0.3\columnwidth]{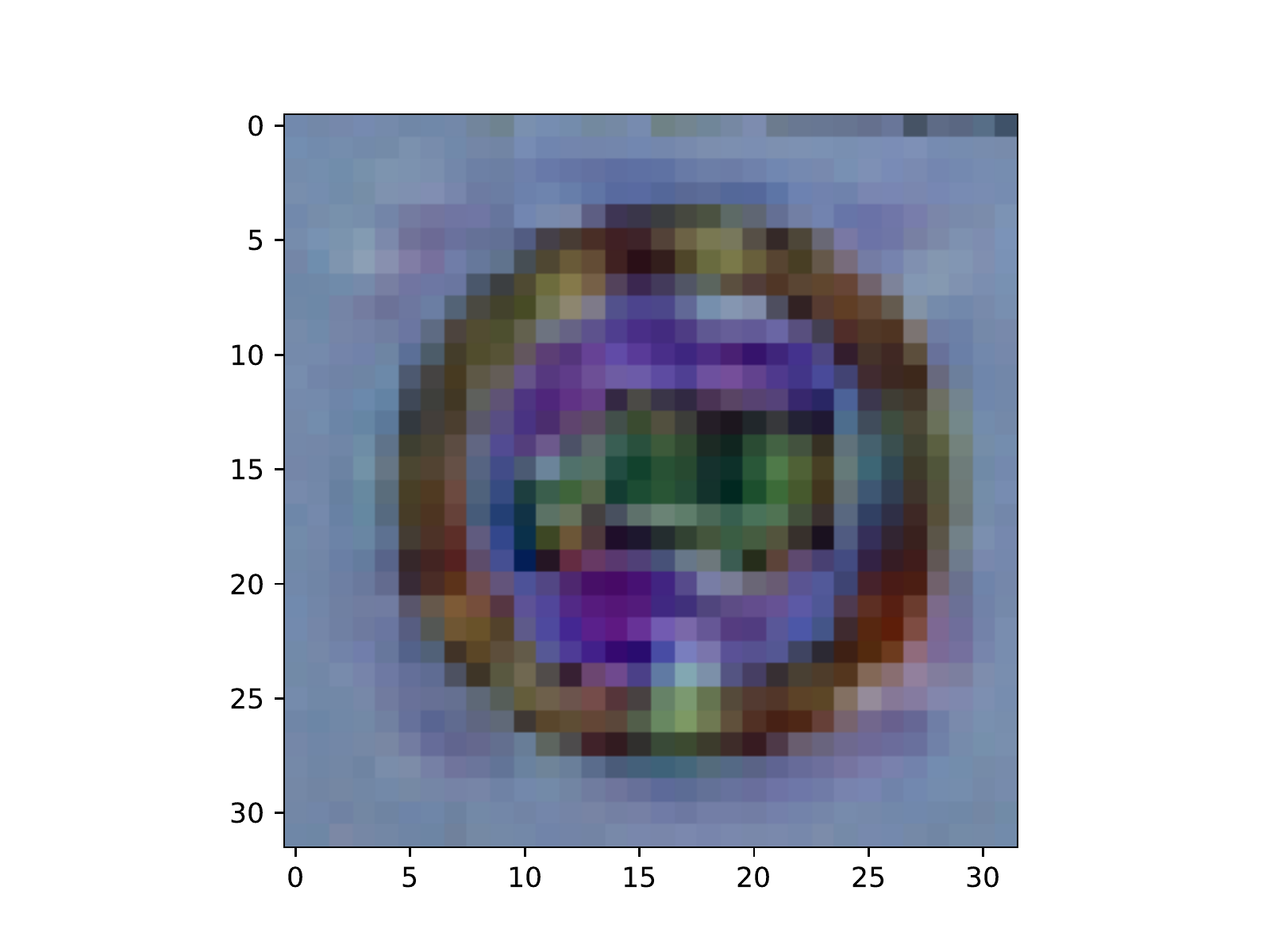}}
\subfigure[PoisonFrog\label{fig:mnn}]{\includegraphics[width=0.3\columnwidth]{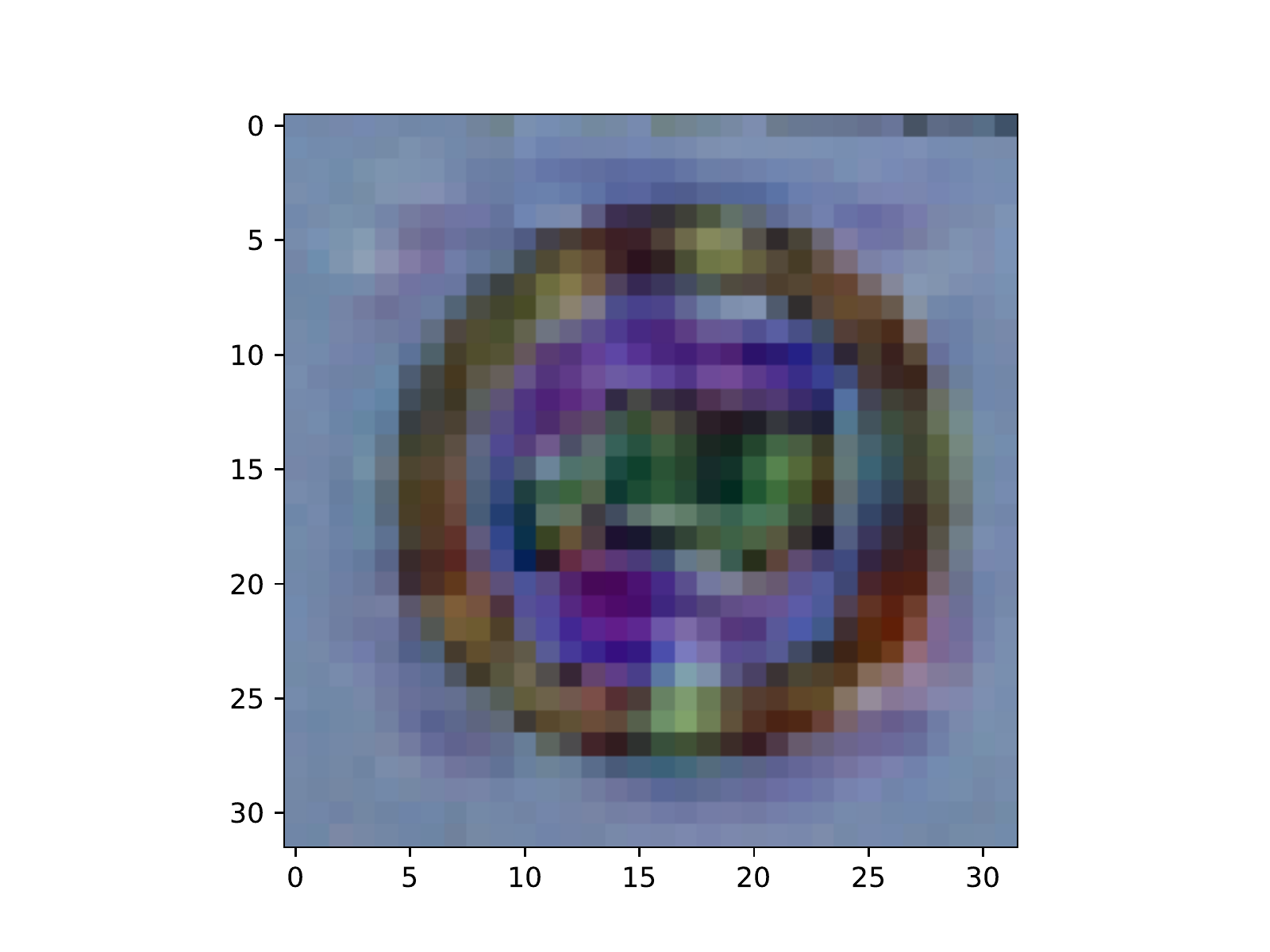}}
\subfigure[Mislabel Attack\label{fig:gbn}]{\includegraphics[width=0.3\columnwidth]{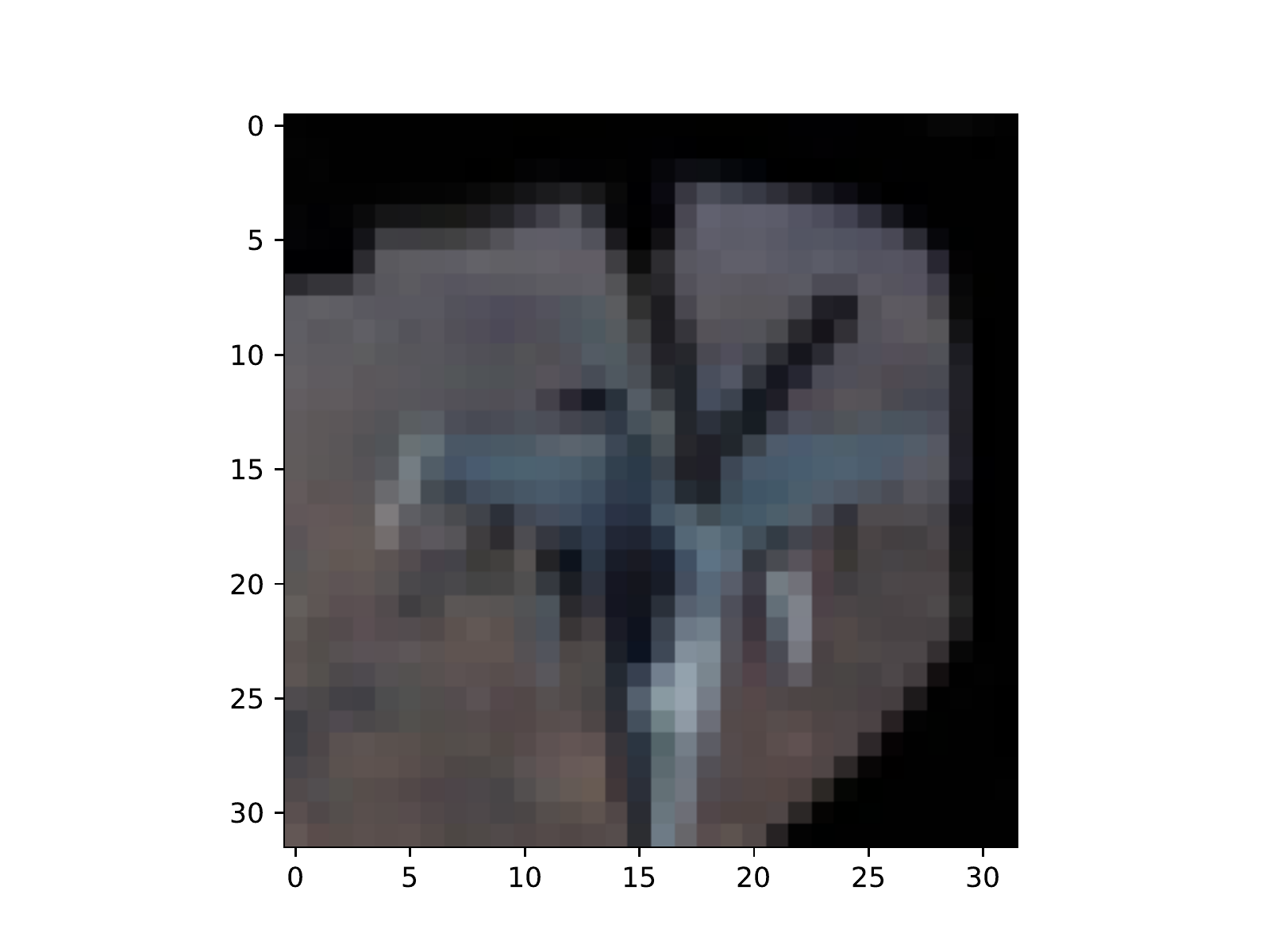}}
\subfigure[BadNets\label{fig:mnb}]{\includegraphics[width=0.3\columnwidth]{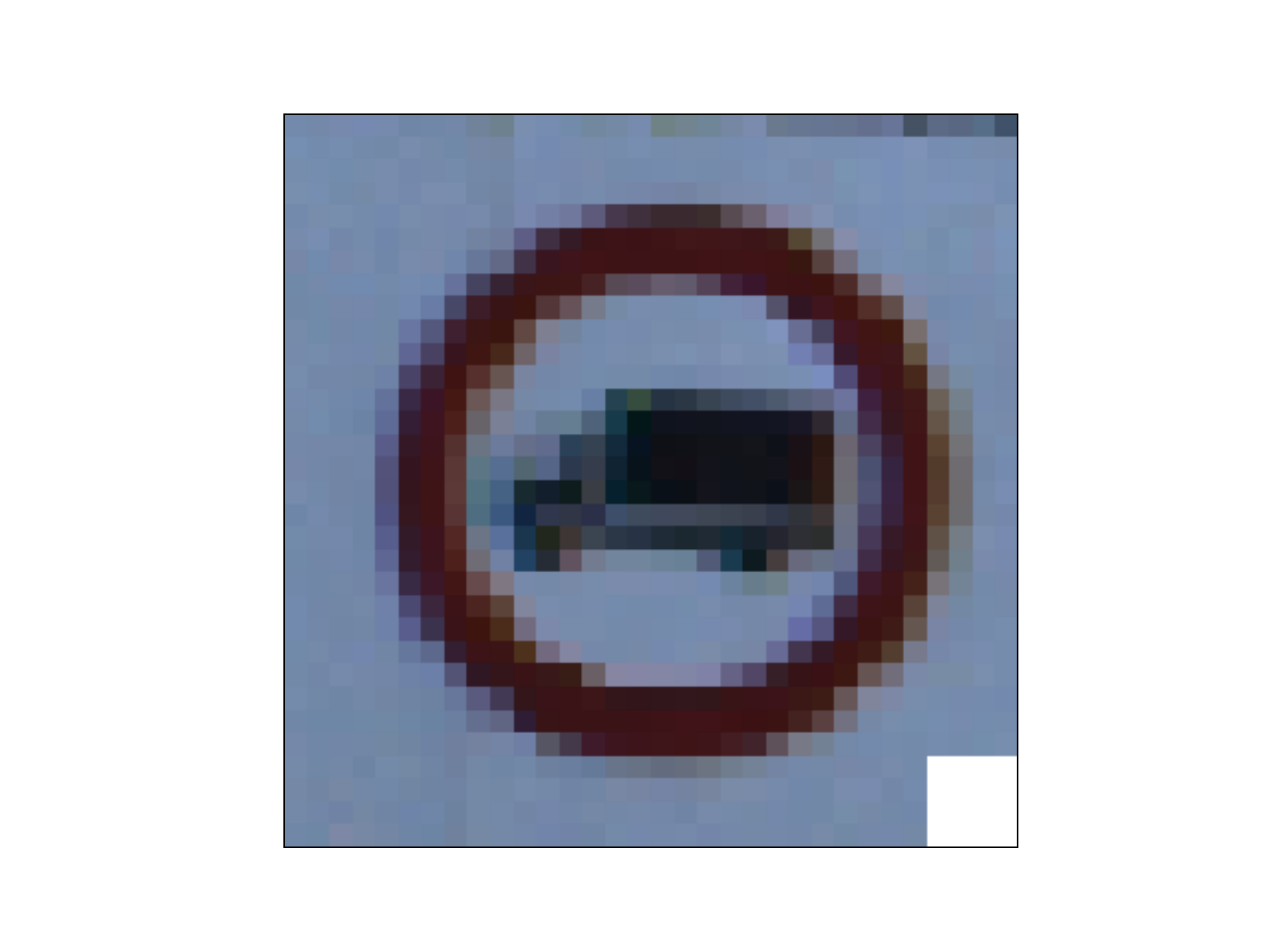}}
\subfigure[Trojan Attack\label{fig:fmnb}]{\includegraphics[width=0.3\columnwidth]{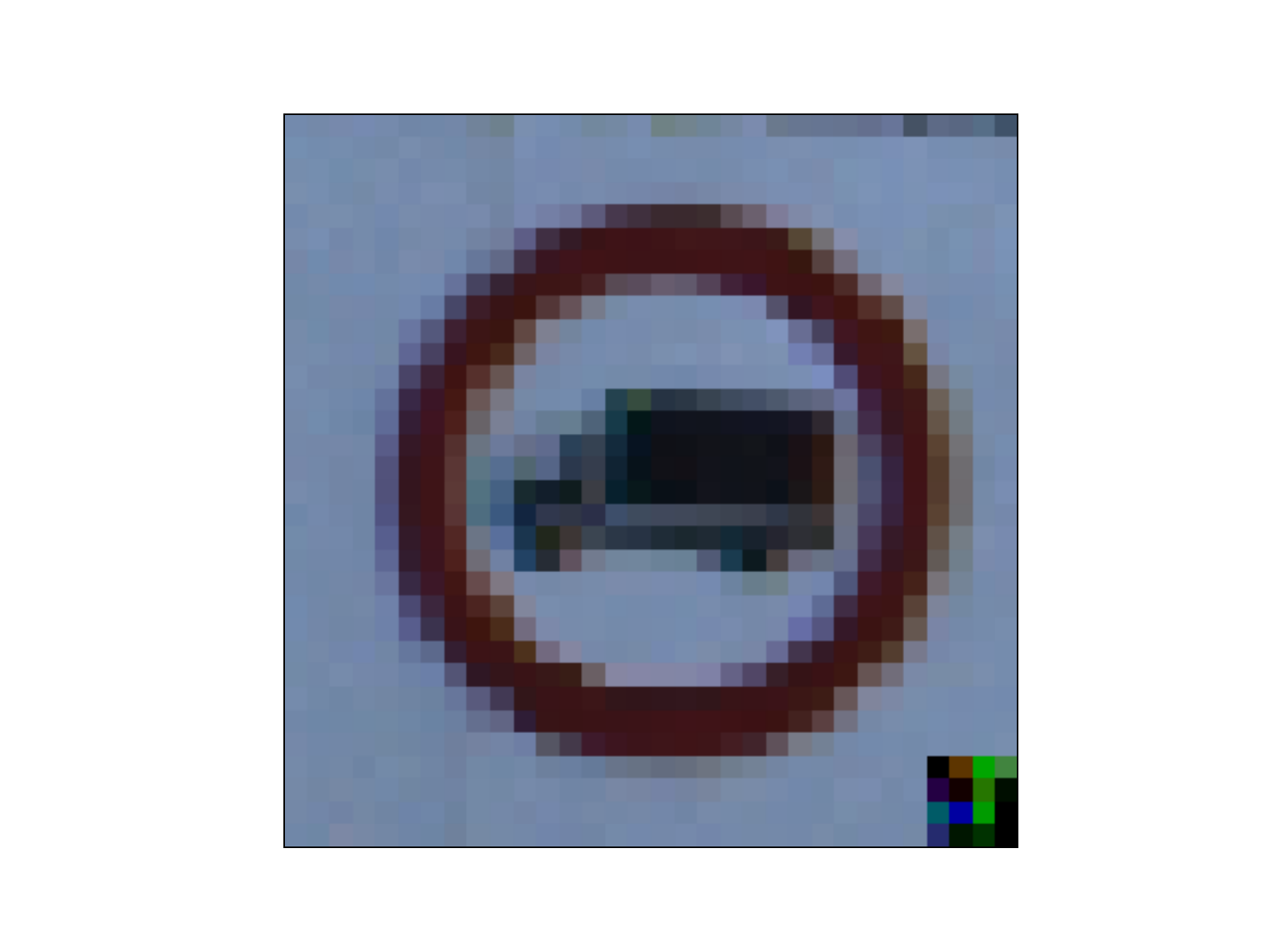}}
\subfigure[Chen et al Attack\label{fig:gb}]{\includegraphics[width=0.3\columnwidth]{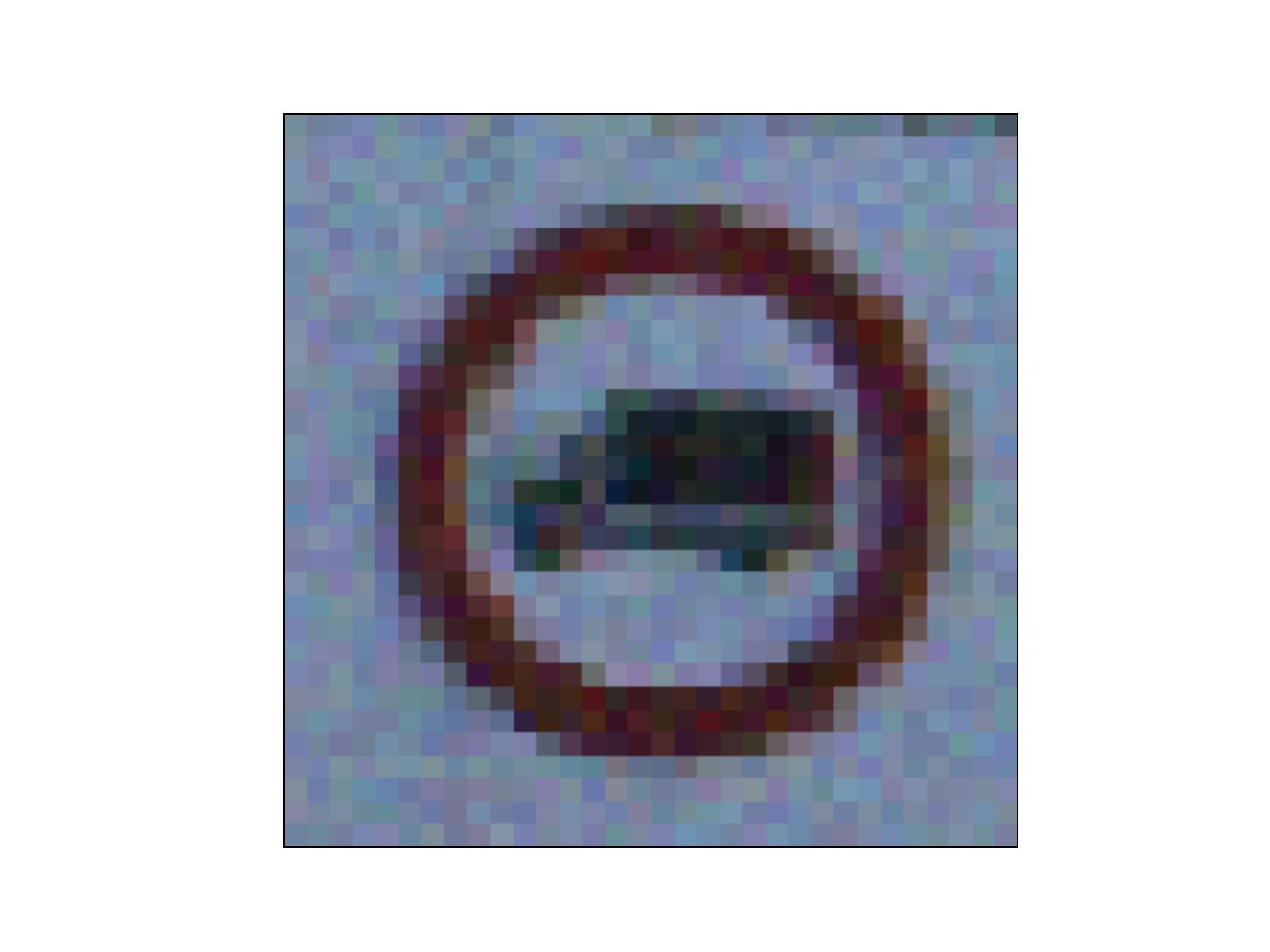}}
\caption{Examples of poisoned images for AP and backdoor attacks.}
\label{fig:sample_back}
\end{figure}

\noindent\textbf{\Huge{$\cdot$}\normalsize  Fig.~\ref{fig:sample_back}: Examples of poisoned images for each attack technique as discussed in Sec~\ref{sec:exp}}

\noindent\textbf{\Huge{$\cdot$}\normalsize  Tables~\ref{table:MNIST1}-\ref{table:training_configuration}: The specific structure and parameters of  models used in our evaluation for various datasets as discussed in Sec~\ref{sec:exp}.} 
The structures of models in the model Pool for MNIST, Fashion-MNIST, GTSRB datasets are shown in Tables~\ref{table:MNIST1},~\ref{table:MNIST2},~\ref{table:MNIST3},~\ref{table:MNIST4},~\ref{table:MNIST5} and~\ref{table:MNIST6}. The training configuration for these models is shown in Table.~\ref{table:training_configuration} 
The models for CIFAR-10 are listed as follows: RESNet.V1.44, RESNet.V1.50, RESNet.V1.56, RESNet.V2.38, RESNet.V2.47, DenseNet-40, DenseNet-121. Note that 
we implemented these models using the mentioned structures and training configurations following prior works~\cite{huang2017densely,he2016deep}.

\begin{table}[H]
\centering
\begin{tabular}{ c c c  } 
\hline
Layer Type & Known Model\\
\hline
Convolution + ReLU&3$\times$3$\times$16\\
Convolution + ReLU&3$\times$3$\times$16\\
Max Pooling&2$\times$2\\
Convolution + ReLU&3$\times$3$\times$32\\
Convolution + ReLU&3$\times$3$\times$32\\
Max Pooling&2$\times$2\\
Fully Connected + ReLU&256\\
Fully Connected + ReLU&256\\
Softmax&10\\
\hline

\end{tabular}
\caption{The Structure of Model I}
\label{table:MNIST1}
\end{table}

\begin{table}[H]
\centering
\begin{tabular}{ c c c  } 
\hline
Layer Type & Known Model\\
\hline
Convolution + ReLU&3$\times$3$\times$32\\
Convolution + ReLU&3$\times$3$\times$32\\
Max Pooling&2$\times$2\\
Convolution + ReLU&3$\times$3$\times$64\\
Convolution + ReLU&3$\times$3$\times$64\\
Max Pooling&2$\times$2\\
Fully Connected + ReLU&200\\
Fully Connected + ReLU&200\\
Softmax&10\\
\hline
\end{tabular}
\caption{The Structure of Model II}
\label{table:MNIST2}

\end{table}

\begin{table}[H]
\centering
\begin{tabular}{ c c c  } 
\hline
Layer Type & Known Model\\
\hline
Convolution + ReLU&5$\times$5$\times$16\\
Convolution + ReLU&5$\times$5$\times$16\\
Max Pooling&2$\times$2\\
Convolution + ReLU&5$\times$5$\times$32\\
Convolution + ReLU&5$\times$5$\times$32\\
Max Pooling&2$\times$2\\
Fully Connected + ReLU&512\\
Softmax&10\\
\hline
\end{tabular}
\caption{The Structure of Model III }
\label{table:MNIST3}

\end{table}

\begin{table}[H]
\centering
\begin{tabular}{ c c c  } 
\hline
Layer Type & Known Model\\
\hline
Convolution + ReLU&3$\times$3$\times$64\\
Convolution + ReLU&3$\times$3$\times$64\\
Max Pooling&2$\times$2\\
Convolution + ReLU&3$\times$3$\times$128\\
Convolution + ReLU&3$\times$3$\times$128\\
Max Pooling&2$\times$2\\
Fully Connected + ReLU&512\\
Fully Connected + ReLU&256\\
Softmax&10\\
\hline
\end{tabular}
\caption{The Structure of Model IV}
\label{table:MNIST4}

\end{table}

\begin{table}[H]
\centering
\begin{tabular}{ c c c  } 
\hline
Layer Type & Known Model\\
\hline
Convolution + ReLU&3$\times$3$\times$32\\
Convolution + ReLU&3$\times$3$\times$32\\
Max Pooling&2$\times$2\\
Convolution + ReLU&3$\times$3$\times$64\\
Convolution + ReLU&3$\times$3$\times$64\\
Max Pooling&2$\times$2\\
Fully Connected + ReLU&512\\
Fully Connected + ReLU&256\\
Fully Connected + ReLU&256\\
Softmax&10\\
\hline
\end{tabular}
\caption{The Structure of Model V}
\label{table:MNIST5}
\end{table}

\begin{table}[H]
\centering
\begin{tabular}{ c c c  } 
\hline
Layer Type & Known Model\\
\hline
Convolution + ReLU&3$\times$3$\times$64\\
Convolution + ReLU&3$\times$3$\times$64\\
Max Pooling&2$\times$2\\
Convolution + ReLU&3$\times$3$\times$128\\
Convolution + ReLU&3$\times$3$\times$128\\
Max Pooling&2$\times$2\\
Fully Connected + ReLU&512\\
Softmax&10\\
\hline
\end{tabular}
\caption{The Structure of Model VI}
\label{table:MNIST6}
\end{table}

\begin{table}[H]
\centering
\begin{tabular}{ c c} 
\hline
Parameter & Models(MNIST,Fashion-MNIST,GTSRB)\\
\hline
Learning Rate&0.01\\
Momentum&0.9\\
Dropout&0.5\\
Batch Size&128\\
Epochs&50\\
\hline
\end{tabular}
\caption{Training Configuration}
\label{table:training_configuration}
\end{table}

\begin{figure}[H]
\centering

\subfigure[MNIST(backdoor)\label{fig:mnb}]{\includegraphics[width=0.45\columnwidth]{ME_changing/MNIST_Backdoor.pdf}}
\subfigure[FashionMNIST
(backdoor)\label{fig:fmb}]{\includegraphics[width=0.45\columnwidth]{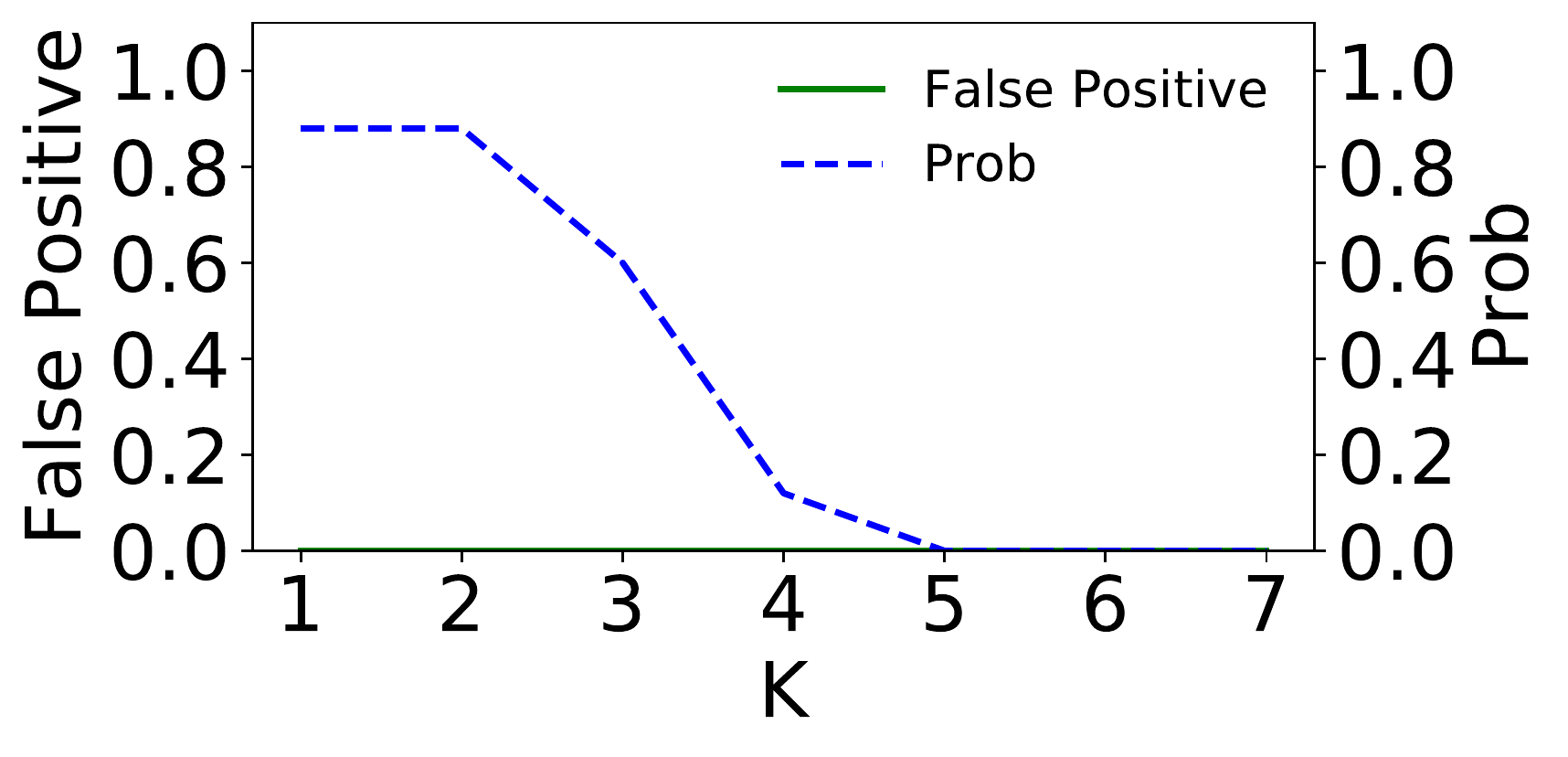}}
\subfigure[GTSRB(backdoor)\label{fig:GTSRBb}]{\includegraphics[width=0.45\columnwidth]{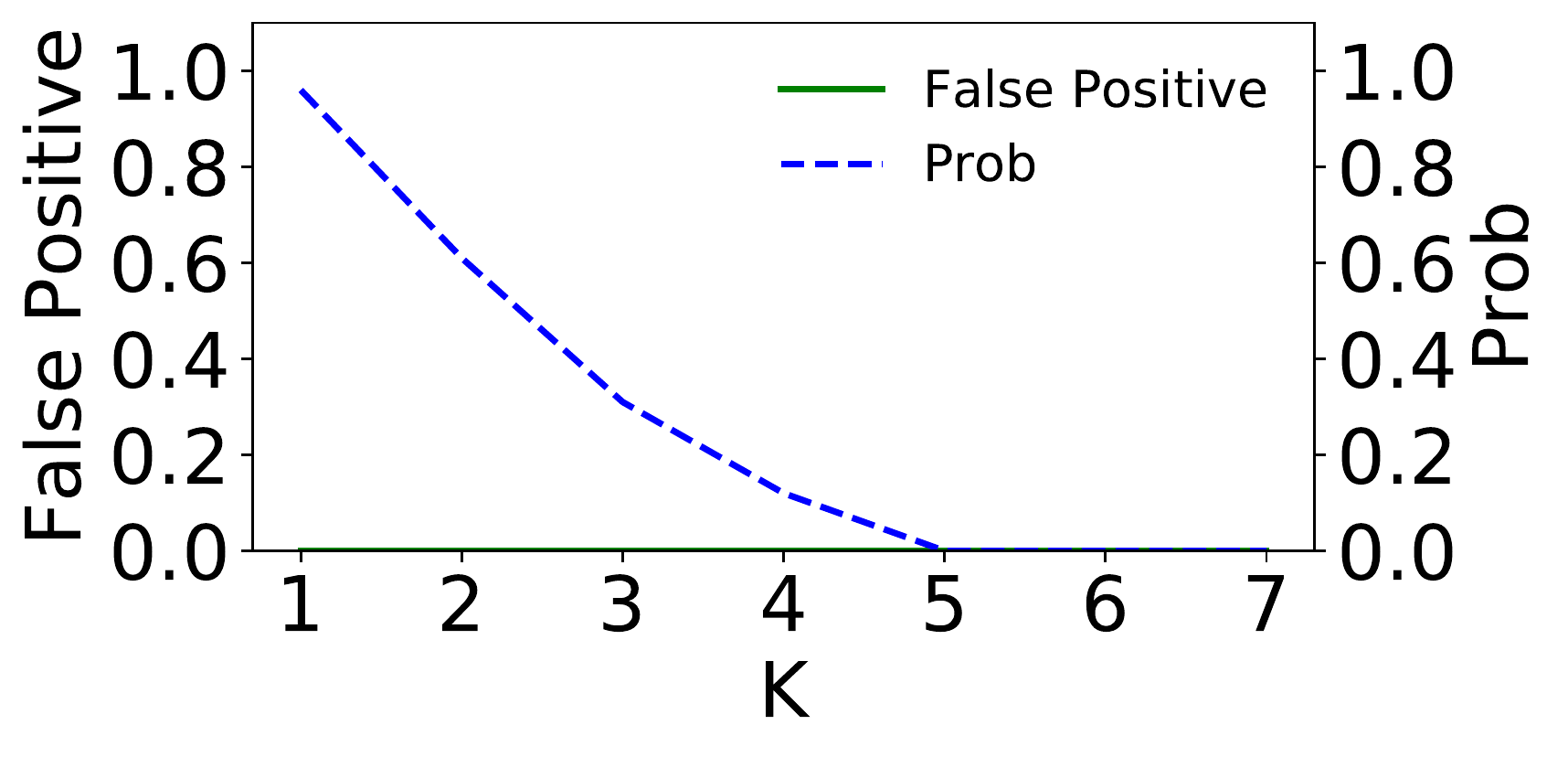}}
\subfigure[CIFAR-10(backdoor)\label{fig:cifarb}]{\includegraphics[width=0.45\columnwidth]{ME_changing/CIFAR-10_Backdoor.pdf}}
\subfigure[MNIST(AP)\label{fig:mnAP}]{
\includegraphics[width=0.45\columnwidth]{ME_changing/MNIST_AP.pdf}}
\subfigure[Fashion-MNIST(AP\label{fig:fmAP})]{\includegraphics[width=0.45\columnwidth]{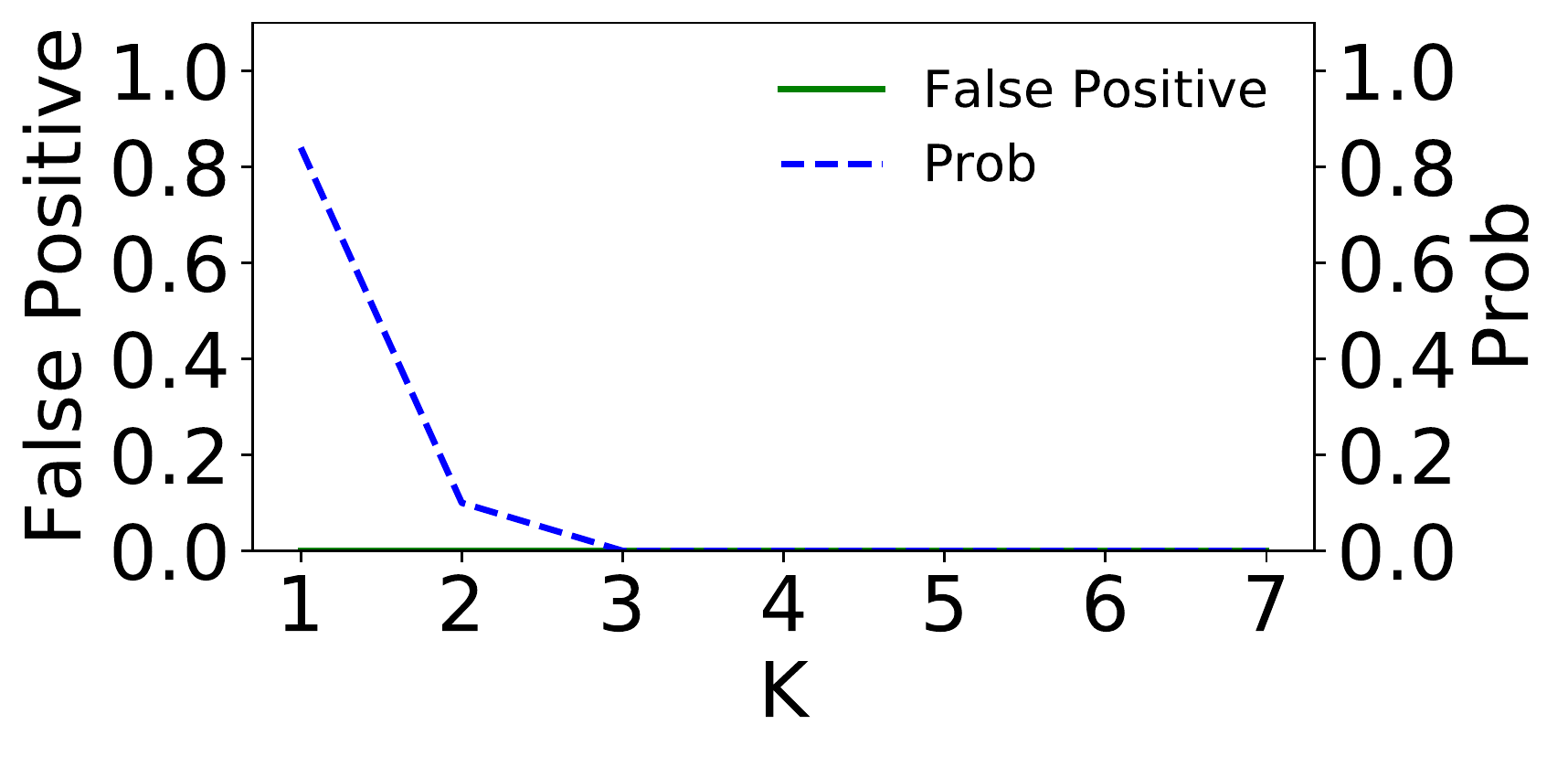}}
\subfigure[GTSRB(AP)\label{fig:GTSRBAP}]{\includegraphics[width=0.45\columnwidth]{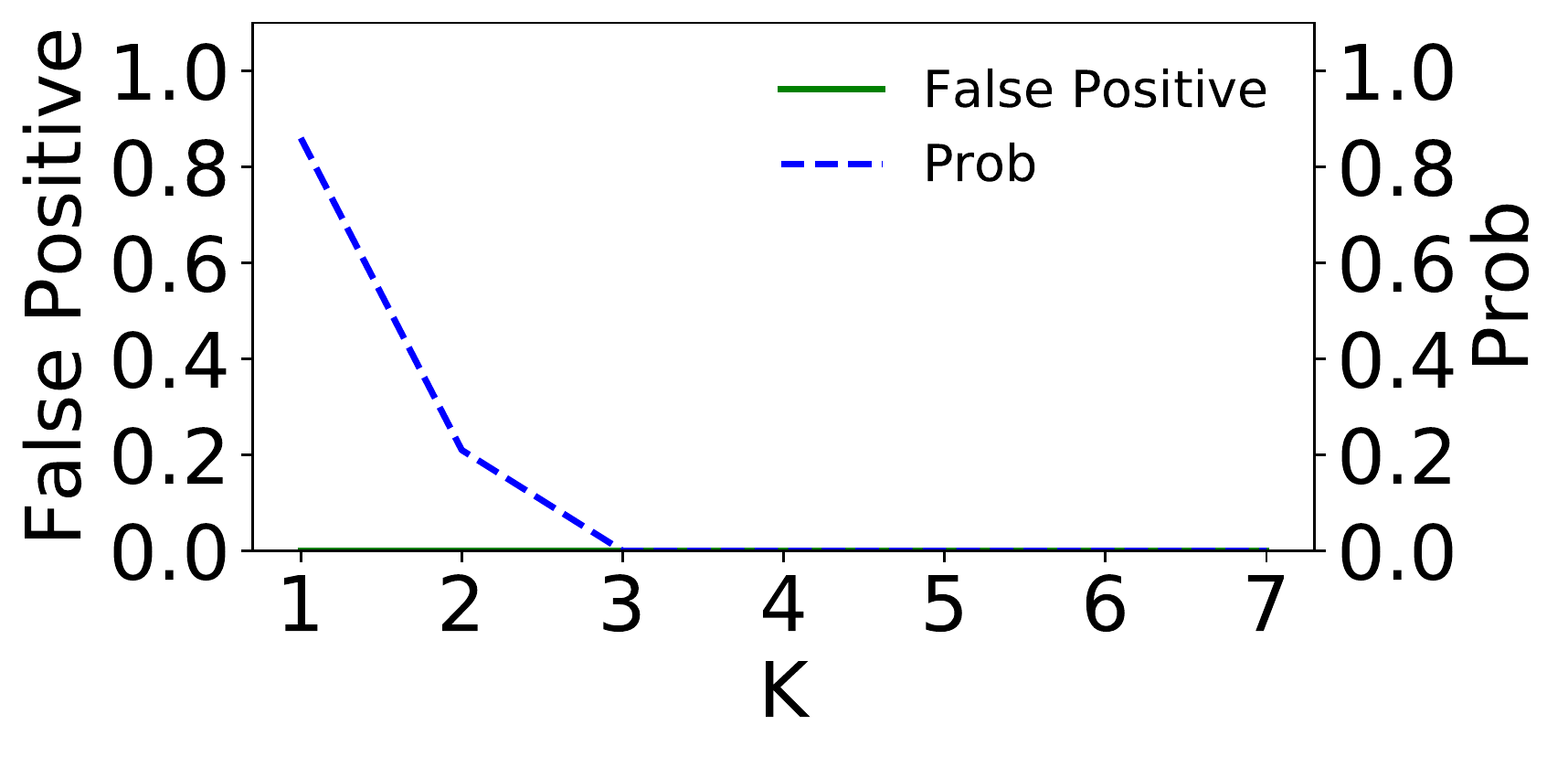}}
\subfigure[CIFAR-10(AP)\label{fig:CIFARAP}]{\includegraphics[width=0.45\columnwidth]{ME_changing/CIFAR_AP.pdf}}
\caption{Overall Detection Performance.}
\label{fig:ME_changing}
\end{figure}

\noindent\textbf{\Huge{$\cdot$}\normalsize  Fig.~\ref{fig:ME_changing}: The value changing pattern of Prob and False Positive rate using different number of models in the model ensemble as discussed in Sec~\ref{sec:exp}.}

\begin{figure}[H]
\centering
\subfigure[MNIST
\label{fig:mb}]{\includegraphics[width=1\columnwidth]{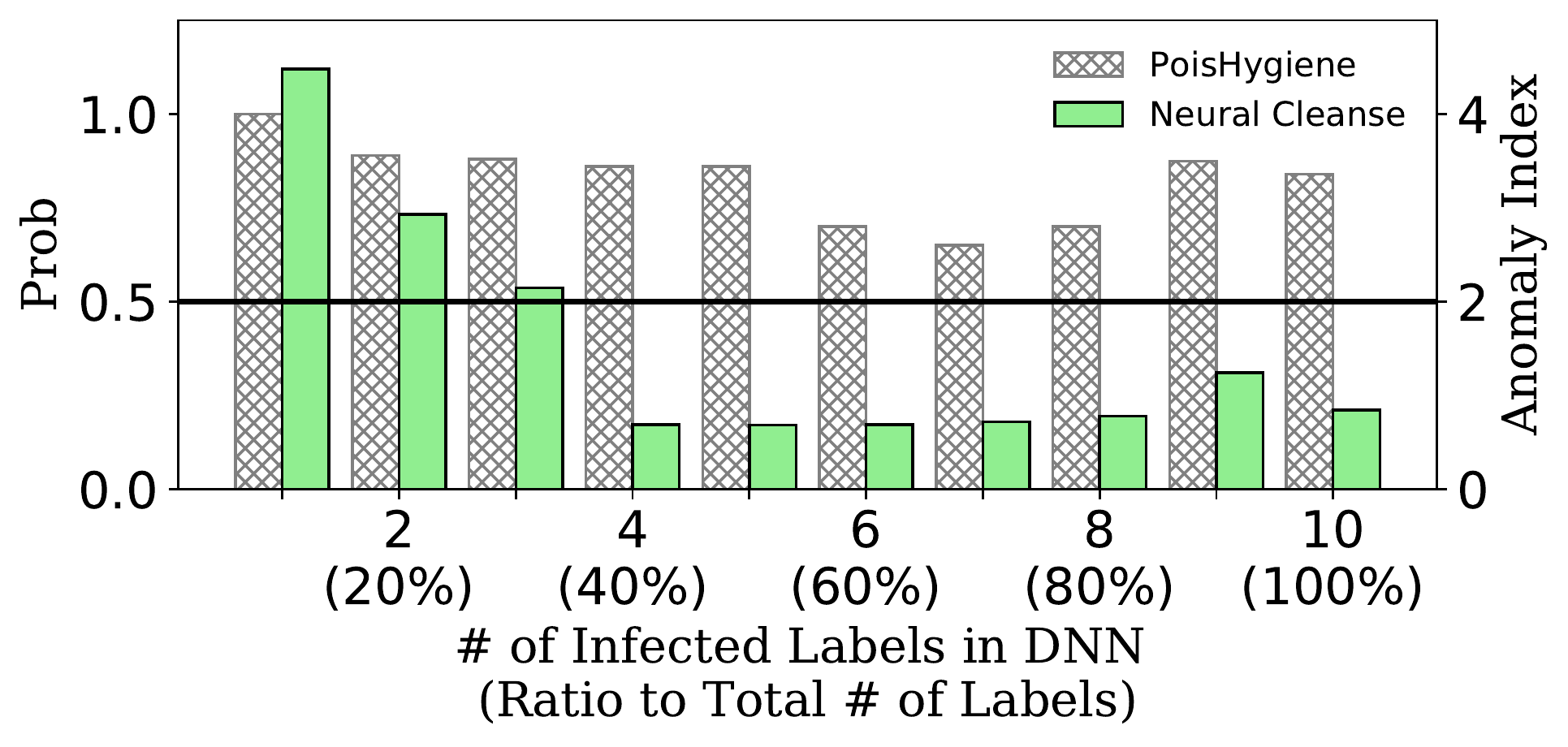}}
\subfigure[FashionMNIST
\label{fig:fmb}]{\includegraphics[width=1\columnwidth]{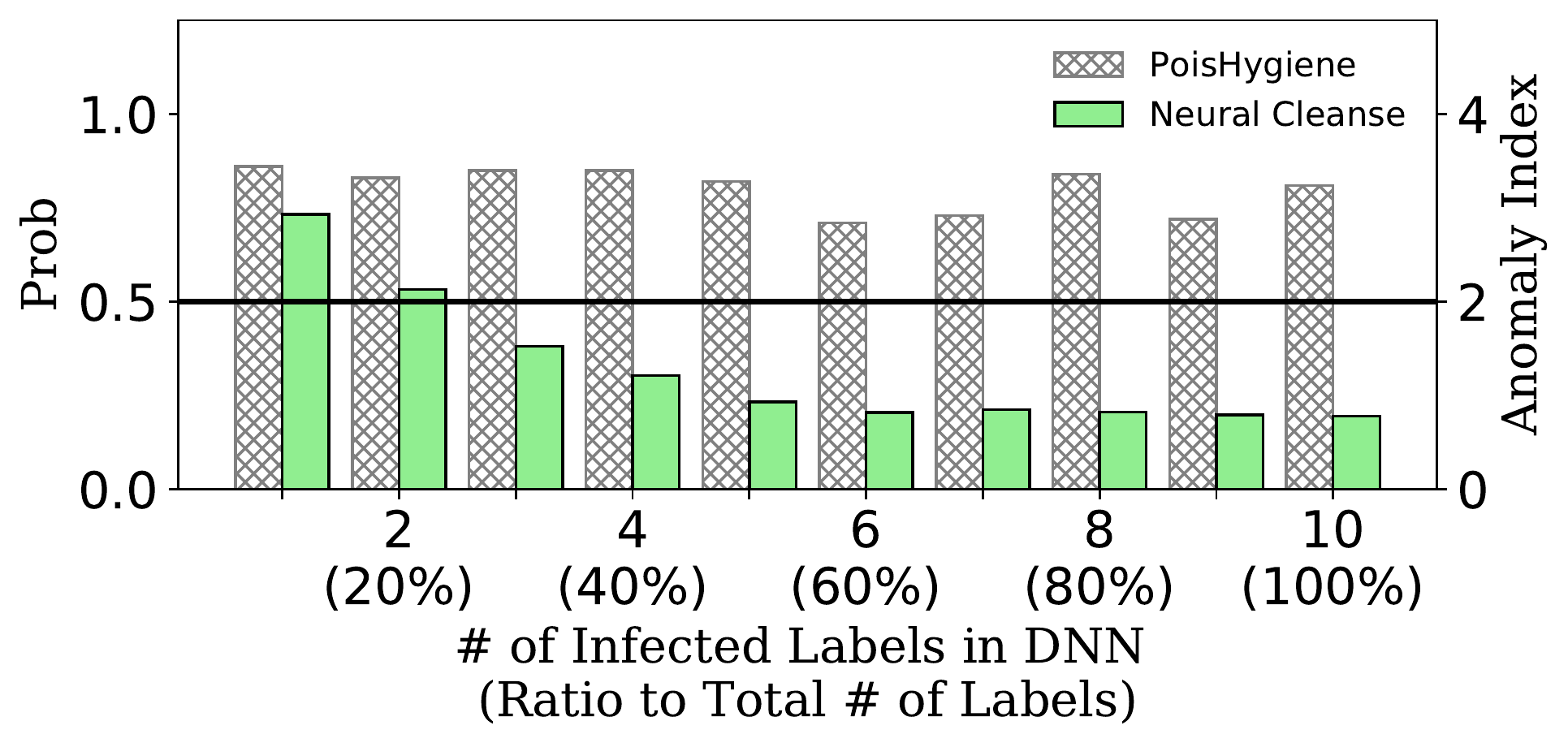}}
\subfigure[CIFAR-10\label{fig:cifarb}]{\includegraphics[width=1\columnwidth]{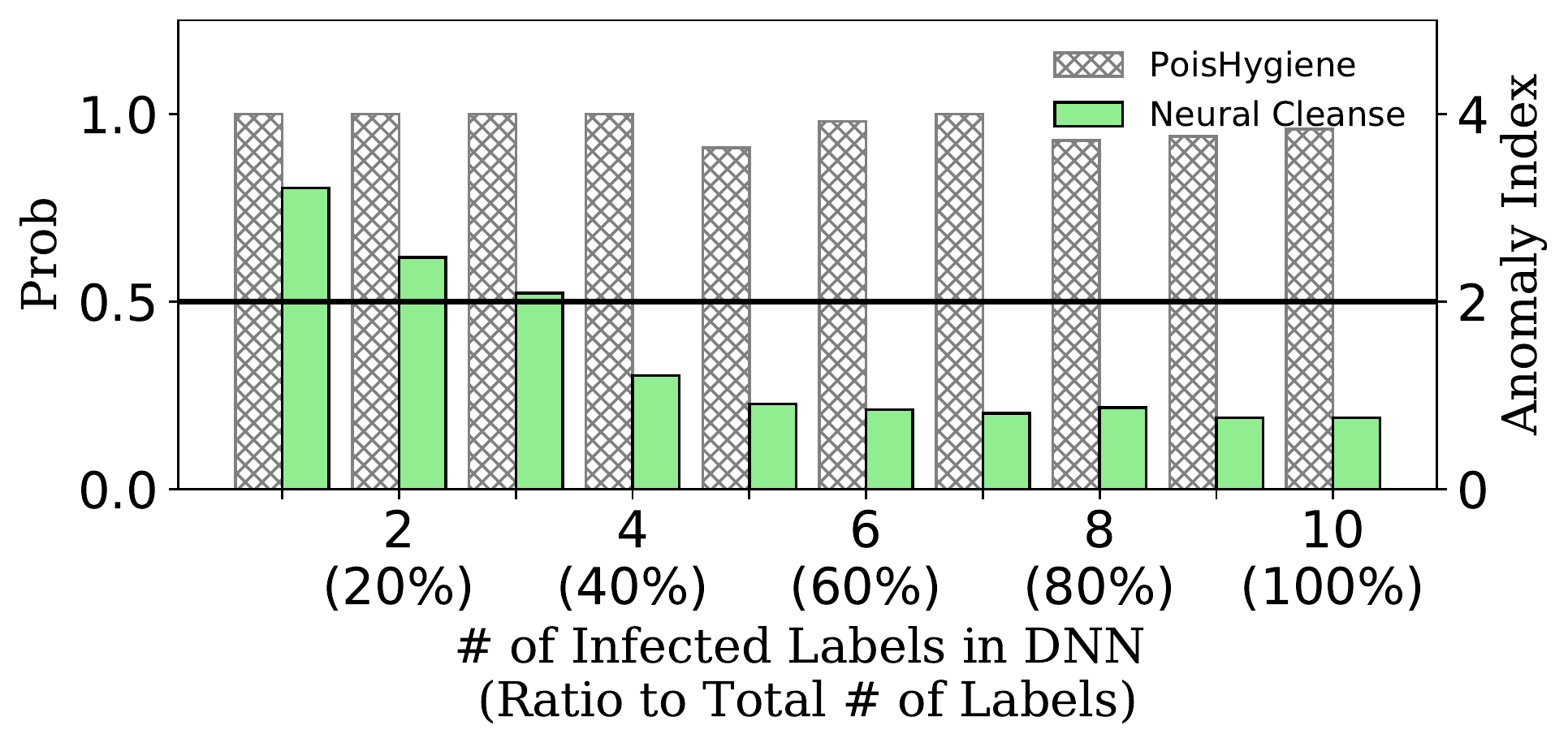}}
\caption{Multi-Label  Detection  Performance  on  various datasets compared to Neural Cleanse.}
\label{fig:mul_compare}
\end{figure}

\noindent\textbf{Fig.~\ref{fig:mul_compare}: Multi-label  detection  performance  on MNIST, Fashion-MNIST, and CIFAR-10 compared to Neural Cleanse as discussed in Sec~\ref{sec:exp}.}

\newpage
\noindent\textbf{Table.~\ref{table:robust_2}: detection performance on MNIST,Fashion-MNIST and CIFAR-10 under complicated triggers scenarios as discussed in Sec~\ref{sec:robustness}.}

\vspace{-2mm}

\begin{table}[H]
\centering
\scalebox{0.4}{
\begin{tabular}{|c|c|c|c|c|}
\hline
\diagbox{\textbf{Dataset}}{Prob(FP Rate)}{\textbf{Trigger shape}}&Circle&Triangle&semi-circle&ellipse\\
\hline
MNIST&$100\%$($0\%$)&$100\%$($0\%$)&$100\%$($0\%$)&$100\%$($0\%$)\\
\hline
Fashion-MNIST&$84\%$($0\%$)&$87\%$(0\%)&$89\%$(0\%)&$87\%$($0\%$)\\
\hline
GTSRB&$93\%$($0\%$)&$92\%$($0\%$)&$93\%$($0\%$)&$93\%$(0\%)\\\hline
CIFAR-10&$100\%$($10\%$)&$100\%$($10\%$)&$100\%$($10\%$))&$100\%$($10\%$)\\
\hline

\end{tabular}
}
\caption{ detection performance on MNIST,Fashion-MNIST and CIFAR-10 under various complicated triggers. The sizes of triggers set under $10\%$ of the entire images.}
\vspace{-4mm}
\label{table:robust_2}
\end{table}

\end{document}